\documentclass[final,1p,times]{elsarticle}

\usepackage{graphicx} 
\usepackage{amsmath,amssymb}

\numberwithin{equation}{section}

\renewcommand{\vec}[1]{\mathbf{#1}}
\newcommand{\nnnl}{\nonumber\\}
\def\beq{\begin{equation}}
\def\eeq{\end{equation}}
\newcommand{\mpi}{M_\pi^2}
\newcommand{\mpin}{M_{\pi^0}^2}
\newcommand{\Order}{\mathcal{O}}
\newcommand{\M}{\mathcal{M}}
\newcommand{\scs}{\, , \,}
\newcommand{\fs}{\, . \,}
\newcommand{\la}{\langle}
\newcommand{\ra}{\rangle}
\newcommand{\bra}{\right\rangle}
\newcommand{\bla}{\left\langle}
\newcommand{\word}[1]{{\mbox{{#1}\,}}}
\newcommand{\eps}{\epsilon}

\journal{Nuclear Physics B}

\begin{document}

\begin{frontmatter}

\title{Cusps in $K\to 3\pi$ decays: a theoretical framework}

\author[Bern]{J\"urg Gasser}\ead{gasser@itp.unibe.ch}
\author[Bonn]{Bastian Kubis}\ead{kubis@hiskp.uni-bonn.de}
\author[Bonn]{Akaki Rusetsky}\ead{rusetsky@hiskp.uni-bonn.de}
\address[Bern]{Albert Einstein Center for Fundamental Physics,
Institute for Theoretical Physics, University of Bern,\\
Sidlerstr.\ 5, CH-3012 Bern, Switzerland}
\address[Bonn]{Helmholtz-Institut f\"ur Strahlen- und Kernphysik and Bethe Center for Theoretical Physics,
               Universit\"at  Bonn, Nussallee 14--16, D-53115 Bonn, Germany}
\begin{abstract}
Based on the analysis of 6.031$\times 10^7 K^\pm\to \pi^0\pi^0\pi^\pm$ decays, 
the NA48/2 collaboration has  recently determined the S-wave $\pi\pi$ 
scattering lengths $a_0 - a_2$ with high 
precision. In addition, the  scattering length  $a_2$ has been 
independently measured, although less precisely so.
The present article discusses in detail one of the  
theoretical frameworks  used in the data analysis.
\end{abstract}


\end{frontmatter}

\section{Introduction}
\label{sec:introduction}

Half a century ago, Budini and Fonda~\cite{budinifonda}
 investigated  the threshold singularities in $K^\pm\to \pi^0\pi^0\pi^\pm$ decays  
and showed that $\pi\pi$ rescattering generates a cusp in the partial decay
rate $d\Gamma/d{E_{\pi^\pm}}$. 
The strength of the cusp   is determined by the amplitude 
for the reaction $\pi^+\pi^-\to\pi^0\pi^0$. Budini and Fonda  provided an analytic 
 formula for the cusp behavior and pointed out that,
 in principle, this decay allows one to measure the $\pi\pi$  S-wave scattering length
$a_0 - a_2$, where the $a_I$ denote the scattering lengths of definite isospin $I=0,\,2$.
 There were only a handful of $K\to 3\pi$  decays available in those days, 
and it was  impossible for the authors to actually determine $a_0$, $a_2$ 
in this manner. The method was then forgotten and rediscovered 
45~years later by Cabibbo in his seminal work~\cite{cabibbo} on 
the interpretation of the cusp detected in data on $K\to 3\pi$ decays, 
collected by the NA48/2 collaboration~\cite{na48k3pi_I}. 

In the last decade,  spectacular progress has been achieved  concerning the knowledge 
of $\pi\pi$ interactions, in theory as well as  in experiment. 
As for theory,
 the scattering lengths  were predicted with
 percent level accuracy~\cite{Colangelo:2000jc,Colangelo:2001df},
\beq\label{eq:a0a2}
a_0=0.220\pm 0.005\, ,\quad a_2=-0.0444\pm 0.0010\, ,\quad a_0-a_2=0.265\pm 0.004\scs
\eeq
within a framework which combines  Roy equations~\cite{roy} and 
chiral perturbation theory~\cite{weinberg,annphysgl}.
On the experimental side,  progress was mainly achieved through the analysis of 
three specific processes. First, pionium decays into two neutral pions allow one to measure 
$|a_0-a_2|$~\cite{pionium_I,pionium_II}. Experimental results have been 
published by the DIRAC collaboration~\cite{DIRAC}. Second, using Watson's theorem 
and numerical solutions of the Roy equations~\cite{numroy_I,numroy_II}, 
it is possible to determine $a_0$ and $a_2$ 
from $K_{\ell 4}$ decays.
 Experiments with spectacular statistics have been carried out in this channel also in the last decade~\cite{ke4_I,ke4_II,ke4_III}.
 Last, the above mentioned cusp in $K^\pm\to\pi^0\pi^0\pi^\pm$ decays has been investigated through the analysis of 6.031$\times 10^7$ such events, and a high  precision value for $a_0 - a_2$  is now available~\cite{na48k3pi_II}. It confirms
 the chiral prediction very nicely. Indeed, combining $K_{e4}$ decays and the result from the cusp analysis, the most 
recent publication by  the NA48/2 collaboration quotes~\cite{ke4_III}
\begin{align}
\label{eq:a0a2experiment}
a_0&=0.2210\pm 0.0047_{\rm stat} \pm 0.0040_{\rm syst}\scs\nonumber\\
a_2&=-0.0429\pm 0.0044_{\rm stat} \pm 0.0028_{\rm syst}\scs\nonumber\\
a_0-a_2&=0.2639\pm0.0020_{\rm stat}\pm 0.0015_{\rm syst}
\fs
\end{align}
It seems fair to say that the precise values Eq.~\eqref{eq:a0a2experiment}  could only be obtained through a 
combined effort of experiment and theory. Indeed, relating experimental data to the scattering lengths is a nontrivial affair~\cite{kaon_2007,ke4_iso,rus_cd09}, 
and a  precise theoretical description 
of the cusp behavior in the amplitude for $K^\pm\to\pi^0\pi^0\pi^\pm$ turns out to be quite difficult.

In their most recent data analysis~\cite{na48k3pi_II}, the NA48/2 collaboration makes use of decay 
amplitudes constructed along two different frameworks. In the first 
one~\cite{cabibbo,cabibboisidori}, the structure of the singularity at the cusp
is investigated using unitarity, analyticity, and the cluster decomposition.
In addition, an approximation scheme is used, which consists in expanding the
decay amplitude in powers of $\pi\pi$ scattering lengths. The latest work~\cite{cabibboisidori} 
retains effects up to order (scattering lengths)$^2$ and
omits radiative corrections. 

The main purpose of the present article is a detailed description of the second method~\cite{cusp_nrqft_I,cusp_nrqft_II,cusp_nrqft_III}. It uses a Lagrangian framework,
which automatically satisfies unitarity and analyticity constraints and allows
one to include electromagnetic contributions in a standard 
manner~\cite{cusp_nrqft_III}. In order to retain the possibility of 
an expansion in powers of scattering lengths, which
is a very convenient concept, a non-relativistic framework is invoked that
has already proven to be useful in the description of bound states~\cite{caswell}, see  
Ref.~\cite{nrqftphysrep} for a review.
 The formalism has recently also found applications in various, mostly cusp-related studies~\cite{FB_Proc}, such as
$\eta\to3\pi$~\cite{Gullstrom,SKDeta3pi} and $\eta'\to\eta\pi\pi$~\cite{etaprime} decays,
as well as in near-threshold pion photo- and electroproduction on the nucleon~\cite{FuhrerPhoto,FuhrerElectro}.
 The amplitudes for   $K\to 3\pi$ decays have been evaluated 
within other frameworks as well. We shall come back to these  in Sect.~\ref{sec:comparison}.

The two main difficulties  we were faced with  to evaluate the pertinent
amplitudes are the following. 
\begin{itemize}
\item
The huge statistics available require a very precise theoretical description of the decay amplitude, 
which allows to determine the scattering lengths from data with high precision as well. 
In particular, it is mandatory 
to include the effects of real and virtual photons 
[``radiative corrections''].
 \item
The decay amplitudes can be calculated only within a certain approximation. It is therefore mandatory to
 set up a power counting scheme that permits to quantify the neglected effects 
in an algebraic manner.
\end{itemize}
As far as we can see, quantum field theory is the only method that allows to satisfy these
two requirements. Independently of the method used, 
one is in addition faced with the problem that 
the decay amplitudes for $K\to3\pi$ are beset with leading Landau 
singularities~\cite{anomalous_I,anomalous_II,anomalous_III,landau_I,ELOP,nakanishi,kaellen}.\footnote{In 
the following, we adhere to the notation used in Refs.~\cite{ELOP,ferroglia} and use
``{\it anomalous threshold}'' as a synonym for ``leading Landau singularity''.}
This does not come as a surprise, because we are dealing with unstable particles here. 

Our article is organized as follows. In Sect.~\ref{sec:covariant}, a covariant, non-relativistic quantum field theory
framework is set up and applied in   Sect.~\ref{sec:pipi} to $\pi\pi$ scattering, 
which plays a crucial role in the analysis of final-state interactions in $K\to 3\pi$ decays. 
In Sect.~\ref{sec:nr_k3pi} we present the Lagrangian for $K\to 3\pi$ decays and determine the tree contributions, 
while one-loop effects (two-loop effects) are treated in Sect.~\ref{sec:oneloop} 
(Sect.~\ref{sec:2loops}). In the latter section we also evaluate the pertinent
two-loop integrals in the non-relativistic theory in some detail. Vertices with six-pion couplings are shown 
to be of no relevance at the present accuracy in Sect.~\ref{sec:6-particles}, 
while we present the structure of the complete amplitude at two-loop order in Sect.~\ref{sec:finalresult}. 
A comparison with other approaches to $K\to 3\pi$  decays is provided in Sect.~\ref{sec:comparison}, 
and a summary and concluding remarks are given in Sect.~\ref{sec:summary}. 
We have relegated  technical aspects of our work to various Appendices. In particular,  in
\ref{app:Ampa2eps4} we provide the complete results for the two-loop amplitude. 
The explicit expressions for the pertinent two-loop integrals are given in  
\ref{app:integralsFF1F2}, and their holomorphic properties are discussed in 
\ref{app:analytic}. In order to make certain that we were not lead astray with the non-relativistic framework, we 
have evaluated several of the emerging two-loop graphs in standard relativistic quantum field theory as well. In  
particular, in \ref{app:relequal}, the case of relativistic two-loop integrals with equal pion masses is considered, 
while loops containing different pion masses are investigated in 
\ref{app:reldiff}. Landau singularities are considered  in 
\ref{app:landau}, and a comparison with the non-relativistic loop integrals
 is performed in 
\ref{app:compare_nrrel}. 
Finally, we comment on the decomposition of the amplitudes in singular and non-singular parts, 
as used in Ref.~\cite{cabibboisidori}, in \ref{app:decomposition}.

\section{Covariant non-relativistic framework}
\label{sec:covariant}
In this section, we construct a non-relativistic effective 
field theory for the two-particle sector,
which will later be used to describe pion--pion scattering. 
In order to simplify the discussion, here we first give a 
formulation for the case of a single self-interacting scalar field 
$\Phi(x)$ with  mass $M$.
In the following, the theory 
will be extended to describe three-particle decays. The main difference
to the conventional non-relativistic formalism
is that our approach as developed here --- despite the fact that we deal with a
{\em non-relativistic} framework --- 
yields {\em explicitly invariant} two-body scattering
amplitudes.\footnote{For an elementary introduction to the 
essentials of the conventional setting in the context of the $\pi\pi$ scattering, 
see, e.g., Ref.~\cite{nrqftphysrep} and references therein.}
The present form turns out to be very convenient to describe
three-body decays, since, in particular, 
 the location of the singularities in the two-body 
subsystems (which are generally not in the rest frame)
coincide with the exact relativistic values to all orders in the
non-relativistic expansion. In the conventional setting, this can be achieved
only perturbatively, order by order.
Such a course of action is not wrong, but very cumbersome and certainly 
not elegant. 

In a non-relativistic theory, one assumes that the momenta of all particles
are much smaller than their masses. The question of the 
{\em domain of applicability} of the non-relativistic description 
is very subtle and depends on the dynamics of the particles 
considered.\footnote{We remind the reader 
that  we are designing an approach with the intention to deal with the rescattering of pions 
in the final state of $K\to 3\pi$ decays. However, as it is easy to check,
the maximal momenta of the pions in this decay 
are of the order of the pion mass and thus the applicability of the 
non-relativistic theory in this case is not {\em a priori} clear.} 
This issue will be discussed in detail in the following sections.
For the moment we merely assume that using the non-relativistic framework
can be justified, and proceed with its precise formulation.

Constructing an effective field theory implies setting certain counting rules.
We formalize the definition of the non-relativistic domain
by introducing a generic small parameter $\epsilon$ and 
postulating the following counting rules in this parameter.
\begin{itemize}
\item
The mass $M$ is to be counted as $\Order(1)$;
\item
all 3-momenta are counted as $|{\bf p}_i|=\Order(\epsilon)$;
\item
consequently, the kinetic energies are counted as 
$T_i=p_i^0-M=\Order(\epsilon^2)$. 
\end{itemize}

The modified approach differs from the conventional one in
two aspects. First, the usual non-relativistic propagator  
has a pole at $p^0=M+{\bf p}^2/2M$ that corresponds to the 
non-relativistic dispersion law. The higher-order corrections to this
dispersion law are treated perturbatively, so that the results of loop calculations with the 
non-relativistic propagators can be made Lorentz-invariant only approximately, at a given order in the expansion in $M^{-1}$.
 In the modified framework, these
higher-order corrections to the one-particle propagator are summed up,
leading to the relativistic dispersion law $p^0=w({\bf p})
\doteq \sqrt{{\bf p}^2+M^2}$. 

The second modification is related to the matching of the 
non-relativistic and relativistic theories.
In the conventional non-relativistic approach, the matching condition
between two-body scattering amplitudes, which
fixes the values of the non-relativistic couplings in terms of the parameters of
the underlying relativistic theory, reads (see, e.g., Ref.~\cite{nrqftphysrep}) 
\beq\label{eq:match1}
\prod_{i=1}^4 (2w({\bf p}_i))^{1/2}
T_{NR}({\bf p}_3,{\bf p}_4;{\bf p}_1,{\bf p}_2) 
=T_R({\bf p}_3,{\bf p}_4;{\bf p}_1,{\bf p}_2)\, ,
\eeq
where the subscripts $NR$ ($R$) stand for the non-relativistic (relativistic) 
amplitudes. The additional factors $(2w({\bf p}_i))^{1/2}$ for each external leg
account for the different normalization of the non-relativistic and relativistic
states.

Since Eq.~\eqref{eq:match1} is not explicitly covariant, the matching condition
is different in different reference frames. For simplicity, we consider
matching at threshold, where the relativistic amplitude is merely a constant
$T_R({\bf p}_3,{\bf p}_4;{\bf p}_1,{\bf p}_2)\bigr|_{\rm thr}=A$ {\em in all reference frames}.
On the other hand, 
the non-relativistic amplitude at threshold in different frames, 
obtained by expanding Eq.~\eqref{eq:match1} in momenta, is given by
\beq\label{eq:TNR}
T_{NR}({\bf p}_3,{\bf p}_4;{\bf p}_1,{\bf p}_2)\biggr|_{\rm thr}
=\frac{A}{4M^2}-\frac{A}{16M^4}\,{\bf P}^2+\ldots\, ,
\eeq
where ${\bf P}$ denotes the total 3-momentum of the particles 1 and 2 at 
threshold. Let us now limit ourselves to the tree approximation and
suppose that one wishes to write down the non-relativistic
effective Lagrangian that reproduces Eq.~\eqref{eq:TNR} in this approximation.
Such a Lagrangian would have to consist of an infinite tower of 
operators, whose couplings are determined by a single constant $A$.
This again means that Lorentz invariance can be taken into account 
only perturbatively, order by order in the expansion in $M^{-1}$.
Once more this procedure, albeit formally correct, looks rather awkward
and renders higher-order calculations cumbersome.

In the particular case considered above, it is clear that
the problem disappears if we arrange the non-relativistic theory 
in a way such that
the overall non-invariant factor on the left-hand side of
Eq.~\eqref{eq:match1} disappears. 
This can be achieved by a non-local
 rescaling of the non-relativistic field $\Phi(x)\to\sqrt{2W}\,\Phi(x)$, where
$W=\sqrt{M^2-\Delta}$\,. In the modified theory, the normalization
of the 1-particle states is given by the relativistic
expression $\langle {\bf p}|{\bf q}\rangle=2w({\bf p})(2\pi)^3\delta^3({\bf p}-{\bf q})$,
whereas the matrix element of the free 
field operator between the vacuum and the
one-particle state is normalized to unity: 
$\langle 0|\Phi(0)|{\bf p}\rangle=1$. The Lagrangian with the rescaled field
is given by
\beq
{\cal L}=\Phi^\dagger 2W( i\partial_t- W) \Phi 
+C(\Phi^\dagger)^2\Phi^2+\ldots\, ,
\eeq
where $C$ denotes the lowest-order ($\Order(1)$) four-particle coupling and 
the ellipsis stands for four-point interactions with derivatives
(these terms count as $\Order(\epsilon^2)$ and higher). 
Note that the above 
Lagrangian is non-local since it contains square roots of a
 differential operator. Heavy-particle number should be conserved in the 
non-relativistic theory by construction (see, e.g., Ref.~\cite{nrqftphysrep}).
The matching condition in tree approximation
yields  $T_{NR}=4C=A$ {\em in all reference frames} or, in other words,
the truncation of the higher-derivative terms in the Lagrangian does not
render the non-relativistic amplitude non-invariant.

The propagator in the rescaled theory is given by 
\beq\label{eq:propnew}
i\langle 0|T\Phi(x)\Phi^\dagger(y)|0\rangle=\int\frac{d^4p}{(2\pi)^4}\,
\frac{{\rm e}^{-ip(x-y)}}{2w({\bf p})(w({\bf p})-p_0-i\eps)}\, .
\eeq
From now on, we shall not display $i\eps$ in the propagators explicitly.
Our next goal will be to demonstrate that
\begin{itemize}
\item
in the above effective theory, the counting rules in $\epsilon$, established
at tree level, are not destroyed by loop corrections;
\item
the amplitude is explicitly Lorentz-invariant also in the presence of loops.
\end{itemize} 
Since all loop diagrams in the non-relativistic approach to the two-particle sector 
can be expressed through the elementary bubble integral
\beq\label{eq:F}
J(P^0,{\bf P})=\int\frac{d^Dl}{(2\pi)^Di}\,
\frac{1}{2w({\bf l})2w({\bf P}-{\bf l})
(w({\bf l})-l^0)(w({\bf P}-{\bf l})-P^0+l^0)}\, ,
\eeq
we concentrate on this integral below. The quantity $J(P^0,{\bf P})$ is
calculated in an arbitrary reference frame 
characterized by the total 4-momentum $P=(P^0,{\bf P})$.
Calculations are done in dimensional regularization, 
$D$ is the number of space-time dimensions and  $d=D-1$.

To begin with, we note that na\"ive power counting predicts $J(P^0,{\bf P})$
to be of $\Order(\epsilon^{d-2})$: each propagator in the integral scales as $\Order(\epsilon^{-2})$, 
while the integration measure, with $d$ momenta and one energy integration, 
contributes a power $\epsilon^{d+2}$.
It can be checked by straightforward calculation that the loop given 
by Eq.~\eqref{eq:F} violates this power counting prediction. This is a well-known phenomenon,
caused by the presence of the heavy mass scale $M$ in the integrand.
In order to circumvent this problem, usual Feynman 
rules in the effective theory have to be supplemented by some additional 
prescription that annihilates the high-energy contribution 
(coming from the integration momenta of order $M$) to the Feynman
integral. We choose a particular prescription
referred to as the ``threshold expansion''~\cite{Beneke} (see also Refs.~\cite{nrqftphysrep,Antonelli:2000mc}
for a focused discussion of the issue within non-relativistic 
effective theories). The prescription defines how the square roots
present in the particle propagators are handled in the calculations. 
According to the prescription, one
 expands the integrand in a Feynman integral in inverse powers of
the heavy scale $M$, integrates the resulting  series term by term in dimensional
regularization, and finally sums up the results.

Let us see how this prescription can be realized in practice. 
We perform
the integration over $l^0$ in Eq.~\eqref{eq:F} by using Cauchy's theorem and rewrite
$J(P^0,{\bf P})$ as
\beq\label{eq:FF}  
J(P^0,{\bf P})=\int\frac{d^dl}{(2\pi)^d}\,
\frac{1}{4w({\bf l})w({\bf P}-{\bf l})}\, 
\frac{1}{w({\bf l})+w({\bf P}-{\bf l})-P^0}\, .
\eeq
Further, performing the shift ${\bf l}\to{\bf l}+{\bf P}/2$,
we transform the integrand by using the identity
\begin{align}\label{eq:identity}
\frac{1}{4w_aw_b}&\biggl\{-\frac{1}{P^0-w_a-w_b}
-\frac{1}{P^0+w_a+w_b}+\frac{1}{P^0+w_a-w_b}
+\frac{1}{P^0-w_a+w_b}\biggr\} \nnnl
&= \frac{1}{2P^0}\,\frac{1}{{\bf l}^2-({\bf l}{\bf P}/P^0)^2-q_0^2}\, ,
\end{align}
where $w_a=w({\bf P}/2+{\bf l})$, $w_b=w({\bf P}/2-{\bf l})$, $q_0^2=s/4-M^2$,
and $s\doteq P^2= (P^0)^2-{\bf P}^2$.

Next, we investigate how the threshold expansion affects the result
of the integration. According to power counting, 
$w_a-M$, $w_b-M=\Order(\epsilon^2)$ and $P^0-2M=\Order(\epsilon^2)$. Consequently,
expanding the 
last three terms in Eq.~\eqref{eq:identity} in powers of $\epsilon$
generates  polynomials in the integration momentum ${\bf l}$. Recalling that
integrals containing only powers of ${\bf l}$ vanish in 
dimensional regularization, we finally conclude that using 
the threshold expansion enables us to rewrite the integral in Eq.~\eqref{eq:FF} as
\beq\label{eq:FFF}
J(P^0,{\bf P})=\int\frac{d^dl}{(2\pi)^d}\,
\frac{1}{2P^0}\,\frac{1}{{\bf l}^2-({\bf l}{\bf P}/P^0)^2-q_0^2}\, .
\eeq
It is seen that the heavy scale $M$ has disappeared as a result of 
the threshold expansion.

In order to perform the integral in Eq.~\eqref{eq:FFF}, we choose
the first axis along the momentum ${\bf P}$, so that 
${\bf P}=(|{\bf P}|,{\bf 0})$ and ${\bf l}=(l_1,{\bf l}_\perp)$. Rescaling
$l_1\to l_1\,P^0/\sqrt{s}$ and doing the momentum integration, 
we finally arrive at
\beq\label{eq:Fs}
J(P^0,{\bf P})\doteq J(s)=\frac{i}{16\pi}\,\biggl(1-\frac{4M^2}{s}\biggr)^{1/2}+\Order(d-3)\, .
\eeq
Note also that $J(s)$ coincides with the imaginary part of the relativistic 
one-loop integral. 

The non-relativistic amplitude in the absence of derivative couplings is given 
by the bubble sum
\beq
T_{NR}({\bf p}_3,{\bf p}_4;{\bf p}_1,{\bf p}_2)=
4C+8C^2 J(s)+16C^3J(s)^2+\ldots\, .
\eeq
The inclusion of derivative couplings is straightforward.
We restrict ourselves
to order $\epsilon^2$, where the real part of the relativistic scattering
amplitude of two identical particles
can be parameterized in terms of two constants $A$ and $B$, which are
related to the S-wave scattering length and the effective range,
\begin{align}\label{eq:T2}
{\rm Re}\, T_R({\bf p}_3,{\bf p}_4;{\bf p}_1,{\bf p}_2)
&= A+B(s-4M^2)+\Order(\epsilon^4)
\nnnl
&= A+B(p_1p_2+p_3p_4-2M^2)+\Order(\epsilon^4)\, .
\end{align}
The term with two derivatives in the non-relativistic effective Lagrangian
can be directly read off from Eq.~\eqref{eq:T2}, leading to
\beq\label{eq:lag2}
{\cal L}^{(2)}=D\Bigl\{\bigl(W\Phi^\dagger W\Phi^\dagger\Phi^2
+\nabla\Phi^\dagger\nabla\Phi^\dagger\Phi^2-M^2(\Phi^\dagger)^2\Phi^2\bigr)
+h.c.\Bigr\}\, ,
\eeq
with $4D=B+C^3/(64\pi^2M^2)$.
Again, this Lagrangian is non-local, and the factor $W$ should be expanded
in actual calculations.

We summarize our main findings up to this point.
\begin{enumerate}
\item
If the threshold expansion is applied, the one-loop integral 
$J(s)$ counts as $\Order(\epsilon)$,
so the loops do not violate power counting.
\item
The insertion of derivative couplings does not violate power counting.
\item
$J(s)$ is explicitly Lorentz-invariant. Since the tree-level amplitude is also
invariant, so is the scattering amplitude at any given
order in the perturbative expansion. This statement is trivial for
non-derivative couplings only; one can easily ensure that it still holds in the
presence of derivative couplings in the scattering amplitude.
\item
According to Eq.~\eqref{eq:Fs}, $J(s=4M^2)=0$. This means in particular that the coupling 
constant $C$ is proportional to the scattering length {\em to all orders 
in perturbation theory}. In general, the coupling constants 
of the non-relativistic theory are expressed through the effective-range
expansion parameters in the two-particle sector. If in the following this non-relativistic
Lagrangian is used 
to evaluate pion--pion rescattering effects in the three-particle decay in perturbation theory, 
the result will be written in terms of these parameters. This property 
constitutes the major advantage of the non-relativistic approach as compared to a
relativistic framework. For example, the decay 
amplitude calculated in chiral perturbation theory is given as an expansion 
in the quark masses, not in the scattering lengths; if one attempts
to extract the values of the scattering lengths from the data on three-particle
decays, the accuracy of the former representation may not suffice.
\end{enumerate}
The generalization of this method to higher orders and to the case
of non-identical particles with different masses can be performed in a 
straightforward manner.

We conclude this section by a remark about the terminology used.  We
still refer to the above framework as ``non-relativistic,'' albeit
the energies and momenta of particles in this approach obey relativistic
dispersion laws. In our naming scheme, ``non-relativistic theory'' denotes a
theory in which {\em explicit antiparticle degrees of freedom} are absent
(respectively, are included in the couplings of the effective Lagrangian),
and the number of particles is preserved in each interaction vertex.

\section{Non-relativistic approach to \boldmath{$\pi\pi$} scattering}
\label{sec:pipi}

\subsection{Lagrangian and scattering amplitude}

Now we apply the modified non-relativistic framework to the $\pi\pi$
scattering amplitudes. The masses of charged and neutral pions
$M_{\pi^+}\doteq M_\pi$ and $M_{\pi^0}$ are taken to be different,
but virtual photons are not included at this stage. 
Due to the inelastic coupling of the $\pi^+\pi^-$ and the $\pi^0\pi^0$ channels,
a consistent power counting
requires the quantity $\Delta_\pi=M_\pi^2-M_{\pi^0}^2$ to be counted 
as $\Order(\epsilon^2)$.
We consider the
following five physical channels in $\pi^a\pi^b\to\pi^c\pi^d$: $(ab;cd)=$
(1)~$(00;00)$, (2)~$(+0;+0)$, (3)~$(+-;00)$, (4)~$(+-;+-)$, (5)~$(++;++)$.
The Lagrangian takes the form
\beq\label{eq:l_pipi}
{\mathcal L}_{\pi\pi}=\sum_\pm\Phi_\pm^\dagger 2W_\pm\bigl(i\partial_t-
W_\pm\bigr)\Phi_\pm 
+\Phi_0^\dagger 2W_0\bigl(i\partial_t- W_0\bigr)\Phi_0
+\sum_{i=1}^5{\mathcal L}_{i}\, ,
\eeq
where $\Phi_i$ is the non-relativistic pion field operator and
$W_\pm=\sqrt{M_\pi^2-\Delta}$, $ W_0=\sqrt{M_{\pi^0}^2-\Delta}$, with
$\Delta$ the Laplacian.
We furthermore introduce the notation 
\newcommand{\Pp}{\mathcal{P}}
\begin{align}
(\Phi_n)_\mu &=(\Pp_n)_\mu\Phi_n\, ,&
(\Phi_n)_{\mu\nu} &=(\Pp_n)_\mu(\Pp_n)_\nu\Phi_n\, ,&
(\Pp_n)_\mu &=(W_n,-i\nabla)\, , \nnnl
(\Phi_n^\dagger)_\mu &=(\Pp^\dagger_n)_\mu\Phi^\dagger_n\, , &
(\Phi_n^\dagger)_{\mu\nu} &= (\Pp^\dagger_n)_\mu(\Pp^\dagger_n)_\nu\Phi_n^\dagger\, , &
(\Pp_n^\dagger)_\mu&=(W_n,i\nabla)\, ,\label{eq:Pn}
\end{align}
for $n=a,\,b,\,c,\,d$, in order to write the interaction Lagrangian in the form 
\begin{align}\label{eq:l_1-5}
{\mathcal L}_{i}&=
x_iC_i \Bigl(\Phi_c^\dagger\Phi_d^\dagger\Phi_a\Phi_b+h.c.\Bigr) \nnnl
&+ x_iD_i\Bigl\{ (\Phi_c^\dagger)_\mu (\Phi_d^\dagger)^\mu\Phi_a\Phi_b
+\Phi_c^\dagger\Phi_d^\dagger (\Phi_a)_\mu (\Phi_b)^\mu
-h_i\Phi_c^\dagger\Phi_d^\dagger\Phi_a\Phi_b +h.c.\Bigr\}
\nnnl
&+ \frac{u_iE_i}{2}
\Bigl\{\Bigl[\Phi_c^\dagger (\Phi_d^\dagger)^\mu- (\Phi_c^\dagger)^\mu\Phi_d^\dagger \Bigr]
\bigl[(\Phi_a)_\mu\Phi_b-\Phi_a (\Phi_b)_\mu\bigr]
 + h.c.\Bigr\} + \ldots~,
\end{align}
with $h_i=\bar s_i-\frac{1}{2}\,(M_a^2+M_b^2+M_c^2+M_d^2)$\,, where $\bar s_i$ denotes
the physical threshold in the $i$th channel. Explicitly,
$h_1=2M_{\pi^0}^2$, $h_2=2M_\pi M_{\pi^0}$,
$h_3=3M_\pi^2-M_{\pi^0}^2$, $h_4=h_5=2M_\pi^2$.
The ellipsis stands for terms of order $\epsilon^4$.
The low-energy constants $C_i$, $D_i$, $E_i$ are matched to the physical scattering
lengths below.
To simplify the resulting expressions, we have furthermore 
introduced the scaling $x_1=x_5=1/4$, $x_2=x_3=x_4=1$, $u_1=u_3=u_5=0$, 
$u_2=u_4=1$.
Finally, note that we do not discuss local six-pion couplings here,
which can potentially play a role in the rescattering of three-pion final states;
we will comment on these in some detail in Sect.~\ref{sec:6-particles}.

Evaluating the non-relativistic scattering amplitude up to order $\epsilon^2$
with the use of the above Lagrangian, we obtain 
\begin{align}\label{eq:NRpipi}
T_{NR}^{\,00}&=2C_{00}+2D_{00}(s-\bar s_{00})
+2C_{00}^2 J_{00}(s) +4C_x^2 J_{+-}(s)
\nnnl
&+2C_{00}^3(J_{00}(s))^2
+8C_{00}C_x^2J_{00}(s)J_{+-}(s)
+8C_x^2C_{+-}(J_{+-}(s))^2+\ldots\, ,
\nnnl
T_{NR}^{\,+0}&=2C_{+0}+2D_{+0}(s-\bar s_{+0})
+E_{+0}(t-u)
+4C_{+0}^2 J_{+0}(s)
+8C_{+0}^3(J_{+0}(s))^2+\ldots\, ,
\nnnl
T_{NR}^{\,x}&=2C_x+2D_x(s-\bar s_x) 
+4C_xC_{+-}J_{+-}(s) + 2C_x C_{00} J_{00}(s) 
+8C_{+-}^2C_x(J_{+-}(s))^2
\nnnl
&+4(C_{+-}C_xC_{00}+C_x^3)J_{+-}(s)J_{00}(s)
+2C_{00}^2C_x(J_{00}(s))^2+\ldots\, ,
\nnnl
T_{NR}^{\,+-}&=2C_{+-}+2D_{+-}(s-\bar s_{+-})
+E_{+-}(t-u)
+4C_{+-}^2 J_{+-}(s) + 2C_x^2 J_{00}(s)
\nnnl
&+8C_{+-}^3(J_{+-}(s))^2
+8C_{+-}C_x^2J_{+-}(s)J_{00}(s)
+2C_x^2C_{00}(J_{00}(s))^2+\ldots\, ,
\nnnl
T_{NR}^{\,++}&=2C_{++}+2D_{++}(s-\bar s_{++})
+2C_{++}^2 J_{++}(s)
+2C_{++}^3 (J_{++}(s))^2
+\ldots\, ,
\end{align}
where $\bar s_i$ denotes the threshold in the pertinent channel 
$\bar s_{00}=4M_{\pi^0}^2$, $\bar s_{+0}=(M_\pi+M_{\pi^0})^2$,
$\bar s_x=\bar s_{+-}=\bar s_{++}=4M_\pi^2$.
Further, in order to make the expressions more transparent, we have modified the notation
according to
\beq
\{ C_1,C_2,C_3,C_4,C_5\}\to\{ C_{00},C_{+0},C_x,C_{+-},C_{++}\} ~,
\eeq
and similarly for $D_i$ and $E_i$.
Finally, $J_{ab}(s)$ denotes the generalization of the loop function $J(s)$
given by Eqs.~\eqref{eq:F} and \eqref{eq:Fs} 
to the case of unequal masses $M_a$ and $M_b$, which is obtained
from Eq.~\eqref{eq:F} by replacing $w({\bf l})\to w_a({\bf l})$, 
$w({\bf P}-{\bf l})\to w_b({\bf P}-{\bf l})$ and 
$w_{a,b}({\bf k})=(M_{a,b}^2+{\bf k}^2)^{1/2}$. This function 
is equal to
\beq\label{eq:Fs_uneq}
J_{ab}(s)=\frac{i}{16\pi}\, v_{ab}(s) ~, \quad 
v_{ab}^2(s) = \frac{4q_{ab}^2(s)}{s} = \frac{\lambda(s,M_a^2,M_b^2)}{s^2} ~,
\eeq
with the usual K\"all\'en function $\lambda(a,b,c)=a^2+b^2+c^2-2(ab+ac+bc)$.

For the purely elastic channels $(+0)$ and $(++)$, Eq.~\eqref{eq:NRpipi}
can be directly compared to the effective-range expansion of the 
relativistic $\pi\pi$ scattering amplitudes
\begin{align}\label{eq:effr}
T_R^i &= 32\pi \sum_{l=0}^\infty (2l+1)t_l^i P_l(z) ~, \quad i=+0,\,++\nnnl
{\rm Re}\, t_0^i &= A_i + B_i q_i^2 + \Order(\eps^4) ~, \quad 
{\rm Re}\, t_1^i = q_i^2 A_i^P + \Order(\eps^4) ~,
\end{align}
where the $t_l^i$ are the partial waves of the respective (physical) channel $i$, 
$A_i$, $B_i$, $A_i^P$ the corresponding S-wave scattering length and effective range
as well as the P-wave scattering length, and $z=\cos\theta$ 
is the cosine of the center-of-mass scattering angle.
In the isospin symmetric limit at $M_\pi=139.6$~MeV, 
which we denote by $\bar C_i$ etc., the matching relations are of the form
\beq \label{eq:elasticMatch}
2\bar C_{+0}=\bar C_{++} = 16\pi a_2 ~, \quad
2\bigg(\bar D_{+0} - \frac{\bar C_{+0}^3}{(32\pi M_\pi)^2}\bigg) = 
\bar D_{++} - \frac{\bar C_{++}^3}{(32\pi M_\pi)^2} = 4\pi b_2 ~, \quad
\bar E_{+0} = 12\pi a_1 ~,
\eeq
with the $I=1$ P-wave scattering length $a_1 = (0.379\pm 0.005)\times 10^{-1}M_\pi^{-2}$
in addition to the S-wave scattering lengths of definite isospin $I=0,\,2$
quoted in Eq.~\eqref{eq:a0a2}~\cite{Colangelo:2000jc,Colangelo:2001df}.
For the remaining channels, the situation is more complicated due to the coupling
of the $\pi^+\pi^-$ and $\pi^0\pi^0$ channels with their different thresholds.
In the isospin limit, with both thresholds coinciding, one has matching relations
also for these similar to Eq.~\eqref{eq:elasticMatch}, and 
the coefficients $\bar C_i$ and $\bar E_i$ are related to the $\pi\pi$
scattering lengths according to
\begin{align}\label{eq:isospin}
\bar C_{00}&=\frac{16\pi}{3}(a_0+2a_2) ~, & 
\bar C_x   &=\frac{16\pi}{3}(a_2-a_0)  ~, &
\bar C_{+-}&=\frac{8\pi}{3}(2a_0+a_2) ~, &
\bar E_{+-}&= 12\pi a_1 \fs
\end{align}
The relations for the $\bar D_i$ are similar to those displayed in Eq.~\eqref{eq:elasticMatch}.
However, taking the pion masses to be different, the threshold behavior for $\pi\pi$
channels of total charge zero is rather of the form (cf.\ Ref.~\cite{ke4_iso})
\beq
32\pi \,t_0^i = \alpha^i + i v_{+-} \beta^i + i v_{00} \gamma^i + v_{+-}v_{00} \delta^i ~
\eeq
(for the case of an S-wave), where $\alpha^i, \ldots, \delta^i$ are real and analytic functions in the low-energy region
on the real axis, $0<s<16\mpin$.
The above parameterization is an analog of the effective-range expansion
for the $\pi\pi$ amplitudes in the presence of multiple thresholds.
Although we do not make use of these expressions in the following, for illustration 
we show the corresponding functions 
for the $(00)$ channel, up to the order displayed in Eq.~\eqref{eq:NRpipi},
i.e.\ $\Order(a^3\eps^2)$, which read
\begin{align}
\alpha^{00} &=  2\bigg(C_{00}+4D_{00} q_{00}^2 - \frac{1}{(16\pi)^2} \big( C_{00}^3v_{00}^2+4C_{+-}C_x^2v_{+-}^2\big)\bigg) ~, \nnnl
\beta^{00} &= \frac{1}{4\pi}C_x^2 ~, \quad
\gamma^{00} = \frac{1}{8\pi}C_{00}^2 ~, \quad
\delta^{00} = -\frac{1}{32\pi^2}C_{00}C_x^2 ~.
\end{align}
We note in passing that recovering the isospin limit from these coefficients is a delicate procedure:
it is obvious that the second term in the $q^2$-expansion of $\alpha^{00}$ will not 
coincide with the effective range in the isospin limit (up to normalization) 
due to the presence of the term $\delta^{00}$.
The $\beta^i$, $\gamma^i$, $\delta^i$ are not independent, though, but also these
can be expressed in terms of the parameters of the effective range expansion.
The constraints that enable us to determine these parameters can also be 
derived from (multi-channel) unitarity, which must be obeyed by the $\pi\pi$
scattering amplitudes. The effective field-theoretical approach that we use
automatically incorporates unitarity order by order.

In order to take isospin breaking in the leading $\pi\pi$ effective range parameters into account, 
we consider the effective $\Order(p^2)$ Lagrangian of chiral perturbation theory
\beq
{\cal{L}}_{\rm eff}= \frac{F^2}{4} \langle \partial_\mu U 
\partial^\mu U^\dagger +
             \chi U^\dagger + \chi^\dagger U \ra+
              e^2C\la QUQU^\dagger\ra ~,
\eeq
with
$\chi=2B\,\, {\rm diag}(m_u,m_d)\,$, $Q={\rm diag}(2,-1)/3\,$,
and $U$ being the standard $2\times 2$ unitary matrix of the pion field.
Further, at this order,
\beq
M_{\pi^0}^2=M^2=B(m_u+m_d) ~,\quad M_{\pi}^2=M^2+\frac{2e^2C}{F^2}\,.
\eeq
Note that evaluating isospin breaking only at $\Order(p^2)$  is consistent with
neglecting virtual photons in the non-relativistic framework. In order to carry
out the matching of the relativistic and non-relativistic theories at higher 
orders, it is necessary to include photons.
At leading order in chiral perturbation theory one finds~\cite{knechturech}
\beq\label{eq:corr}
C_{00,+0,++}=\bar C_{00,+0,++}(1-\eta)~,\quad
C_x=\bar C_x(1+\eta/3) ~,\quad
C_{+-}=\bar C_{+-}(1+\eta) ~,
\eeq
where
$\eta=\Delta_\pi/M_{\pi}^2=6.5\times 10^{-2}$.  Equation~\eqref{eq:corr} shows that
the threshold amplitudes, which occur in the scattering length expansion, 
are affected by substantial isospin breaking corrections.
Isospin-breaking corrections in the effective ranges and P-wave scattering lengths
(or, more precisely, in the non-relativistic couplings $D_i$ and $E_i$)
do not contribute at the accuracy considered here.  
They have been investigated in Ref.~\cite{SKDeta3pi} and were found to be tiny.

Generalizations to higher loop orders in $\pi\pi$ scattering are straightforward. 
However, they are irrelevant 
for the calculation of $K\to 3\pi$ decays to two loops, as performed below;
in fact, the corrections of $\Order(\bar C_i^3)$ in the matching relations for the 
effective ranges Eq.~\eqref{eq:elasticMatch} are already beyond the order needed
for our purposes, as they constitute $\Order(a^3)$ effects and only enter $K\to3\pi$
at three loops.

\subsection{Convergence of the non-relativistic expansion}
\label{sec:pipiconv}

\begin{figure}
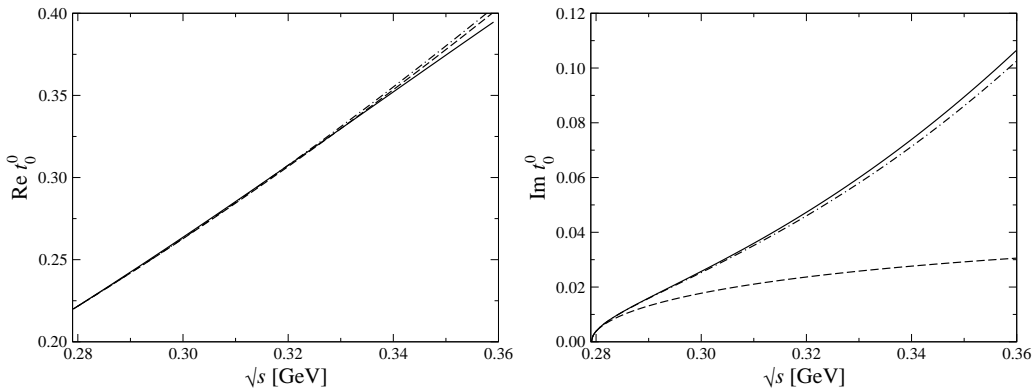

\centering
\includegraphics[width=0.495\linewidth]{ReT00.eps} \hfill
\includegraphics[width=0.495\linewidth]{ImT00.eps}
\caption{Convergence of the $\ell=0$, $I=0$ $\pi\pi$ partial wave in the non-relativistic expansion;
the left panel shows the real, the right panel the imaginary part. 
In both cases, the full lines represent the Roy equation solution~\cite{numroy_I} matched
to chiral perturbation theory~\cite{Colangelo:2000jc,Colangelo:2001df}.
In the left panel, the dashed line is the non-relativistic tree-level amplitude, 
while the dash-dotted curve represents the two-loop representation.
In the right panel, the dashed curve is the one-loop amplitude according to Eq.~\eqref{eq:NRpipi}, 
while the dash-dotted is the improved one-loop representation, see main text.}
\label{fig:T00conv}
\end{figure}

We wish to briefly discuss the convergence of the non-relativistic representation of the $\pi\pi$
amplitude as given in Eq.~\eqref{eq:NRpipi}.  As an example, we show real and imaginary part of 
the $I=0$ S-wave amplitude in Fig.~\ref{fig:T00conv} over the kinematic range accessible
in $K\to 3\pi$ decays, i.e.\ for $4\mpi\leq s\leq(M_K-M_\pi)^2$.  
We compare to the Roy equation solution~\cite{numroy_I} matched to 
chiral perturbation theory~\cite{Colangelo:2000jc,Colangelo:2001df}.
We observe that at such low energies, already the tree amplitude in Eq.~\eqref{eq:NRpipi},
consisting merely of scattering length and effective range term, gives a very good description
of the real part, see Fig.~\ref{fig:T00conv} (left);
including the two-loop corrections hardly changes the amplitude at all.  
This is easily understood by simplifying the $I=0$ S-wave two-loop amplitude 
with the  matching relations analogous to Eq.~\eqref{eq:elasticMatch} according 
to
\beq
{\rm Re}\,t_0^0 = a_0 + b_0 q^2 +  \frac{(a_0)^3q^4}{\mpi(\mpi+q^2)} ~,
\eeq
so it differs from the tree amplitude only by terms of $\Order(a^3q^4)$ and higher.
A further improvement
requires the introduction of a shape parameter $\propto c_0 q^4$.
On the other hand, the description of the imaginary part, given purely by the 
one-loop contributions in Eq.~\eqref{eq:NRpipi}, 
misses the phenomenological amplitude completely, see Fig.~\ref{fig:T00conv} (right).
This is also easy to explain: even in the sense of (only) perturbative unitarity,
the imaginary part of $\Order(a^2\eps)$ corresponds to a real part in the 
scattering length approximation.  We have to improve the imaginary part to 
$\Order(a^2\eps^3)$ and partial $\Order(a^2\eps^5)$ by the replacement
$C_i \to C_i + D_i(s-\bar s_i)$ in the one-loop contributions of Eq.~\eqref{eq:NRpipi};
this improved one-loop representation is then a very good approximation to 
the true imaginary part at the energies relevant for $K\to3\pi$, see also Fig.~\ref{fig:T00conv} (right)

Finally, we wish to anticipate the use of the $\pi\pi$ amplitudes discussed 
above in the cusp analysis of $K\to3\pi$.  
The decay rates will be expressed in terms of the couplings 
$C_i$, $D_i$, $E_i$, $\ldots$ that are related to the physical 
$\pi\pi$ scattering amplitudes.
Once these quantities have been determined from data
(in practice, one may decide to use some of the parameters, for instance effective ranges or P-waves, 
as input, employing their theoretically predicted values), 
they can be related to the S-wave scattering 
lengths using the corrections displayed in Eq.~\eqref{eq:corr}. 
With radiative corrections applied~\cite{cusp_nrqft_III}, these relations must be adapted accordingly.

\section{Non-relativistic approach to $K^+\to 3\pi$ decays}
\label{sec:nr_k3pi}

\subsection{Kinematics}

In this and in the following sections,  we develop in detail a  non-relativistic
framework for $K\to 3\pi$ decays by means of the particular channels 
$K^+(P_K)\to\pi^0(p_1)\pi^0(p_2)\pi^+(p_3)$ and
$K^+(P_K)\to\pi^+(p_1)\pi^+(p_2)\pi^-(p_3)$.
The method can straightforwardly be extended to neutral kaon decays -- for early 
applications of the method, see Refs.~\cite{cusp_nrqft_I,cusp_nrqft_II,cusp_nrqft_III}.

The kinematical variables are defined as follows,
\begin{align}\label{eq:kin1}
s_i&\doteq (P_K-p_i)^2\, ,\quad M_i^2\doteq p_i^2\scs\nnnl
3s_0^0&\doteq M_K^2+M_\pi^2+2M_{\pi^0}^2\, ,\quad
3s_0^+\doteq M_K^2+3M_\pi^2\, ,\quad T_i\doteq p_i^0-M_i\, ,\quad i=1,2,3\, , 
\end{align}
with
\beq
s_1+s_2+s_3=M_K^2+\sum_i M_i^2\fs
\eeq
Further, 
assuming that $i$, $j$, $k$ are all different, we set
\beq\label{eq:DeltaandQ}
\Delta_i^2 \doteq \frac{\lambda(M_K^2,M_i^2,(M_j+M_k)^2)}{4M_K^2}\, ,
\quad
{\bf Q}_i \doteq {\bf p}_j+{\bf p}_k\, ,\quad
Q_i^0\doteq p_j^0+p_k^0\, .
\eeq
In the center-of-mass frame $P_K=(M_K,{\bf 0})$\scs
\beq\label{eq:v}
p_i^0=\frac{M_K^2+M_i^2-s_i}{2M_K}\, ,\quad
{\bf p}_i^2=\frac{\lambda(M_K^2,M_i^2,s_i)}{4M_K^2}\, .
\eeq
As usual, the label $i=3$ is always assigned to the ``odd'' particle, i.e.
to the $\pi^+$ in the neutral mode $K^+\to\pi^0\pi^0\pi^+$ and to the
$\pi^-$ in the charged mode $K^+\to\pi^+\pi^+\pi^-$. Note that the values of the masses 
$M_i$ are channel dependent.

\subsection{Lagrangian and  tree amplitude}

A non-relativistic approach to describe decays $K\to 3\pi$ can be 
justified if the typical kinetic energies $T_i$ of the decay products 
are much smaller than the masses. This can be achieved by considering a world 
where the strange quark mass is taken to be smaller than its actual value.
Then, a consistent counting scheme arises if one 
introduces a formal parameter $\epsilon$ (the same as in the two-particle case)
and counts $T_i$ as a term 
of order $\epsilon^2$, the pion momenta as order $\epsilon$, 
whereas the pion and kaon masses are counted as $\Order(1)$.
From $\sum_iT_i=M_K-\sum_iM_i$, one concludes that the difference
$M_K-\sum_iM_i$ is then a quantity of order $\epsilon^2$ as well. 
In addition, as mentioned in the previous section, 
the pion mass difference $\Delta_\pi$ must also be counted as 
$\Order(\epsilon^2)$. The effective field theory 
framework, which we construct below, enables us
to obtain a systematic expansion of the amplitudes in $\epsilon$.
For sufficiently small $m_s$, the expansion in $\epsilon$ is expected to work
very well.

Together with $\epsilon$, our theory has another expansion parameter,
namely a characteristic size of the $\pi\pi$ threshold parameters,
which we denote generically by $a$.  In particular, the amplitudes in the
non-relativistic framework are given in form of an expansion in several
low-energy couplings $C_i$, $D_i$, $E_i$, which can be expressed in terms of the
threshold parameters of the relativistic $\pi\pi$ scattering amplitude.  We
expect the expansion in $a$ to converge rapidly because of the smallness of
the scattering lengths.  These two expansions are correlated:
because one-loop integrals are of order $\epsilon$, adding a pion loop
generated by a four-pion vertex increases both the order in $a$ and in
$\epsilon$ by one.  A consistent power counting is achieved: to a given
order in $a$ and in $\epsilon$, a well-defined finite number of diagrams
contributes.

Increasing $m_s$ to its physical value again, 
convergence in the $\epsilon$-expansion
is not {\it a priori} evident, 
because $T_i/M_i$ can become as large as 0.4, and the
corresponding maximal momentum ${\bf {|p_i|}}$ is then 
not much smaller than the pion mass.
 However, let us note that the non-relativistic framework is only 
used to correctly reproduce
the non-analytic behavior
of the decay amplitudes in the kinematical variables $s_1$, $s_2$, $s_3$, and to
thus provide a parameterization consistent with unitarity and analyticity
 -- a trivial polynomial part in the amplitudes can be removed by a 
redefinition of the couplings in the Lagrangian.
In addition, from the analysis of the experimental data one  
knows~\cite{cabibboisidori} that in the
whole physical region the real part of the decay 
amplitude can be well approximated by a polynomial in  $s_1$, $s_2$, $s_3$ with
a maximum degree $2$. We interpret this fact as an experimental indication 
for a good convergence of the $\epsilon$-expansion for the 
quantities we are interested in.
In the following, when we also study the behavior of the decay amplitude in the complex energy plane
away from the real axis, we therefore understand
the low-energy region (or, equivalently, the non-relativistic region)
to be determined as a strip enclosing the real axis 
from $s_i=(M_j+M_k)^2$ to $s_i=(M_K-M_i)^2$, and going slightly beyond the
boundaries. The width of the strip 
should be smaller than the hard scale set by the pion mass squared.

We now proceed with the construction of the 
Lagrangian framework. Aside from the Lagrangian ${\mathcal L}_{\pi\pi}$
displayed in Eq.~\eqref{eq:l_pipi} 
that describes $\pi\pi$ final state interactions, 
we need the Lagrangian ${\mathcal L}_K$ which generates genuine $K\to 3\pi$ decays, such that the complete 
Lagrangian is
\beq\label{eq:L_total} 
{\mathcal L} = {\mathcal L}_K+{\mathcal L}_{\pi\pi}\fs
\eeq
At order (scattering lengths)$^2$, the amplitudes are given by
\begin{align}\label{eq:defexpand0}
{\mathcal M}_{00+} &={\mathcal M}_N^{\rm {tree}}
                  +{\mathcal M}_N^{\rm {1-loop}}
                  +{\mathcal M}_N^{\rm {2-loops}} \qquad
[K^+\to\pi^0\pi^0\pi^+] ~, \nnnl
{\mathcal M}_{++-}&=\underbrace{{\mathcal M}_C^{\rm {tree}}}_{{\mathcal L}_K}
                  +\underbrace{{\mathcal M}_C^{\rm {1-loop}}}
_{{\mathcal L}_K\times {\mathcal L}_{\pi\pi}}
                  +\underbrace{{\mathcal M}_C^{\rm {2-loops}}}
_{{\mathcal L}_K\times{\mathcal L}_{\pi\pi}\times {\mathcal L}_{\pi\pi}} \qquad
[K^+\to\pi^+\pi^+\pi^-]\scs
\end{align}
with obvious notation. The Lagrangian ${\mathcal L}_K$ is now chosen such that the tree-level amplitudes 
up to and including $\Order(\epsilon^4)$ become 
 (cf.\ Ref.~\cite{cabibboisidori}, Eqs.~(4.6), (4.7))
\begin{align}\label{eq:par}
{\cal M}_N^{\rm tree}(s_1,s_2,s_3) &=
X_0+X_1(s_3-s_0^0)+X_2(s_3-s_0^0)^2+X_3(s_1-s_2)^2\, ,
\nnnl
{\cal M}_C^{\rm tree}(s_1,s_2,s_3) &=
Y_0+Y_1(s_3-s_0^+)+Y_2(s_3-s_0^+)^2+Y_3(s_1-s_2)^2\, .
\end{align}
We assume $T$-invariance and a hermitian ${\mathcal L}_K$, as a result of which the couplings $X_i$, $Y_i$ are real.
Expressing $s_i$ through $p_i^0$ with the use of Eq.~\eqref{eq:v},
these expressions are  equivalent to
\begin{align}\label{eq:par1}
{\cal M}_N^{\rm tree}(s_1,s_2,s_3) &=
G_0+G_1(p_3^0-M_\pi)+G_2 (p_3^0-M_\pi)^2+
G_3(p_1^0-p_2^0)^2\, ,
\nnnl
{\cal M}_C^{\rm tree}(s_1,s_2,s_3) &=
H_0+H_1(p_3^0-M_\pi)+H_2 (p_3^0-M_\pi)^2+
H_3(p_1^0-p_2^0)^2\, ,
\end{align}
where
\begin{align}\label{eq:GH}
G_0&=X_0+\left((M_K-M_\pi)^2-s_0^0\right)X_1+\left((M_K-M_\pi)^2-s_0^0\right)^2X_2\, ,
\nnnl
G_1&=-2M_KX_1-4M_K\left((M_K-M_\pi)^2-s_0^0\right)X_2\, ,
\quad
G_2=4M_K^2 X_2\, ,\quad G_3=4M_K^2 X_3\, ,
\nnnl
H_0&=Y_0+\left((M_K-M_\pi)^2-s_0^+\right)Y_1+\left((M_K-M_\pi)^2-s_0^+\right)^2Y_2\, ,
\nnnl
H_1&=-2M_KY_1-4M_K\left((M_K-M_\pi)^2-s_0^+\right)Y_2\, ,
\quad
H_2=4M_K^2 Y_2\, ,\quad  H_3=4M_K^2 Y_3\, .
\end{align}
From the expressions Eq.~\eqref{eq:par1} one may read off the pertinent Lagrangian,
\begin{align}\label{eq:lag_K}
{\cal L}_K&=K^\dagger 2W_K(i\partial_t- W_K)K
+\frac{G_0}{2}\left(K^\dagger\Phi_+\Phi_0^2+h.c.\right)
+\frac{G_1}{2}\left(K^\dagger(W_+-M_\pi)\Phi_+\Phi_0^2+h.c.\right)
\nnnl
&+\frac{G_2}{2}\left(K^\dagger(W_+-M_\pi)^2\Phi_+\Phi_0^2+h.c.\right)
+G_3\left(K^\dagger\Phi_+(W_0^2\Phi_0\Phi_0- W_0\Phi_0 W_0\Phi_0) +h.c.\right)
\nnnl
&+\frac{H_0}{2}\left(K^\dagger\Phi_-\Phi_+^2+h.c.\right)
+\frac{H_1}{2}\left(K^\dagger(W_--M_\pi)\Phi_-\Phi_+^2+h.c.\right)
\\
&+\frac{H_2}{2}\left(K^\dagger(W_--M_\pi)^2\Phi_-\Phi_+^2+h.c.\right)
+H_3\left(K^\dagger\Phi_-(W_+^2\Phi_+\Phi_+- W_+\Phi_+W_+\Phi_+) +h.c.\right)+\ldots\, ,\nonumber
\end{align}
where $K$ denotes the non-relativistic field for the $K^+$ meson, 
$W_K=(M_K^2-\Delta)^{1/2}$, and the ellipsis stands for 
higher-order terms in $\epsilon$. Note that, by construction, the above Lagrangian contains all
terms up to and including $\Order(\epsilon^4)$, allowed by the symmetries
in the non-relativistic effective theory.

\section{$K^+\to 3\pi$ decays: pion rescattering to one loop}
\label{sec:oneloop}
\begin{figure}
\centering
\includegraphics[width=0.35\linewidth]{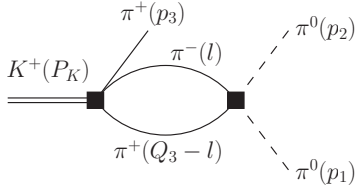}
\caption{One-loop graph with derivative vertices (denoted by filled squares), 
with $Q_3^\mu=(p_1+p_2)^\mu$.}
\label{fig:1loop_der}
\end{figure}

Here, we discuss terms of $ \Order(a)$, generated by one-loop graphs of the type ${\mathcal L}_K\times {\mathcal L}_{\pi\pi}$,
see Fig.~\ref{fig:1loop_der} for a specific example. 
The technique used for the calculation of these loops is described
in Sect.~\ref{sec:covariant} for the case of non-derivative couplings.
The presence of the latter does not change the procedure fundamentally, as we demonstrate 
for one specific example.
Consider the diagram shown in Fig.~\ref{fig:1loop_der}. 
We restrict ourselves to
the part of the amplitude proportional to the coupling $H_1$, which can be written as 
\beq
{\cal M}_N^{H_1}(s_1,s_2,s_3)=
2H_1\big(C_x+D_x(s_3-\bar s_x)\big)
\int\frac{d^dl}{(2\pi)^d}\,
\frac{w({\bf l})-M_\pi}{2w({\bf l})2w({\bf Q}_3-{\bf l})
\left(w({\bf l})+w({\bf Q}_3-{\bf l})-Q_3^0\right)}\scs
\eeq
with $Q_3^\mu=(p_1+p_2)^\mu$. Rewriting the numerator of the integrand as
\beq
w({\bf l})-M_\pi=\left(Q_3^0/2-M_\pi\right)+\frac{1}{2}\left(w({\bf l})+w({\bf Q}_3-{\bf l})-Q_3^0\right)
+\frac{1}{2}\big(w({\bf l})-w({\bf Q}_3-{\bf l})\big)\, ,
\eeq
we see that only the first term yields a non-vanishing contribution 
after the integration, since the second term after expansion in momenta
leads to  dimensionally regularized no-scale integrals, and the third term is
antisymmetric with respect to ${\bf l}\to {\bf Q}_3-{\bf l}$. We finally obtain
\beq
{\cal M}_N^{H_1}(s_1,s_2,s_3) = 
2H_1 \bigg(\frac{Q_3^0}{2}-M_\pi\bigg)\,\bigl(C_x+D_x(s_3-\bar s_x)\bigr)
J_{+-}(s_3)\, ,
\eeq
where the function $J_{+-}(s_3)$ is given in Eq.~\eqref{eq:Fs_uneq}.

The generalization to other derivative couplings is obvious.
 The complete one-loop representation for
the decay amplitudes ${\cal M}_N$ and ${\cal M}_C$, 
up-to-and-including terms of $\Order(a \,\epsilon^5)$, reads
\begin{align}\label{eq:1loop1}
{\cal M}^{\rm (1-loop)}_N(s_1,s_2,s_3)
&= B_{N1}(s_3)J_{+-}(s_3)+B_{N2}(s_3)J_{00}(s_3)
+\bigl\{B_{N3}(s_1,s_2,s_3)J_{+0}(s_1)+(s_1\leftrightarrow s_2)\bigr\}\, ,
\nnnl
{\cal M}^{\rm (1-loop)}_C(s_1,s_2,s_3)&=
B_{C1}(s_3)J_{++}(s_3)
+\bigl\{B_{C2}(s_1,s_2,s_3)J_{+-}(s_1) 
+B_{C3}(s_1)J_{00}(s_1)+(s_1\leftrightarrow s_2)\bigr\} \, ,
\end{align}
where
\begin{align}\label{eq:1loop2}
B_{N1}(s_3)&=2\bigl[ C_x+D_x(s_3-\bar s_x)+F_x(s_3-\bar s_x)^2\bigr]
\biggl\{H_0
+H_1\Bigl(\frac{Q_3^0}{2}-M_\pi\Bigr) \nnnl &
+H_2\biggl[\Bigl(\frac{Q_3^0}{2}-M_\pi\Bigr)^2+\frac{{\bf Q}_3^2}{12}\,
\biggl(1- \frac{4M_\pi^2}{s_3}\biggr)\biggr]
+H_3\biggl[\Bigl(\frac{Q_3^0}{2}-p_3^0\Bigr)^2+\frac{{\bf Q}_3^2}{12}\,
\Bigl(1- \frac{4M_\pi^2}{s_3}\Bigr)\biggr]\biggr\}\, ,
\nnnl
B_{N2}(s_3)&=\bigl[C_{00}+D_{00}(s_3-\bar s_{00})+F_{00}(s_3-\bar s_{00})^2\bigr]
\biggl\{ G_0
+G_1\bigl(p_3^0-M_\pi\bigr) \nnnl &
+G_2\bigl(p_3^0-M_\pi\bigr)^2
+G_3\frac{{\bf Q}_3^2}{3}\Bigl(1-\frac{4M_{\pi^0}^2}{s_3}\Bigr)\biggr\}\, ,
\nnnl
B_{N3}(s_1,s_2,s_3)&=2 \bigl[C_{+0}+D_{+0}(s_1-\bar s_{+0})+F_{+0}(s_1-\bar s_{+0})^2\bigr]
\biggl\{ G_0 
+G_1\biggl[\frac{Q_1^0}{2}\Bigl(1+\frac{\Delta_\pi}{s_1}\Bigr)-M_\pi\biggr]
\nnnl
&+G_2\biggl[\biggl(\frac{Q_1^0}{2}\Bigl(1+\frac{\Delta_\pi}{s_1}\Bigr)
-M_\pi\biggr)^2+\frac{{\bf Q}_1^2}{12s_1^2}\,\lambda\bigl(s_1,M_\pi^2,M_{\pi^0}^2\bigr)\biggr]
\nnnl
&+G_3\biggl[\biggl(\frac{Q_1^0}{2}\Bigl(1-\frac{\Delta_\pi}{s_1}\Bigr)
-p_1^0\biggr)^2+\frac{{\bf Q}_1^2}{12s_1^2}\,\lambda\bigl(s_1,M_\pi^2,M_{\pi^0}^2\bigr)\biggr]\biggr\}
\nnnl
&-\frac{1}{3}\,E_{+0}\frac{q_{23}^2(s_1)}{M_K}\,
\Bigl[s_3-s_2+\frac{\Delta_\pi}{s_1}\bigl(M_K^2-M_{\pi^0}^2\bigr)\Bigr]
\nnnl
&\quad \times
\biggr\{G_1+G_2\biggl[\Bigl(1+\frac{\Delta_\pi}{s_1}\Bigr)Q_1^0-2M_\pi\biggr]
+G_3\biggl[2p_1^0-\Bigl(1-\frac{\Delta_\pi}{s_1}\Bigr)Q_1^0\biggr]\biggr\}
+O\bigl(\Delta_\pi^2\bigr) \, ,
\end{align}
and
\begin{align}\label{eq:1loop3}
B_{C1}(s_3)&=\bigl[C_{++}+D_{++}(s_3-\bar s_{++})+F_{++}(s_3-\bar s_{++})^2\bigr]
\biggl\{ H_0 +H_1\bigl(p_3^0-M_\pi\bigr)
\nnnl &
+H_2\bigl(p_3^0-M_\pi\bigr)^2
+H_3\frac{{\bf Q}_3^2}{3}\Bigl(1-\frac{4M_\pi^2}{s_3}\Bigr)\biggr\}\, ,
\nnnl
B_{C2}(s_1,s_2,s_3)&=2\bigl[C_{+-}+D_{+-}(s_1-\bar s_{+-})+F_{+-}(s_1-\bar s_{+-})^2\bigr]
\biggl\{ H_0
+H_1\biggl[\frac{Q_1^0}{2}-M_\pi\biggr] \nnnl &
+H_2\biggl[\biggl(\frac{Q_1^0}{2}-M_\pi\biggr)^2
+\frac{{\bf Q}_1^2}{12}\,\Bigl(1-\frac{4M_\pi^2}{s_1}\Bigr)\biggr]
+H_3\biggl[\biggl(\frac{Q_1^0}{2}-p_1^0\biggr)^2
+\frac{{\bf Q}_1^2}{12}\,\Bigl(1-\frac{4M_\pi^2}{s_1}\Bigr)\biggr]\biggr\}
\nnnl
&-\frac{1}{3}\,E_{+-}\frac{q_{23}^2(s_1)}{M_K}\,(s_3-s_2)
\biggr\{H_1+H_2\bigl[Q_1^0-2M_\pi\bigr]+H_3\bigl[2p_1^0-Q_1^0\bigr]\biggr\}
\, ,
\nnnl
B_{C3}(s_1)&=\bigl[C_x+D_x(s_1-\bar s_x)+F_x(s_1-\bar s_x)^2 \bigr]
\biggl\{ G_0 +G_1\bigl(p_1^0-M_\pi\bigr)
\nnnl &
+G_2\bigl(p_1^0-M_\pi\bigr)^2
+G_3\frac{{\bf Q}_1^2}{3}\Bigl(1-\frac{4M_{\pi^0}^2}{s_1}\Bigr)\biggr\}\, .
\end{align}
Here we have added terms of $\Order(\epsilon^4)$ ($\propto F_i$, in a canonical
extension of our notation) to the $\pi\pi$ amplitudes,
without having them formally introduced on the Lagrangian level. 
They contribute  terms of order $(s-\bar s)^2$ 
to the S-waves.
Only some terms of $\Order(\Delta_\pi^2)$ have been neglected in Eq.~\eqref{eq:1loop2}.
Note in particular that there are no contributions of $\pi\pi$ D-waves at this order.

\section{$K^+\to 3\pi$ decays: pion rescattering to two loops} 
\label{sec:2loops}

\subsection{The diagrams}

\begin{figure}
\centering
\includegraphics[width=0.9\linewidth]{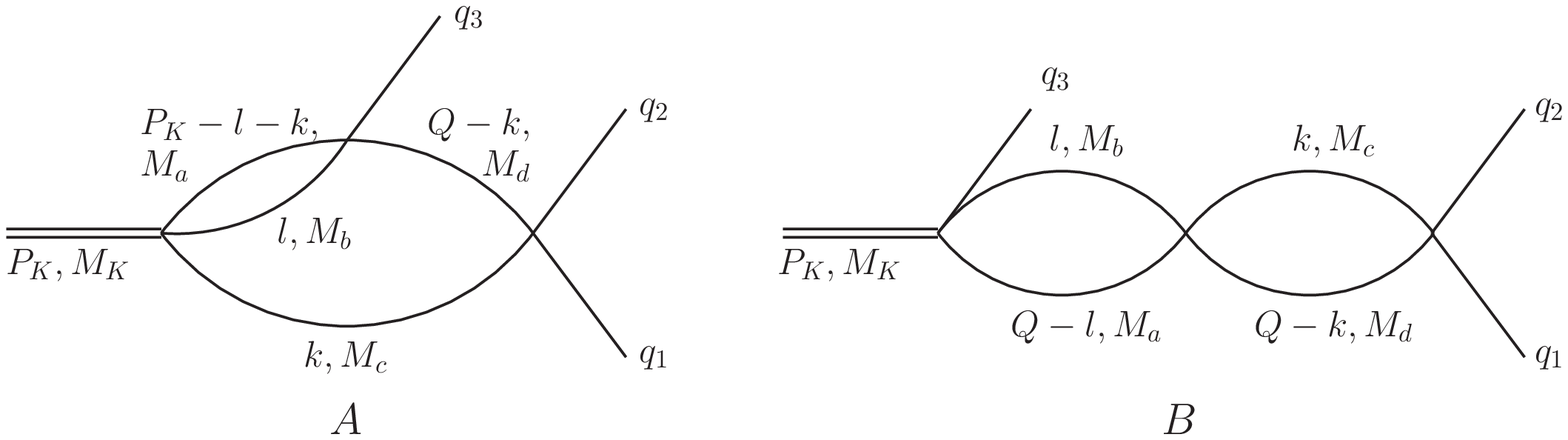}
\caption{Two topologically distinct non-relativistic two-loop graphs 
describing the final-state $\pi\pi$ rescattering in the decay $K\to 3\pi$, with
$Q^\mu=q_1^\mu+q_2^\mu$. 
The positions of the leading Landau singularities of graph~A are discussed in \ref{app:landau}.}
\label{fig:2loop_text}
\end{figure}

There are two topologically distinct two-loop graphs that describe
pion--pion rescattering in the final state, see Fig.~\ref{fig:2loop_text}.
In order to 
ease notation, we set $Q^\mu\doteq q_1^\mu+q_2^\mu$,
$s\doteq Q^2$. Further,
\beq
Q^0=\frac{M_K^2+s-q_3^2}{2M_K}\, ,\quad\quad
{\bf Q}^2=\frac{\lambda(M_K^2,s,q_3^2)}{4M_K^2}\, .
\eeq
In addition, throughout this chapter we consider the case of the
non-derivative couplings only. In this case, in the rest frame of the
kaon it is possible to express both diagrams shown in 
Fig.~\ref{fig:2loop_text} in terms of a single 
variable $s$ and one has
\beq\label{eq:2loop_AB}
{\cal M}_{N,C}^{\rm 2-loops}(s)={\cal M}_{N,C}^A(s)+{\cal M}_{N,C}^B(s)\, .
\eeq
The diagram in  Fig.~\ref{fig:2loop_text}B, apart from the prefactor 
containing coupling constants, is given by a trivial product of two one-loop
diagrams, which were already given in Eq.~\eqref{eq:Fs_uneq},
\beq
{\cal M}_{N,C}^B(s)\propto J_{ab}(s)J_{cd}(s)\, .
\eeq
Obviously, in the non-relativistic framework ${\cal M}_{N,C}^B(s)$ is therefore 
ultraviolet finite and of order $\epsilon^2$. 

In the remaining part of this section we discuss the evaluation of the
non-trivial contribution ${\cal M}_{N,C}^A(s)$, which 
stems from Fig.~\ref{fig:2loop_text}A. It is in particular shown that 
-- up to a low-energy polynomial and an imaginary part which does not
contribute at the accuracy we are working -- 
${\cal M}_{N,C}^A(s)$ is given by the function $F(M_a,M_b,M_c,M_d;s)$
defined in Eq.~\eqref{eq:Fint}. So, the reader
not interested in the details of the derivation may directly proceed to
Eq.~\eqref{eq:Fint}.

\subsection{Evaluation of the generic two-loop function}

The quantity ${\cal M}_{N,C}^A(s)$
is proportional to the generic two-loop function
\begin{align}
{\cal M}(s)
&=\int\frac{d^D l}{(2\pi)^Di}\,\frac{d^D k}{(2\pi)^Di}\,
\frac{1}{2w_a(-{\bf l}-{\bf k})}\, 
\frac{1}{w_a(-{\bf l}-{\bf k})-M_K+l^0+k^0}\,
\frac{1}{2w_b({\bf l})}\, 
\frac{1}{w_b({\bf l})-l^0}\,
\nnnl
&\times\frac{1}{2w_c({\bf k})}\, 
\frac{1}{w_c({\bf k})-k^0}\,
\frac{1}{2w_d({\bf Q}-{\bf k})}\, 
\frac{1}{w_d({\bf Q}-{\bf k})-Q^0+k^0}\, .
\end{align}
Performing the integration over the fourth components of the momenta, 
we can rewrite the above
integral in the rest frame of the kaon as
\beq\label{eq:twoloop_F} 
{\cal M}(s)
=\int \frac{d^dk}{(2\pi)^d}\, \frac{1}{2w_c({\bf k})2w_d({\bf Q}-{\bf k})}\,
\frac{J_{ab}(M_K-w_c({\bf k}),-{\bf k})}{w_c({\bf k})+w_d({\bf Q}-{\bf k})-Q_0}\, ,
\eeq
where $J_{ab}(L_0,{\bf L})$ denotes the inner one-loop integral
\beq\label{eq:F_L}
J_{ab}(L_0,{\bf L})=
\int \frac{d^dl}{(2\pi)^d}\, \frac{1}{2w_a({\bf L}-{\bf l})2w_b({\bf l})}\,
\frac{1}{w_a({\bf L}-{\bf l})+w_b({\bf l})-L_0}\, .
\eeq
In the calculation of the inner integral, we proceed as in the case of equal
masses, see Eqs.~\eqref{eq:F}--\eqref{eq:Fs}: we first shift the integration variable according to
\beq\label{eq:shift}
{\bf l}\to {\bf l}+\frac{1}{2}(1+\delta_L){\bf L}
\, ,\quad\quad
\delta_L=-\frac{M_a^2-M_b^2}{L^2}
\, ,\quad 
L^2=(L^0)^2-{\bf L}^2\, .
\eeq
Further, using the identity
\begin{align}\label{eq:trick}
&\frac{1}{4w_aw_b}\biggl\{-\frac{1}{L_0-w_a-w_b}-\frac{1}{L^0+w_a+w_b}
+\frac{1}{L^0+w_a-w_b}+\frac{1}{L^0-w_a+w_b}\biggr\} \nnnl
&=
\frac{1}{2L^0}\,\frac{1}{{\bf l}^2-({\bf l}{\bf L}/L^0)^2-k_0^2}\, ,
\end{align}
where
\beq
w_a=\sqrt{M_a^2+\biggl(\frac{1}{2}\,(1-\delta_L){\bf L}-{\bf l}\biggr)^2}\, ,
\quad
w_b=\sqrt{M_b^2+\biggl(\frac{1}{2}\,(1+\delta_L){\bf L}+{\bf l}\biggr)^2}\, ,
\quad
k_0^2=\frac{\lambda(L^2,M_a^2,M_b^2)}{4L^2}\, ,
\eeq
applying threshold expansion and rescaling the component $l_1$
(see the discussion after Eq.~\eqref{eq:FFF}), we find
\beq
J_{ab}(L^0,{\bf L})=\frac{1}{2\sqrt{L^2}}\int\frac{d^dl}{(2\pi)^d}\,
\frac{1}{{\bf l}^2-k_0^2}\, .
\eeq
Finally, performing the ${\bf l}$ integration, we reproduce Eq.~\eqref{eq:Fs_uneq} in the
limit $d\to 3$.

In the two-loop diagram 
Eq.~\eqref{eq:twoloop_F} the components of the momentum $L^\mu$,
 corresponding to the one-loop subdiagram, are equal to
$L^0=M_K-w_c({\bf k})$ and ${\bf L}=-{\bf k}$. One may further ensure that
$k_0^2=A_k(\Delta^2-{\bf k}^2)$, where the quantities $\Delta^2$ and $A_k$ are given by
\begin{align}
\Delta^2&=\frac{\lambda(M_K^2,M_c^2,(M_a+M_b)^2)}{4M_K^2}\, ,
\quad
A_k=\frac{M_K^2}{2(M_K^2+M_c^2)-(M_a+M_b)^2-s_k}
\biggl(1-\frac{(M_a-M_b)^2}{s_k}\biggr)\, ,
\nnnl
s_k&=M_K^2+M_c^2-2M_Kw_c({\bf k})\, .
\end{align} 
Rescaling
${\bf l}\to (A_k)^{1/2}{\bf l}$, we finally arrive at
\beq\label{eq:Fab_k}
J_{ab}(M_K-w_c({\bf k}),-{\bf k})=\frac{A_k^{d/2-1}}{2\sqrt{s_k}}
\int\frac{d^dl}{(2\pi)^d}\,\frac{1}{{\bf l}^2+{\bf k}^2-\Delta^2}\, .
\eeq
Now we have to insert the above expression into Eq.~\eqref{eq:twoloop_F}
and calculate the two-loop integral. To this end, we
rewrite the denominators in the outer loop, using again 
Eq.~\eqref{eq:trick} with $w_a,w_b\to w_d,w_c$ and $L^\mu\to Q^\mu$.
The integral over the last three terms 
does not vanish any more in dimensional regularization, since these are multiplied by the inner loop,
which is not a low-energy polynomial. It can however be easily checked that,
expanding everything but the non-polynomial factor 
$({\bf l}^2+{\bf k}^2-\Delta^2)^{-1}$ coming from the inner loop, one obtains a result of the form
$P(s)=\tilde P(s)(-\Delta^2-i\eps)^{d-2}$, 
where $\tilde P(s)$ is a low-energy polynomial in 
the variable $s$ with real coefficients and a
simple pole in $d-3$. 
Consequently,
\beq\label{eq:barM}
 {\cal M}(s)
=\frac{1}{2Q^0}\int\frac{d^dl}{(2\pi)^d}\,\frac{d^dk}{(2\pi)^d}\,
\frac{A_k^{d/2-1}}{2\sqrt{s_k}}\,\frac{1}{{\bf l}^2+{\bf k}^2-\Delta^2}
\,\frac{1}{{\bf p}^2-({\bf p}{\bf Q}/Q^0)^2-q_0^2}
+P(s)
\doteq \bar {\cal M}(s)+P(s)\, ,
\eeq
where ${\bf k}={\bf p}+\frac{1}{2}\,(1+\delta){\bf Q}$, and
\beq
q_0^2=\frac{\lambda(s,M_c^2,M_d^2)}{4s}\, ,\quad\quad
\delta=\frac{M_c^2-M_d^2}{s}\, .
\eeq

\subsection{Renormalization}

At this stage, it is appropriate to discuss the freedom in the definition
of ${\cal M}(s)$. For example, one may add an arbitrary
 low-energy polynomial of $s$ with {\em real} coefficients
to ${\cal M}(s)$ -- this would amount to a renormalization of the $K\to3\pi$
vertices $G_i$, $H_i$ 
in the non-relativistic effective Lagrangian. One may use this freedom
to remove the {\em real part} of the polynomial $P(s)$, 
which is of order $a^2$ and which contains an
ultraviolet pole at $d=3$.\footnote{Renormalization of the couplings $G_i$, $H_i$ first occurs
at two-loop order in our framework.}  
On the contrary, the imaginary part
can not be removed in this manner. So if the imaginary part of $P(s)$ were
divergent at $d\to 3$, it would constitute a major problem for the validity
of the framework. However, this does not happen: as can be easily seen,
the imaginary part is ultraviolet-finite. Moreover, at the accuracy we are 
working, one may neglect this imaginary part altogether, because its 
contribution to the decay rate starts at order $a^3$, that is beyond the
scope of the present paper.

To summarize, the two-loop function ${\cal M}(s)$ at the accuracy we are 
working can be replaced everywhere by
\beq \label{eq:F-barM}
F(M_a,M_b,M_c,M_d;s)=\bar{\cal M}(s)-\mbox{Re}\,\bar{\cal M}(s_t)\, ,
\eeq
 where 
$s_t=(M_c+M_d)^2$. Here, we used the above-mentioned freedom to normalize
the real part of $F(M_a,M_b,M_c,M_d;s)$ to zero at $s=s_t$. The difference
between $F$ and ${\cal M}$ is a low-energy polynomial. The real part of this
polynomial can be removed by renormalization, and the imaginary part
is ultraviolet-finite and
 does not contribute at the required precision.

\subsection{Integral representation}

The rest of this section is devoted to the calculation of the function 
$F$. After shifting the integration variable in Eq.~\eqref{eq:barM}
according to
\beq
{\bf k}\to {\bf k}
+\biggl(
\frac{{\bf k}{\bf Q}}{{\bf Q}^2}
\biggl(\frac{Q^0}{\sqrt{s}}-1\biggr)+\frac{1+\delta}{2}\biggr)\,{\bf Q}\, ,
\eeq
we arrive at
\beq\label{eq:nowexpand}
\bar {\cal M}(s)
=\frac{1}{2\sqrt{s}}
\int\frac{d^dl}{(2\pi)^d}\,\frac{d^dk}{(2\pi)^d}\,
\frac{1}{({\bf k}^2-q_0^2)}
\,\frac{N(x)}
{({\bf l}^2+\frac{(1+\delta)^2}{4}\,{\bf Q}^2+x-\Delta^2)}\, ,
\eeq
where
\begin{align}
x&={\bf k}^2+\frac{({\bf k}{\bf Q})^2}{s}+\frac{{\bf k}{\bf Q}}{\sqrt{s}}\,
Q^0(1+\delta)\, ,
\qquad
N(x) =\frac{A_k^{d/2-1}(s_k(x))}{2\sqrt{s_k(x)}}\, ,
\nnnl
s_k(x)&=M_K^2+M_c^2-2M_K
\biggl(M_c^2+\frac{(1+\delta)^2}{4}\,{\bf Q}^2+x\biggr)^{1/2}\, .
\end{align}
In order to evaluate the integral in Eq.~\eqref{eq:nowexpand}, we expand
the numerator $N(x)$ in the variable $x$ and integrate term by term. Using
Feynman parameterization to combine the two denominators, we obtain
\begin{align}
\bar {\cal M}(s) &= \frac{1}{2\sqrt{s}}\,
\sum_{n=0}^\infty\frac{1}{n!}\,\frac{d^n}{dx^n}N(x)\biggr|_{x=0}\,J_n(s)\, ,
\nnnl
J_n(s)& =\frac{\Gamma(2-d)}{(4\pi)^d}\int_0^1dy\,
y^{-d/2}\biggl(1+\frac{y{\bf Q}^2}{s}\biggr)^{-1/2}f_d^{(n)}(y,s)(g(y,s)-i\eps)^{d-2}\, ,
\end{align}
where $g(y,s)$ is defined by 
\beq\label{eq:D2}
g(y,s) = -(1-y)q_0^2-y\Delta^2+\frac{\frac{1}{4}\,y(1-y){\bf Q}^2(1+\delta)^2}{1+\frac{y{\bf Q}^2}{s}}\, ,
\eeq
and the first few coefficients
in the expansion are given by
\begin{align}
f_d^{(0)}(y,s)&= 1\, , \qquad
f_d^{(1)}(y,s)= \frac{dg(y,s)}{2(1-d)}\biggl(1
+\frac{{\bf Q}^2\alpha}{ds}\biggr)+\gamma\, ,
\\
f_d^{(2)}(y,s)&= -\frac{(2+d)g^2(y,s)}{4(1-d)}\biggl(1
+\frac{2{\bf Q}^2\alpha}{ds}+\frac{3{\bf Q}^4\alpha^2}{d(d+2)s^2}\biggr)
+\frac{dg(y,s)}{2(1-d)}\biggl( 2\gamma
+\frac{{\bf Q}^2(2\alpha\gamma+\beta^2s)}{ds}\biggr)+\gamma^2\, . \nonumber
\end{align}
Here,
\beq
\alpha =\frac{1-y}{1+\frac{y{\bf Q}^2}{s}}\, ,\quad
\beta = \frac{\alpha Q^0}{\sqrt{s}}\,(1+\delta)\, 
\frac{1}{\sqrt{1+\frac{y{\bf Q}^2}{s}}}\, ,
\quad
\gamma =-\frac{{\bf Q}^2(Q^0)^2}{2s}\,y(1+\delta)^2\,
\frac{1-\frac{y}{2}\,\left(1-\frac{{\bf Q}^2}{s}\right)}
{\left(1+\frac{y{\bf Q}^2}{s}\right)^2}\, .
\eeq
In order to eliminate the singularity at $y=0$ when $d\to 3$, we first 
integrate by parts and omit the surface term, which is a low-energy polynomial
 in
$s$. One may further verify that in the limit $d\to 3$, {\em up to a 
low-energy polynomial}, the function $F$ is given by the following 
integral representation
\beq\label{eq:Fint}
F(M_a,M_b,M_c,M_d;s) = \frac{1}{256\pi^3\sqrt{s}}\,
\int_0^1\frac{dy}{\sqrt{y}}\,\,{\cal F}(y,s)
\Bigl(\ln g(y,s)
-\ln g(y,s_t)\Bigr)\, .
\eeq
The function ${\cal F}(y,s)$ is given by an infinite sum 
\beq\label{eq:calFn}
{\cal F}(y,s)=\sum_{n=0}^\infty{\cal F}_n(y,s)\, ,\quad
{\cal F}_n(y,s)=\frac{4}{n!}\,\,\frac{d^n}{dx^n}N(x)\biggr|_{x=0}
\,\,\frac{d}{dy}\biggl(\frac{g(y,s)f^{(n)}(y,s)}{\sqrt{1+\frac{y{\bf Q}^2}{s}}}\biggr)
=\Order(\epsilon^{2n+2})\, ,
\eeq
and $f^{(n)}(y,s)=f_d^{(n)}(y,s)\bigr|_{d=3}$. In particular,
for $n=0$ we find
\beq\label{eq:F_lead}
{\cal F}_0(y,s)=\frac{\lambda^{1/2}(s_0,M_a^2,M_b^2)}{s_0}\,
\frac{1}{\sqrt{\Delta^2-\frac{(1+\delta)^2}{4}\,{\bf Q}^2}}\,
\frac{d}{dy}\biggl(\frac{g(y,s)}{\sqrt{1+\frac{y{\bf Q}^2}{s}}}\biggr) ~,
\eeq
where
\beq\label{eq:s_0}
s_0 =M_K^2+M_c^2-2M_K\sqrt{M_c^2+\frac{{\bf Q}^2}{4}\,(1+\delta)^2} ~.
\eeq
The series in $n$ for the function $F(M_a,M_b,M_c,M_d;s)$  
converges rapidly:
\begin{figure}
\centering
\includegraphics[width=0.55\linewidth]{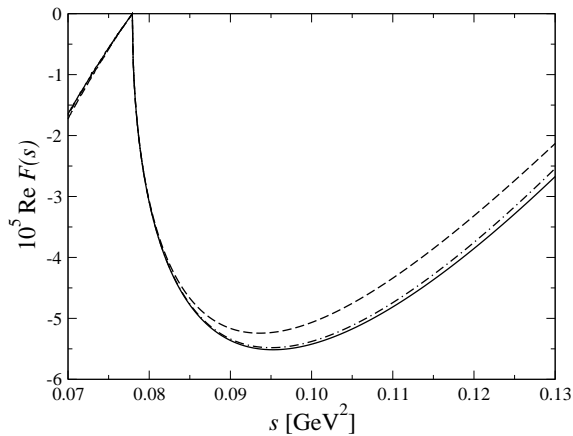}
\caption{Successive approximations for ${\rm Re}\,F(M_\pi,M_\pi,M_\pi,M_\pi;s)\doteq {\rm Re}\,F(s)$
in the equal-mass case. 
We show the leading-order (maximum $n=0$, dashed),  
next-to-leading-order ($n=1$, dash-dotted), and 
next-to-next-to-leading-order ($n=2$, full line) approximations, see Eq.~\eqref{eq:calFn}.}
\label{fig:arctan}
\end{figure}
we display the successive approximations with maximum $n=0,\,1,\,2$
for the equal-mass case in Fig.~\ref{fig:arctan}. It is seen that there is almost no 
difference between the two approximations with maximum $n=1$ and $n=2$.

The function $F(M_a,M_b,M_c,M_d;s)$ given by Eq.~\eqref{eq:Fint}
starts at $\Order(\epsilon^2)$ and thus does not violate power counting.
 Note also that, expanding
the function ${\cal F}$ in powers of $\frac{y{\bf Q}^2}{s}=\Order(\epsilon^2)$,
 it is
possible to perform the integrals in $y$ analytically at each order.
The result at next-to-leading order is given in \ref{app:integralsFF1F2}.
Analytic properties of this function are considered in \mbox{\ref{app:analytic}}, and the comparison
to the relativistic approach is discussed in \ref{app:compare_nrrel}.

\subsection{Threshold behavior}

Finally, we display the singularity structure 
of the function $F(M_a,M_b,M_c,M_d;s)$ near the cusp.
It can be shown that
\beq\label{eq:LO_sing}
F(M_a,M_b,M_c,M_d;s)= \frac{i\lambda^{1/2}(s,M_c^2,M_d^2)}{16\pi s}
\frac{i\lambda^{1/2}(s_{0t},M_a^2,M_b^2)}{16\pi s_{0t}}+\Order(q_0^2)\, ,
\eeq
where $s_{0t}$ denotes the function $s_0(s)$ in Eq.~\eqref{eq:s_0},
evaluated at $s=s_t$.
Hence in the vicinity of the cusp, the two-loop diagram is given as a product of
two factors: the first factor describes the cusp emerging
in the outer loop, while the second factor is the inner loop evaluated
at the threshold $s=4M_\pi^2$. Thus, the above two-loop diagram satisfies the 
threshold theorem~\cite{budinifonda,cabibboisidori,cusp_nrqft_I,cusp_nrqft_II,cusp_nrqft_III}.

\section{Three-pion couplings}
\label{sec:6-particles}

\begin{figure}
\centering
\includegraphics[width=0.35\linewidth]{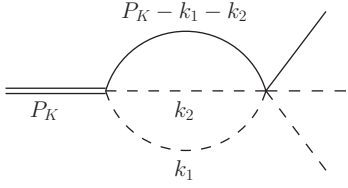}
\caption{A non-relativistic two-loop graph describing the
decay $K^+\to\pi^0\pi^0\pi^+$ that involves a 
six-particle vertex. The intermediate state contains the particles 
$\pi^0(k_1)$, $\pi^0(k_2)$, and $\pi^+(P_K-k_1-k_2)$.}
\label{fig:6particle}
\end{figure}

Pion rescattering in the final state also contains a contribution from
diagrams of the type shown in Fig.~\ref{fig:6particle}. These diagrams
contain a vertex describing the interaction of six pions at the same
point. In the non-relativistic theory, such interactions are generated 
by a Lagrangian of the type
\beq
{\cal L}_{3\pi}=\frac{1}{4}\,F_0\Phi_+^\dagger(\Phi_0^\dagger)^2\Phi_+\Phi_0^2
+\frac{1}{4}\,F'_0\Phi_-^\dagger(\Phi_+^\dagger)^2\Phi_-\Phi_+^2+\ldots\, ,
\eeq
where the ellipsis stands for terms with space derivatives.

For demonstration, let us consider the diagram with a $\pi^0\pi^0\pi^+$ 
intermediate state, depicted in Fig.~\ref{fig:6particle}. The contribution
of this diagram to the decay amplitude ${\cal M}_N$ is a constant, 
which in the center-of-mass frame of the kaon $P_K=(M_K,{\bf 0})$ is given by
\begin{align}
{\cal M}_N^{3\pi}(s_1,s_2,s_3) &=\frac{1}{2}\,G_0F_0
\int\frac{d^dk_1}{(2\pi)^d}\,\int\frac{d^dk_2}{(2\pi)^d}\,\frac{1}{2w_0({\bf k}_1)}\,\frac{1}{2w_0({\bf k}_2)}
\nnnl
&\times
\frac{1}{2w({\bf k}_1+{\bf k}_2)}\,
\frac{1}{w_0({\bf k}_1)+w_0({\bf k}_2)+w({\bf k}_1+{\bf k}_2)-M_K}\, .
\end{align}
Obviously, the diagram is a constant.  Its real part, which is in fact divergent for $d\to3$, can therefore
be absorbed in a redefinition of the coupling constant $G_0$, and will not be considered further.
We concentrate on the finite imaginary part, for which 
we may take $d\to 3$ and
rewrite the above expression in the relativistically covariant manner 
\begin{align}
{\rm Im}\,{\cal M}_N^{3\pi}(s_1,s_2,s_3)
&=\frac{i}{128\pi^5}\,G_0F_0\int\prod_{i=1}^3 d^4k_i \,\theta(k_i^0)
\,\delta(k_i^2-M_i^2)\,\delta^{(4)}(P_K-k_1-k_2-k_3)
\nnnl
&=\frac{i\,G_0F_0(M_K-3M_\pi)^2}{768\sqrt{3}\pi^2} +
\Order(\Delta_\pi,(M_K-3M_\pi)^3)\, .
\end{align}
Similar to Ref.~\cite{cabibboisidori}, for the rough estimate of the magnitude of 
the coupling $F_0$ we use matching to chiral perturbation theory. At lowest order this results in
\beq
F_0=\frac{M_\pi^2}{8F_\pi^4}+\ldots\, ,
\eeq
where the ellipsis stands for higher chiral orders, and $F_\pi = 92.2~{\rm MeV}$ 
is the pion decay constant.
Since also the imaginary part of the diagram in Fig.~\ref{fig:6particle}
is a constant, it is possible to eliminate this contribution
by allowing the constant $G_0$ to have a small imaginary part 
\beq
\frac{{\rm Im}\, G_0}{{\rm Re}\, G_0}\simeq
\frac{(M_K-3M_\pi)^2}{768\sqrt{3}\pi^2}\,\frac{M_\pi^2}{8F_\pi^4}
\simeq 1.5\cdot 10^{-5}\, . \label{eq:ImG0}
\eeq
Note that Eq.~\eqref{eq:ImG0} does not contain effects from the $\pi^+\pi^+\pi^-$ 
intermediate state, which would contribute a similar term $\propto H_0/G_0$.
As the effect turns out to be so small, we will neglect it below and 
continue to assume that the couplings in the non-relativistic kaon effective
Lagrangian are real. Note that our result is in qualitative agreement with
Eq.~(80) of Ref.~\cite{cabibboisidori}, although the exact coefficients differ.

We wish to stress that the non-relativistic theory without
six-particle couplings is in general not self-consistent (see, e.g., Ref.~\cite{Braaten}).
The reason for this is that there exist divergent two-loop graphs for
the process $3\pi\to 3\pi$ that require the introduction of a six-particle
counterterm in the Lagrangian. These diagrams arise, however, at order $a^4$
in the expansion in (small) $\pi\pi$ scattering lengths and thus are beyond the
scope of the present article.

\section{$K^+\to 3\pi$ decay amplitudes to two loops: final result}
\label{sec:finalresult}
In order to find the representation of the decay amplitudes at
$\Order(a^2)$, we draw all possible graphs for
$K^+\to\pi^0\pi^0\pi^+$ and $K^+\to\pi^+\pi^+\pi^-$
at one- and two-loop order, comprising the topologies discussed in detail
in Sects.~\ref{sec:oneloop} and \ref{sec:2loops};
for an explicit display of the different pion intermediate states, see Refs.~\cite{cusp_nrqft_I,cusp_nrqft_II}.
Further, in the graphs of type
Fig.~\ref{fig:2loop_text}A we retain only the non-analytic piece
$F(M_a,M_b,M_c,M_d;s)$, whereas the (infinite) polynomial is included
in the tree-level couplings $G_i$, $H_i$. The choice of the particular
representation Eq.~\eqref{eq:Fint} is equivalent to the choice of a
renormalization prescription. Note also that the resulting modification
of $G_i$, $H_i$ is of order $a^2$, so that if one uses the modified couplings 
also to calculate one- and two-loop diagrams, the results change at
$\Order(a^3)$ and $\Order(a^4)$, respectively, that is beyond the accuracy we consider.

The amplitudes up to and including two loops are given by
\begin{align}\label{eq:defexpand}
{\mathcal M}_{00+} &={\mathcal M}_N^{\rm {tree}}
                  +{\mathcal M}_N^{\rm {1-loop}}
                  +{\mathcal M}_N^{\rm {2-loops}}+\ldots \qquad
[K^+\to\pi^0\pi^0\pi^+] ~, \nnnl
{\mathcal M}_{++-}&={\mathcal M}_C^{\rm {tree}}
                  +{\mathcal M}_C^{\rm {1-loop}}
                  +{\mathcal M}_C^{\rm {2-loops}}+\ldots \qquad
[K^+\to\pi^+\pi^+\pi^-] ~.
\end{align}
Our amplitudes are
normalized such that the decay rates are given by
\beq
d\Gamma=\frac{1}{2M_K}(2\pi)^4\delta^{(4)}(P_f-P_i)
{|\mathcal M|}^2\prod_{i=1}^3
\frac{d^3{\bf p}_i}{2(2\pi)^3p_i^0}\,.
\eeq
The tree-level contribution to the 
$K^+\to\pi^0\pi^0\pi^+$ and $K^+\to\pi^+\pi^+\pi^-$ decay amplitudes 
up to $\Order(\eps^4)$ is given by Eq.~\eqref{eq:par1}. 
The one-loop amplitude fully consistent up to $\Order(a\eps^5)$
is displayed in Eqs.~\eqref{eq:1loop1}, \eqref{eq:1loop2}, and \eqref{eq:1loop3}.
Finally, the two-loop contributions to the amplitude, which are by far
the lengthiest and most complicated part, are given explicitly in \ref{app:Ampa2eps4}.
These are complete up to $\Order(a^2\eps^4)$, but contain partial terms
of $\Order(a^2\eps^6)$ and $\Order(a^2\eps^8)$, retained for the following reason.
As we have seen in Sect.~\ref{sec:pipiconv}, no reasonable description of the $\pi\pi$
S-wave scattering amplitudes is possible without including the effective ranges.
We have therefore kept all linear energy dependences in the S-wave interactions,
generating in combination terms up to $\Order(a^2\eps^8)$.
In the two-loop graphs, we neglect however terms induced by couplings quadratic in energy 
($\propto G_{2/3},\,H_{2/3}$, $\pi\pi$ shape parameters) as well as higher orders 
in P- (or even D-)waves.

Prior to using this representation in the fit, one eliminates
the coupling constants $C_i$, $D_i$, $E_i$ in favor of the $\pi\pi$ 
threshold parameters through the matching conditions discussed in Sect.~\ref{sec:pipi}.
This representation depends on the eight real 
$K\to 3\pi$ coupling constants $H_i$, $G_i$ and the threshold parameters for
$\pi\pi$ scattering, which in the end ought to be determined from a 
fit to the experimental data.
It is this amplitude, fully documented here for the first time, that, amended
with the radiative corrections provided in Ref.~\cite{cusp_nrqft_III}, has 
been used in the analysis of the NA48/2 $K^+\to\pi^0\pi^0\pi^+$ data~\cite{na48k3pi_II},
which in combination with results from $K_{e4}$ decays led
to the determination of $\pi\pi$ scattering lengths quoted in Eq.~\eqref{eq:a0a2experiment}.

\section{Comparison with other approaches}
\label{sec:comparison}

The low-energy expansion proposed here is closely related to
early work performed in the 1960s by many authors, 
who used $S$-matrix methods to investigate the
production of particles in the threshold region, 
in particular also in $K\to3\pi$ decays. We refer the interested reader to the article
 by Anisovich and Ansel'm~\cite{anisovich} for a review. The framework presented here may  be considered an
effective field theory realization of these approaches. The most notable differences are: 
\begin{enumerate}
\item
In these early approaches, the occurrence of a cusp in the decay rates went unnoticed. 
\item
It is explicitly stated in Ref.~\cite{anisovich} that there are no leading Landau singularities 
in the graphs of the type Fig.~\ref{fig:2loop_text}A, which does not agree with what we find. 
As a result, our expressions for the two-loop integrals do not agree with the ones presented in these early articles. 
\item
Our approach allows for a straightforward inclusion of the effects of real and virtual photons, 
see Ref.~\cite{cusp_nrqft_III} 
for the actual evaluation of the pertinent matrix elements,  
and Ref.~\cite{na48k3pi_II} for applications in the data analysis of $K\to 3\pi$ decays.
\end{enumerate}

A comparison of the present framework with the seminal articles of Cabibbo and Isidori~\cite{cabibbo,cabibboisidori} 
was already provided in Ref.~\cite{cusp_nrqft_I}, and we refer to that article for details. 
Here we only recall that the decomposition of the decay amplitudes~\cite{cabibbo,cabibboisidori} 
into the form ${\cal M}={\cal M}_0+v_{cd}(s){\cal M}_1$, with $M_{0,1}$ holomorphic, 
is valid near the threshold only, as is detailed in \ref{app:decomposition}. 
Furthermore, the leading Landau singularities found here were missed in  Ref.~\cite{cabibbo,cabibboisidori}.
However, the analysis in Ref.~\cite{na48k3pi_II} shows that, at the accuracy 
considered, these effects are immaterial for the main problem at hand,
the extraction of the $\pi\pi$ scattering lengths from the cusp strength.
Various aspects of  $K\to 3\pi$ decays are presented in the recent articles 
Refs.~\cite{nehme,bb1,bb2,gps,tarasovk3piA,tarasovk3piB,tarasovk3piC,isidorirad}. 
A comparison of these works with the present approach is provided in Refs.~\cite{cusp_nrqft_I,cusp_nrqft_II,cusp_nrqft_III}, 
see also Ref.~\cite{na48k3pi_II} for a careful comparison of the results of Refs.~\cite{tarasovk3piA,tarasovk3piB,tarasovk3piC} 
with those of Ref.~\cite{cusp_nrqft_III}. 
We refer the interested reader to these articles for details.

A dispersive approach has recently been undertaken to $K\to 3\pi$ (and $\eta\to 3\pi$) 
decays~\cite{zdrahal_I,Kampf:2011wr}, 
and an analytical expression for the decay amplitudes for $K_L\to 3\pi$ valid up to two loops 
in chiral perturbation theory is already available.
On the other hand, it seems to us that the effects of real and virtual photons 
are very difficult to incorporate without a Lagrangian framework. 
The main difference to the  framework presented here is the fact 
that quark mass effects are treated in a perturbative manner in Refs.~\cite{zdrahal_I}.

\section{Summary and conclusions}
\label{sec:summary}

We summarize the main findings of our investigations as follows.
\begin{enumerate}
\item
$\pi\pi$ rescattering generates a cusp in the decay distribution of $K^\pm\to\pi^0\pi^0\pi^\pm$ decays 
and allows one to measure $\pi\pi$ S-wave scattering 
lengths $a_0$ and $a_2$~\cite{budinifonda}. This cusp has been investigated in recent years by the NA48/2 collaboration in great 
detail~\cite{na48k3pi_I,ke4_III,na48k3pi_II}, and the two scattering lengths were determined in this manner.
\item
The analysis of the NA48/2 collaboration relies on two theoretical 
frameworks~\cite{cabibboisidori,cusp_nrqft_I,cusp_nrqft_II}, 
which relate the decay distribution to the scattering lengths. 
In this article, we spell out details of the method that 
underlies the matrix elements worked 
out in Refs.~\cite{cusp_nrqft_I,cusp_nrqft_II} and used in Ref.~\cite{na48k3pi_II} for the data analysis.
\item
We investigate the decays
$K\to 3\pi$ within an approach
which is  based on  non-relativistic effective Lagrangians.
It enables one to systematically calculate the decay amplitudes
$K^+\to\pi^0\pi^0\pi^+$ and $K^+\to\pi^+\pi^+\pi^-$ in a double expansion
in small momenta of the decay products and in the threshold parameters
of the $\pi\pi$ scattering (scattering lengths, etc.). 
Furthermore, it allows one to take into account
radiative corrections in a standard manner~\cite{cusp_nrqft_III}.
\item
Because it is a Lagrangian framework, the strictures of unitarity, analyticity and of 
the cluster decomposition are automatically taken into account. 
Furthermore, the amplitudes contain the Landau singularities automatically.
\item
In this article, we provide the amplitudes for $K^\pm\to \pi^0\pi^0\pi^\pm$ and for
$K^\pm\to \pi^\pm\pi^+\pi^-$ in closed form, at two-loop order, up to and including terms of order $a^2\eps^4$,
with the most important terms of order $a^2\eps^6$ and $a^2\eps^8$ also retained.
 The extension to $K_L\to 3\pi$ decays is straightforward. 
\end{enumerate}
In conclusion, the framework presented here is very suitable to investigate the 
beautiful set of data on $K\to 3\pi$ decays collected by the NA48/2 collaboration 
-- it has all the bells and whistles 
provided by quantum field theory.  We expect that the 
empirical parameterization of the Dalitz plot distribution published recently~\cite{na48k3pi_III} 
would even allow one to study the convergence of the $a,\epsilon$ expansion and to narrow down 
the emerging theoretical uncertainty in the scattering lengths $a_{0,2}$. This endeavour is, however,
beyond the scope of this work.

\section*{Acknowledgments}

We thank Gilberto Colangelo for collaboration at an early stage of this work,
Sebastian Schneider for some independent checks,
and Martin Zdr\'ahal for useful discussions as well as extensive comments that
helped us improve this manuscript.
We enjoyed extensive conversations with and support from Brigitte Bloch-Devaux, Luigi DiLella, 
and Italo Mannelli concerning applications of the results presented here in the data analysis. 
We thank Thomas Hahn for generous help with using {\it LoopTools}, 
which we used to check some of our results on relativistic amplitudes.
Partial financial support by the Helmholtz Association through funds provided
to the virtual institute ``Spin and strong QCD'' (VH-VI-231), 
the project ``Study of Strongly Interacting Matter''
(HadronPhysics2, grant No.~227431) under the 7th Framework Programme of the EU, 
and by DFG (SFB/TR 16, ``Subnuclear Structure of Matter'') 
is gratefully acknowledged. 
This work was  supported  by the Swiss
National Science Foundation,  and by EU MRTN--CT--2006--035482
(FLAVIA{\it net}). 
 One of us (J.G.) is grateful to the Alexander von Humboldt--Stiftung and to 
the Helmholtz--Gemeinschaft for the award of  a prize 
 that allowed him to stay at the HISKP at the University of Bonn, 
where part of this work was performed. 
He also thanks the HISKP for the warm hospitality during these stays.
A.R.\ acknowledges  support
by the Georgia National Science Foundation (grant No.~GNSF/ST08/4-401).

\newpage

%--------------------------------------------------------------------------------

\begin{appendix}

\renewcommand{\thefigure}{\arabic{figure}}
\renewcommand{\thetable}{\arabic{table}}

%--------------------------------------------------------------------------------

\section{\boldmath{$K^+\to3\pi$} amplitudes up to \boldmath{$\Order(a^2\eps^4)$}}
\label{app:Ampa2eps4}

We have calculated the two-loop contributions to the decay amplitudes for
$K^+ \to \pi^0\pi^0\pi^+$ and $K^+\to \pi^+\pi^+\pi^-$ including derivative couplings, 
along the lines explained in detail in Sect.~\ref{sec:2loops}.  
The results we find, complete at order $a^2\epsilon^4$ and containing additional 
terms of  $\Order(a^2\epsilon^6)$, $\Order(a^2\epsilon^8)$ (see the comments in Sect.~\ref{sec:finalresult}), 
read
\beq\label{eq:rep}
\M_{N,C}^{\textrm{2-loops}}
=\M_{N,C}^A(s_1,s_2,s_3)
+\M_{N,C}^B(s_1,s_2,s_3)\,,
\eeq
where
\allowdisplaybreaks{
\begin{align}
\M_N^A
%(1)
&=4 H'''_{+-}(s_3) C_{+-}\big(\tilde s_3^{+-}\big) C_x(s_3) 
     F_+\bigl(M_\pi,M_\pi,M_\pi,M_\pi;s_3\bigr)  \nnnl
&\quad -4 \Bigl[ \frac{1}{2}H_1 C_{+-}\big(\tilde s_3^{+-}\big) + 2M_K H'''_{+-}(s_3)D_{+-} \Bigr] C_x(s_3) 
    \frac{\vec{Q}_3^2}{Q_3^0} \,
     F^{(1)}_+\bigl(M_\pi,M_\pi,M_\pi,M_\pi;s_3\bigr)  \nnnl
&\quad +4 M_K H_1 D_{+-} C_x(s_3) \frac{\vec{Q}_3^4}{(Q_3^0)^2} \,
     F^{(2)}_+\bigl(M_\pi,M_\pi,M_\pi,M_\pi;s_3\bigr)  \nnnl
%(2)
&+2 G''(s_3)C_x\big(\tilde s_3^{+-}\big)C_x(s_3)
     F_+\bigl(M_{\pi^0},M_{\pi^0},M_\pi,M_\pi;s_3\bigr)  \nnnl
&\quad +2 \bigl[G_1 C_x\big(\tilde s_3^{+-}\big) -2M_K G''(s_3) D_x \bigr] C_x(s_3)\frac{\vec{Q}_3^2}{Q_3^0} \,
     F^{(1)}_+\bigl(M_{\pi^0},M_{\pi^0},M_\pi,M_\pi;s_3\bigr)  \nnnl
&\quad -4 M_K G_1 D_x C_x(s_3) \frac{\vec{Q}_3^4}{(Q_3^0)^2} \,
     F^{(2)}_+\bigl(M_{\pi^0},M_{\pi^0},M_\pi,M_\pi;s_3\bigr)  \nnnl
%(3)
&+2 H''(s_3) C_{++}\big(\tilde s_3^{+-}\big) C_x(s_3) 
    F_+\bigl(M_\pi,M_\pi,M_\pi,M_\pi;s_3\bigr)  \nnnl
&\quad +2 \bigl[H_1 C_{++}\big(\tilde s_3^{+-}\big) -2M_K H''(s_3) D_{++} \bigr] C_x(s_3) \frac{\vec{Q}_3^2}{Q_3^0} \,
    F^{(1)}_+\bigl(M_\pi,M_\pi,M_\pi,M_\pi;s_3\bigr)  \nnnl
&\quad -4M_K H_1 D_{++} C_x(s_3) \frac{\vec{Q}_3^4}{(Q_3^0)^2}\,F^{(2)}_+\bigl(M_\pi,M_\pi,M_\pi,M_\pi;s_3\bigr) \nnnl
%(4)
&+4 G'''_{+0}(s_3)C_{+0}\big(\tilde s_3^{00}\big) C_{00}(s_3)
    F_+\bigl(M_\pi,M_{\pi^0},M_{\pi^0},M_{\pi^0};s_3\bigr)  \nnnl
&\quad -4 \Bigl[\frac{1}{2}G_1 C_{+0}\big(\tilde s_3^{00}\big) + 2M_K G'''_{+0}(s_3) D_{+0} \Bigr] C_{00}(s_3) \frac{\vec{Q}_3^2}{Q_3^0} \,
    F^{(1)}_+\bigl(M_\pi,M_{\pi^0},M_{\pi^0},M_{\pi^0};s_3\bigr)  \nnnl
&\quad +4M_K G_1 D_{+0} C_{00}(s_3) \frac{\vec{Q}_3^4}{(Q_3^0)^2}\,
   F^{(2)}_+\bigl(M_\pi,M_{\pi^0},M_{\pi^0},M_{\pi^0};s_3\bigr) \nnnl
%(5)
&+\biggl\{
  4 H'''_{+-}\big(s_1^+\big) C_x\big(\tilde s_1^{+0}\big)
    \bigl[C_{+0}(s_1)+E_{+0}^+(s_1,s_2,s_3)\bigr]  
    F_0\bigl(M_\pi,M_\pi,M_\pi,M_{\pi^0};s_1\bigr)  \nnnl
&\quad -4 \biggl[ \Bigl(\frac{1}{2}H_1 C_x\big(\tilde s_1^{+0}\big)
+ 2M_K H'''_{+-}\big(s_1^+\big) D_x \Bigr)C_{+0}(s_1)\frac{\vec{Q}_1^2}{Q_1^0} 
+ 2H'''_{+-}\big(s_1^+\big) C_x\big(\tilde s_1^{+0}\big) E_{+0}(s_1,s_2,s_3)\biggr]  \nnnl & \qquad \qquad  \times
    F^{(1)}_0\bigl(M_\pi,M_\pi,M_\pi,M_{\pi^0};s_1\bigr)  \nnnl
&\quad +4 M_K H_1 D_x C_{+0}(s_1) \frac{\vec{Q}_1^4}{(Q_1^0)^2} \,
    F^{(2)}_0\bigl(M_\pi,M_\pi,M_\pi,M_{\pi^0};s_1\bigr) \nnnl
%(6)
&+4 G'''_{+0}(s_1^-) C_{+0}\big(\tilde s_1^{0+}\big)
    \bigl[C_{+0}(s_1) -E_{+0}^-(s_1,s_2,s_3)\bigr]  
    F_0\bigl(M_{\pi^0},M_\pi,M_{\pi^0},M_\pi;s_1\bigr)  \nnnl
&\quad -4 \biggl[ \Bigl(\frac{1}{2}G_1 C_{+0}\big(\tilde s_1^{0+}\big) 
+ 2M_K G'''_{+0}(s_1^-) D_{+0} \Bigr)C_{+0}(s_1)\frac{\vec{Q}_1^2}{Q_1^0} 
  -2G'''_{+0}(s_1^-) C_{+0}\big(\tilde s_1^{0+}\big) E_{+0}(s_1,s_2,s_3) \biggr] \nnnl & \qquad \qquad \times
    F^{(1)}_0\bigl(M_{\pi^0},M_\pi,M_{\pi^0},M_\pi;s_1\bigr)  \nnnl
&\quad +4 M_K G_1 D_{+0} C_{+0}(s_1) \frac{\vec{Q}_1^4}{(Q_1^0)^2} \,
    F^{(2)}_0\bigl(M_{\pi^0},M_\pi,M_{\pi^0},M_\pi;s_1\bigr)  \nnnl
%(7)
&+2 G''(s_1^+) C_{00}\big(\tilde s_1^{+0} \big)
    \bigl[C_{+0}(s_1) +E_{+0}^+(s_1,s_2,s_3)\bigr] 
    F_0\bigl(M_{\pi^0},M_{\pi^0},M_\pi,M_{\pi^0};s_1\bigr)  \nnnl
&\quad +2 \biggl[ \bigl( G_1 C_{00}\big(\tilde s_1^{+0} \big)  - 2M_K G''(s_1^+) D_{00} \bigr)C_{+0}(s_1)\frac{\vec{Q}_1^2}{Q_1^0} 
  - 2 G''(s_1^+) C_{00}\big(\tilde s_1^{+0} \big) E_{+0}(s_1,s_2,s_3)\biggr] \nnnl & \qquad \qquad \times
    F^{(1)}_0\bigl(M_{\pi^0},M_{\pi^0},M_\pi,M_{\pi^0};s_1\bigr) \nnnl
&\quad  - 4 M_K G_1 D_{00} C_{+0}(s_1) \frac{\vec{Q}_1^4}{(Q_1^0)^2} \,
    F^{(2)}_0\bigl(M_{\pi^0},M_{\pi^0},M_\pi,M_{\pi^0};s_1\bigr)
+(s_1\leftrightarrow s_2)\biggr\} ~,  \label{eq:0Aapp} \\[5mm]
\M_N^B
&=4 H'(s_3) C_x(s_3) C_{+-}(s_3) J^2_{+-}(s_3) 
+ G(s_3) C_{00}(s_3)^2 J^2_{00}(s_3) 
+2 G(s_3) C_x(s_3)^2 J_{00}(s_3)J_{+-}(s_3)  \nnnl
&+ 2 H'(s_3) C_x(s_3)C_{00}(s_3)
    J_{+-}(s_3)J_{00}(s_3)  
+\Bigl\{ 4 G'(s_1) C_{+0}(s_1)^2 J^2_{+0}(s_1) 
+(s_1\leftrightarrow s_2)\Bigr\} ~, \label{eq:0Bapp} 
\\[5mm]
\M_C^A
%(1)
&=2 G''(s_3) C_x\big(\tilde s_3^{++}\big)
    C_{++}(s_3)
    F_-\bigl(M_{\pi^0},M_{\pi^0},M_\pi,M_\pi;s_3\bigr)   \nnnl
&\quad+2 \bigl[ G_1 C_x\big(\tilde s_3^{++}\big) -2M_K G''(s_3) D_x  \bigr] C_{++}(s_3) \frac{\vec{Q}_3^2}{Q_3^0} \,
    F^{(1)}_-\bigl(M_{\pi^0},M_{\pi^0},M_\pi,M_\pi;s_3\bigr)   \nnnl
&\quad -4 M_K G_1 D_x C_{++}(s_3)  \frac{\vec{Q}_3^4}{(Q_3^0)^2} \,
F^{(2)}_-\bigl(M_{\pi^0},M_{\pi^0},M_\pi,M_\pi;s_3\bigr)   \nnnl
%(2)
&+4 H'''_{+-}(s_3) C_{+-}\big(\tilde s_3^{++}\big)
    C_{++}(s_3)
    F_-\bigl(M_\pi,M_\pi,M_\pi,M_\pi;s_3\bigr)  \nnnl
&\quad -4 \Bigl[ \frac{1}{2}H_1 C_{+-}\big(\tilde s_3^{++}\big) + 2M_K H'''_{+-}(s_3) D_{+-} \Bigr] C_{++}(s_3) \frac{\vec{Q}_3^2}{Q_3^0} \,
    F^{(1)}_-\bigl(M_\pi,M_\pi,M_\pi,M_\pi;s_3\bigr)  \nnnl
&\quad +4M_K H_1 D_{+-} C_{++}(s_3) \frac{\vec{Q}_3^4}{(Q_3^0)^2} \,
F^{(2)}_-\bigl(M_\pi,M_\pi,M_\pi,M_\pi;s_3\bigr)  \nnnl
%(3)
&+\biggl\{
  4 H'''_{+-}(s_1)C_{+-}\big(\tilde s_1^{+-}\big) 
    \bigl[C_{+-}(s_1)-E_{+-}(s_1,s_2,s_3)\bigr] 
    F_+\bigl(M_\pi,M_\pi,M_\pi,M_\pi;s_1\bigr)  \nnnl
&\quad -4 \biggl[ \Bigl(\frac{1}{2}H_1 C_{+-}\big(\tilde s_1^{+-}\big)  
+ 2M_K H'''_{+-}(s_1) D_{+-} \Bigr)C_{+-}(s_1)\frac{\vec{Q}_1^2}{Q_1^0} \nnnl
&\qquad \qquad 
  - 2H'''_{+-}(s_1)C_{+-}\big(\tilde s_1^{+-}\big)  E_{+-}(s_1,s_2,s_3) \biggr] 
    F^{(1)}_+\bigl(M_\pi,M_\pi,M_\pi,M_\pi;s_1\bigr)  \nnnl
&\quad +4M_K H_1 D_{+-} C_{+-}(s_1) \frac{\vec{Q}_1^4}{(Q_1^0)^2} \,
F^{(2)}_+\bigl(M_\pi,M_\pi,M_\pi,M_\pi;s_1\bigr)  \nnnl
%(4)
&+2 G''(s_1) C_x\big(\tilde s_1^{+-}\big)
    \bigl[C_{+-}(s_1)-E_{+-}(s_1,s_2,s_3)\bigr] 
    F_+\bigl(M_{\pi^0},M_{\pi^0},M_\pi,M_\pi;s_1\bigr)  \nnnl
&\quad +2 \biggl[ \bigl(G_1 C_x\big(\tilde s_1^{+-}\big)  - 2M_K G''(s_1) D_x \Bigr)C_{+-}(s_1)\frac{\vec{Q}_1^2}{Q_1^0} \nnnl
&\qquad\qquad
  +2G''(s_1) C_x\big(\tilde s_1^{+-}\big) E_{+-}(s_1,s_2,s_3)\biggr]
    F^{(1)}_+\bigl(M_{\pi^0},M_{\pi^0},M_\pi,M_\pi;s_1\bigr)  \nnnl
&\quad -4M_K G_1 D_x C_{+-}(s_1) \frac{\vec{Q}_1^4}{(Q_1^0)^2} \,
F^{(2)}_+\bigl(M_{\pi^0},M_{\pi^0},M_\pi,M_\pi;s_1\bigr)  \nnnl
%(5)
&+2 H''(s_1) C_{++}\big(\tilde s_1^{-+}\big)
    \bigl[C_{+-}(s_1)+E_{+-}(s_1,s_2,s_3)\bigr] 
    F_+\bigl(M_\pi,M_\pi,M_\pi,M_\pi;s_1\bigr)  \nnnl
&\quad+2  \biggl[ \bigl(H_1 C_{++}\big(\tilde s_1^{-+}\big)  
- 2M_K H''(s_1) D_{++} \Bigr)C_{+-}(s_1)\frac{\vec{Q}_1^2}{Q_1^0} \nnnl &\qquad\qquad
  - 2H''(s_1) C_{++}\big(\tilde s_1^{-+}\big)E_{+-}(s_1,s_2,s_3) \biggr]
    F^{(1)}_+\bigl(M_\pi,M_\pi,M_\pi,M_\pi;s_1\bigr)  \nnnl
&\quad  -4M_KH_1 D_{++} C_{+-}(s_1) \frac{\vec{Q}_1^4}{(Q_1^0)^2} \,
F^{(2)}_+\bigl(M_\pi,M_\pi,M_\pi,M_\pi;s_1\bigr)  \nnnl
%(6)
&+4 G'''_{+0}(s_1) C_{+0}\big(\tilde s_1^{00}\big)
    C_x(s_1) 
    F_+\bigl(M_\pi,M_{\pi^0},M_{\pi^0},M_{\pi^0};s_1\bigr) \nnnl
&\quad -4 \Bigl[\frac{1}{2} G_1 C_{+0}\big(\tilde s_1^{00}\big) +2M_K G'''_{+0}(s_1) D_{+0}\Bigr] C_x(s_1) \frac{\vec{Q}_1^2}{Q_1^0} \,
    F^{(1)}_+\bigl(M_\pi,M_{\pi^0},M_{\pi^0},M_{\pi^0};s_1\bigr) \nnnl
&\quad +4M_K G_1 D_{+0} C_x(s_1) \frac{\vec{Q}_1^4}{(Q_1^0)^2} \,
F^{(2)}_+\bigl(M_\pi,M_{\pi^0},M_{\pi^0},M_{\pi^0};s_1\bigr) 
+ (s_1\leftrightarrow s_2)\biggr\} ~, \label{eq:+Aapp} \\[5mm]
\M_C^B
&= H(s_3) C_{++}(s_3)^2 J^2_{++}(s_3) 
+\Bigl\{ 
  4 H'(s_1) C_{+-}(s_1)^2 J^2_{+-}(s_1) 
+ G(s_1) C_x(s_1) C_{00}(s_1) J^2_{00}(s_1) \nnnl
&+2 \Big[ H'(s_1) C_x(s_1) 
+ G(s_1) C_{+-}(s_1) \Big] C_x(s_1)  J_{+-}(s_1) J_{00}(s_1) 
+(s_1\leftrightarrow s_2)\Bigr\} ~. \label{eq:+Bapp}
\end{align}}\noindent
We have used the following abbreviations:
\begin{align}
C_n(s_i) &= C_n + D_n \big(s_i - \bar s_n\big) ~, \quad
\tilde s_i^{cd} = M_c^2+M_i^2-s_i+\frac{M_K}{Q_i^0}\big(s_i+2\vec{Q}_i^2 -M_c^2+M_d^2 \big) ~,\nnnl
E_{+-}(s_1,&s_2,s_3) = E_{+-} \, \frac{s_1(s_3-s_2)}{2M_KQ_1^0} ~,\quad
E_{+0}(s_1,s_2,s_3) = E_{+0} \biggl[\frac{s_1(s_3-s_2-\Delta_\pi)}{2M_KQ_1^0}+\Delta_\pi\biggr] ~,\nnnl
E_{+0}^{\pm}(s_1,&s_2,s_3) = E_{+0} \biggl[\frac{(s_1\pm\Delta_\pi)(s_3-s_2-\Delta_\pi)}{2M_KQ_1^0}+\Delta_\pi\biggr] ~,
\nnnl
G(s_i) &= G_0 + G_1 \left(p_i^0-M_\pi\right) + G_2 \left(p_i^0-M_\pi\right)^2 
+ G_3 \frac{\vec{Q}_i^2}{3}\Big(1-\frac{4M_{\pi^0}^2}{s_i}\Big) ~,\nnnl
H(s_i) &= H_0 + H_1 \left(p_i^0-M_\pi\right) + H_2 \left(p_i^0-M_\pi\right)^2 
+ H_3 \frac{\vec{Q}_i^2}{3}\Big(1-\frac{4M_\pi^2}{s_i}\Big) ~,\nnnl
G'(s_i) &= G_0 
+ G_1\biggl(\frac{Q_i^0}{2}\Bigl(1+\frac{\Delta_\pi}{s_i}\Bigr)-M_\pi\biggr)
+ G_2\biggl[\biggl(\frac{Q_i^0}{2}\Bigl(1+\frac{\Delta_\pi}{s_i}\Bigr)
-M_\pi\biggr)^2+\frac{{\bf Q}_i^2}{12s_i^2}\,\lambda\bigl(s_i,M_\pi^2,M_{\pi^0}^2\bigr)\biggr]
\nnnl & \quad
+G_3\biggl[\biggl(\frac{Q_i^0}{2}\Bigl(1-\frac{\Delta_\pi}{s_i}\Bigr)
-p_i^0\biggr)^2+\frac{{\bf Q}_i^2}{12s_i^2}\,\lambda\bigl(s_i,M_\pi^2,M_{\pi^0}^2\bigr)\biggr] ~, \nnnl
H'(s_i) &= H_0 + H_1\Bigl(\frac{Q_i^0}{2}-M_\pi\Bigr)
+H_2\biggl[\Bigl(\frac{Q_i^0}{2}-M_\pi\Bigr)^2+\frac{\vec{Q}_i^2}{12}\,
\Bigl(1- \frac{4M_\pi^2}{s_i}\Bigr)\biggr] \nnnl
& \quad + H_3\biggl[\Bigl(\frac{Q_i^0}{2}-p_i^0\Bigr)^2+\frac{\vec{Q}_i^2}{12}\,
\Bigl(1- \frac{4M_\pi^2}{s_i}\Bigr)\biggr] ~,\nnnl
G''(s_i) &= G_0 + G_1 \bigg( \frac{s_i}{2Q_i^0} - M_\pi \bigg) ~, \qquad
G'''_{ab}(s_i) = G_0 + G_1 \bigg( \frac{1}{2}(M_K-M_a-M_b) - \frac{s_i}{4Q_i^0}\bigg) ~, \nnnl
G''\big(s_i^+\big) &= G_0 + G_1 \bigg( \frac{s_i + \Delta_\pi}{2Q_i^0} - M_\pi \bigg) ~, ~~ 
G'''_{ab}\big(s_i^-\big) = G_0 + G_1 \bigg( \frac{1}{2}(M_K-M_a-M_b) - \frac{s_i- \Delta_\pi}{4Q_i^0}\bigg) ~,\nnnl
H''(s_i) &= H_0 + H_1 \bigg( \frac{s_i}{2Q_i^0} - M_\pi \bigg) ~, \qquad
H'''_{ab}(s_i) = H_0 + H_1 \bigg( \frac{1}{2}(M_K-M_a-M_b) - \frac{s_i}{4Q_i^0}\bigg) ~, \nnnl
H'''_{ab}\big(s_i^+\big) &= H_0 + H_1 \bigg( \frac{1}{2}(M_K-M_a-M_b) - \frac{s_i+ \Delta_\pi}{4Q_i^0}\bigg) ~.
\end{align}
We remark that the representations of the ``double bubbles'' $\M_{N/C}^B$ are even \emph{complete} to 
$\Order(a^2\epsilon^6)$ if the polynomials $C_n(s_i)$ are amended by shape parameter terms 
$\ldots+F_n(s_i-\bar s_n)^2$: P-wave contributions only start at $\Order(a^2\epsilon^8)$.
In contrast, in the ``genuine'' two-loop graphs $\M_{N/C}^A$, terms $\propto E_n F^{(2)}$ have not been included,
although they contribute in principle at $\Order(a^2\epsilon^6)$.
We remark that the two-loop amplitudes for the $K_L\to 3\pi$ decay channels at the same accuracy
can be retrieved from Ref.~\cite{SKDeta3pi}, where the $\eta\to 3\pi$ amplitudes are discussed within
the same formalism, with the obvious replacements as described in Ref.~\cite{cusp_nrqft_II}.

$F_i(\ldots;s)$, $F^{(1)}_i(\ldots;s)$, $F^{(2)}_i(\ldots;s)$ 
stand for the integrals $F(\ldots;s)$, $F^{(1)}(\ldots;s)$, $F^{(2)}(\ldots;s)$, 
evaluated at $\vec{Q}^2=\lambda(M_K^2,M_{\pi^i}^2,s)/4M_K^2$, with $i=\pm,0$.
The first few terms in an expansion in $\epsilon$
of the analytic expression for these two-loop functions are displayed 
in \ref{app:integralsFF1F2}.

\section{The loop functions $F$, $F^{(1)}$, and $F^{(2)}$}
\label{app:integralsFF1F2}

The functions 
$F_i(\ldots;s)$, $F^{(1)}_i(\ldots;s)$, $F^{(2)}_i(\ldots;s)$
introduced in \ref{app:Ampa2eps4} are explicitly given by
\begin{align}
F(M_a,M_b,M_c,M_d,s)   &=  \mathcal{N}\,\biggl[2\nu\, f_1+\rho\, f_0 -\frac{3\vec{Q}^2}{10s}(\rho\,f_1-2q_0^2\,f_0)\nnnl 
&+\mathcal{K}
(X_3f_3+X_2f_2+X_1f_1+X_0f_0)\biggr]+\Order(\epsilon^6) ~,\nnnl
F^{(1)}(M_a,M_b,M_c,M_d,s) &= \frac{\mathcal{N}}{10}{(1+\delta)}\bigl[ (10\nu-\rho)f_1 + (5\rho+2q_0^2)f_0 \bigr] + \Order(\epsilon^4)~, \nnnl
F^{(2)}(M_a,M_b,M_c,M_d,s) &= \frac{\mathcal{N}}{2} \biggl[ - \frac{1}{\vec{Q}^2}
\Big(2\nu^2 f_3 + 3\nu\rho f_2 +(\rho^2-2\nu q_0^2) f_1 - \rho q_0^2 f_0 \Big) \label{eq:functionF1app}\\
+&  \frac{(1+\delta)^2}{4} \Big( \nu f_3 + (\rho-2\nu) f_2 + (4\nu-2\rho-q_0^2) f_1 +2(\rho+q_0^2) f_0 \Big) \biggr] +\Order(\epsilon^4) ~,\nonumber
\end{align}
with
\begin{align}
\mathcal{N}&=\frac{1}{256\pi^3\sqrt{s}}\, 
\frac{\lambda^{1/2}(s_0,M_a^2,M_b^2)}{s_0\sqrt{\Delta^2-\frac{(1+\delta)^2}{4}\,\vec{Q}^2}} ~, \nnnl
\mathcal{K}&=
\biggl[\frac{1}{2(M_K^2+M_c^2)-(M_a+M_b)^2-s_0} + \frac{1}{s_0-(M_a-M_b)^2} - \frac{2}{s_0} \biggr] 
\frac{M_K^2}{s_0-M_K^2-M_c^2} ~, \nnnl
f_0&=4\bigl(v_1+v_2-\bar v_{2}+h\bigr) ~, \quad 
f_1=
\frac{4}{3}\,\bigl(y_1(v_1-1)+y_2(v_2-1)-\bar y_{2}(\bar v_{2}-1) +h \bigr) ~,\nnnl
f_2&=-\frac{1}{5\nu}\,(3\rho f_1-q_0^2f_0) ~,\quad
f_3=\frac{1}{7\nu^2}\,\big[3(\nu q_0^2+\rho^2)f_1-\rho q_0^2 f_0\big] ~,\nnnl
h&=\frac{1}{2}\ln\biggl(\frac{1+\vec{Q}^2/s}{1+\bar{\vec{Q}}^2/s_t}\biggr)  ~,\quad
\vec{Q}^2=\frac{\lambda(M_K^2,q_3^2,s)}{4M_K^2} ~,
\quad \bar{\vec{Q}}^2=\vec{Q}^2(s_t) ~, \quad s_t=(M_c+M_d)^2 ~, \nnnl
v_i&=\sqrt{-y_i}\,\arctan\frac{1}{\sqrt{-y_i}}  ~,\quad i=1,2 ~;\quad
\bar v_{2}=\sqrt{-\bar y_{2}}\,\arctan\frac{1}{\sqrt{-\bar y_{2}}} ~, \quad 
y_{1,2} = \frac{-\rho \mp \sqrt{D}}{2\nu} ~, \nnnl
\bar y_{2}&=y_2(s_t) ~,  \quad
\nu= -\frac{\vec{Q}^2}{s}\,(M_c^2+\Delta^2) ~,\quad
\rho =q_0^2-\Delta^2+\frac{\vec{Q}^2}{s}\, M_c^2 ~,\quad
D = \rho^2+4\nu q_0^2 ~,\nnnl
X_0&=(R-H\rho) q_0^2 \, , \quad X_1=H(\rho^2-2\nu q_0^2)-R(2\rho+q_0^2)\, ,\quad X_2=3H\nu\rho-3R\Big(\nu-\frac{\rho}{2}\Big)\, ,\nnnl
X_3&=2\nu(H\nu+R)\, ,\quad
H=-\frac{3}{2}\,\biggl(1+\frac{{\bf Q}^2}{3s}\biggr)\, ,\quad
R=\frac{{\bf Q}^2Q_0^2}{2s}\,(1+\delta)^2\, , \nnnl
s_0&=M_K^2+M_c^2-2M_K\biggl( M_c^2+\frac{\vec{Q}^2(1+\delta)^2}{4}\biggr)^{1/2} ~, \quad
q_0^2 =\frac{\lambda(s,M_c^2,M_d^2)}{4s} ~, \nnnl
\Delta^2&=\frac{\lambda(M_K^2,M_c^2,(M_a+M_b)^2)}{4M_K^2} ~,\quad
\delta=\frac{M_c^2-M_d^2}{s} ~,\label{eq:verybig}
\end{align}
compare also Fig.~\ref{fig:2loop_text}A.
The $\arctan$ is understood to be evaluated according to 
\begin{equation}\label{eq:functionF3}
\arctan x = \frac{1}{2i}\ln\frac{1+ix}{1-ix} ~,
\end{equation}
and $s$ is given a small positive imaginary part in all  arguments, $s \to s+i\epsilon$.

\section{Holomorphic properties of $F$}\label{app:analytic}

The  loop functions $F$, $F^{(1)}$, and  $F^{(2)}$ 
 in \ref{app:Ampa2eps4} are generated by the two-loop graphs displayed 
in Fig.~\ref{fig:2loop_text}A. The explicit expressions in \ref{app:integralsFF1F2} 
are valid on the upper rim
of the real $s$-axis. Here, we show how to  analytically continue them to the whole 
non-relativistic region.
The holomorphic properties of these loop functions play a role in connection with the 
decomposition of the amplitude as proposed by Cabibbo 
and Isidori~\cite{cabibboisidori}, see also the discussion in \ref{app:decomposition}.

The region of holomorphicity  could be established directly
from the explicit lowest-order expressions given in \ref{app:integralsFF1F2} -- 
higher-order terms in $\epsilon$ emerge from the Taylor expansion of
low-energy polynomials and do not change the analytic structure. 
 Here, we follow a different path and derive 
the singularity structure from the integral representation Eq.~\eqref{eq:Fint}. To ease notation, we set in this section
\beq\label{eq:defF}
F(s)\doteq F(M_a,M_b,M_c,M_d;s)\fs
\eeq

\subsection{Holomorphic properties from the integral representation}
\label{app:holomorphicintegral}
First we note that the prefactor ${\cal F}(y,s)$ and the function $g(y,s)$ 
in the integral representation Eq.~\eqref{eq:Fint} 
are low-energy polynomials
in $(s-(M_c+M_d)^2)$ and in $y$. Because the argument of the logarithm
does not vanish for
$0\leq y\leq 1$ when $s$ approaches the real axis from above at $\mbox{Re}\,s>s_t$, 
$F(s)$ is analytic in the part of the non-relativistic region located
in the upper half of the complex $s$-plane. To analyze the singularities that may occur
during the continuation  to the lower
rim of the real axis, we  note that  $g(y,s)$  has two zeros in the low-energy region, 
which we denote by $y_1(s)$ and $y_2(s)$,
\beq\label{eq:ABCD}
g(y,s)=\frac{\nu}{1+\frac{y {\bf Q}^2}{s}}\,(y-y_1(s))(y-y_2(s))\, ,
\eeq
where $y_{1,2}(s)$ and $\nu$ are defined in Eq.~\eqref{eq:verybig} 
[the sign convention is chosen so that $y_1(s)=0$ at $s=s_t=(M_c+M_d)^2$]. 
During the analytic continuation, singularities 
may occur at the following values $\bar s$ 
of $s$.
\begin{itemize}
\item[i)] 
$y_{1,2}(\bar s)=0~\mbox{or}~1$. This may generate endpoint singularities.
\item[ii)]
If the integration contour runs between $y_1(s)$ and $y_2(s)$, 
a pinch singularity may occur whenever  $y_1(\bar s)=y_2(\bar s)$.
\item[iii)]
If $|\lim_{s\to\bar s}y_1(s)|\to \infty$ or $|\lim_{s\to\bar s}y_2(s)|\to \infty$, 
dragging the contour along,
the integral in Eq.~\eqref{eq:Fint} may diverge at $\bar s$ and lead to a singularity
in $F(s)$. 
\end{itemize}
We now discuss these possibilities in turn.
\subsubsection{Endpoint singularity}
The quantity $y_1(s)$ [$y_2(s)$] vanishes linearly [tends to a constant] as $s\to s_t$,
as a result of which the derivative $dF(s)/ds$ diverges at threshold. We conclude 
that $F(s)$ is singular at $s=s_t$. The case $y_i(\bar s)=1$ does not occur in the low-energy region.

\subsubsection{Pinch singularity}\label{app:subsectionpinch}
Concerning point ii), we note that $D(\bar s)=0$ is a necessary and sufficient condition 
for the equality $y_1(\bar s)=y_2(\bar s)$ to hold, 
see Eq.~\eqref{eq:verybig}. The quantity $D(s)$ may be factorized in the following manner, 
\beq
D(s)=\frac{P(s,q_3^2,M_K^2;M_a,M_b,M_c,M_d)P_L(s,q_3^2,M_K^2;M_a,M_b,M_c,M_d)}{64M_K^2M_d^2s^2}\fs
\eeq
Here, $P$ and $P_L$ denote two polynomials of second order in $s$. 
Whereas $P$  has no 
roots in the low-energy region, the roots of $P_L$, explicitly given by
\begin{align}
P_L&=\lambda(s, M_d^2,M_c^2)\lambda\big(q_3^2,(M_a+M_b)^2,M_d^2\big) -Q^2(s,q_3^2,M_K^2;M_a,M_b,M_c,M_d)\scs\nnnl
Q&=2M_d^2(M_K^2-s-q_3^2)+\Big(q_3^2-(M_a+M_b)^2+M_d^2\Big)(s+M_d^2-M_c^2) \scs
\label{eq:PL+P}
\end{align}
denote the positions of the potential leading Landau singularities of the relativistic two-loop graph,
obtained from Fig.~\ref{fig:2loop_text}A by replacing all propagators with the relativistic ones~\cite{passeraetal}. 
We refer the interested reader to \ref{app:landau} for a discussion of 
this point. We conclude  that the positions of the 
potential pinch singularities 
in the non-relativistic graph Fig.~\ref{fig:2loop_text}A are identical to the location 
of the leading Landau singularities of the corresponding  relativistic Feynman diagram.

\begin{figure}
\includegraphics[width=\linewidth]{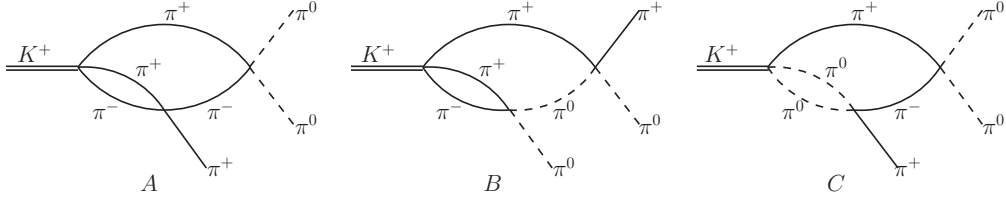}
\caption{Two-loop graphs contributing to $K^+\to \pi^+\pi^0\pi^0$.
Their holomorphic properties are discussed in~\ref{app:analytic} (\ref{app:relequal} and \ref{app:reldiff}) 
in the case where the propagators are the non-relativistic (the relativistic) ones.}
\label{fig:analytic}
\end{figure}
The relevant roots of the equation 
$D(s)=0$ can be classified according to the values of the
 masses $M_a$, $M_b$, $M_c$, $M_d$, and $M_3=(q_3^2)^{1/2}$. 
To simplify the discussion, we consider the following three cases, visualized in Fig.~\ref{fig:analytic}A--C, 
see also Fig.~\ref{fig:2loop_text}.
\begin{enumerate}
\item\label{item:s1}
\underline{Fig.~\ref{fig:analytic}A:} \\[1mm]
The equal-mass case with
$M_a=M_b=M_c=M_d=M_3=M_\pi$. Here,
$D(s)$ has a second-order zero on the real axis, at
$s=s_1=(M_K^2-M_\pi^2)/2$. 
\item\label{item:s1pm}
\underline{Fig.~\ref{fig:analytic}B:} \\[1mm]
$M_a=M_b=M_c=M_\pi,~M_d=M_3=M_{\pi^0}$. Here,
$D(s)$ has two real roots  $s_{1-}$ and $s_{1+}$, 
\beq\label{eq:s1pm0}
s_{1\pm}= \frac{1}{2}\left[M_K^2+2M_{\pi^0}^2-3M_\pi^2\right]
\pm\frac{1}{2M_\pi}
\left[(M_{\pi}^2-M_{\pi^0}^2)\lambda(M_K^2,M_\pi^2,4M_\pi^2)\right]^{1/2}\fs
\eeq
\item \label{item:anomthresh}
\underline{Fig.~\ref{fig:analytic}C: }\\[1mm]
$M_a=M_b=M_{\pi^0},~M_c=M_d=M_3=M_\pi$. Here,
$D(s)$ has complex-conjugated roots at $s=s_a,\,s_a^*$ 
in the complex $s$-plane, with
\beq\label{eq:sa0}
s_a=\frac{1}{2}\left[M_K^2+3M_\pi^2-4M_{\pi^0}^2\right]
-\frac{i}{2M_{\pi^0}}\left[(M_\pi^2-M_{\pi^0}^2)\lambda\big(M_K^2,M_\pi^2,4M_{\pi^0}^2\big)\right]^{1/2}\fs
\eeq
\end{enumerate}
Note that the roots differ from each other by terms of  $\Order(\eps^2)$.

\begin{figure}
\centering
\includegraphics[width=5cm,angle=0]{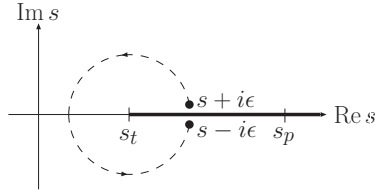}
\caption{The path in the complex $s$-plane from the upper rim of the cut to
the lower rim. The physical function at real values of $s$
coincides with the boundary value of the analytic function $F(s)$ on the
upper rim.}
\label{fig:circle_s}
\end{figure}
Let us now continue $F(s)$ to the lower rim of the real axis, 
along a generic path near  threshold, displayed in Fig.~\ref{fig:circle_s}. 
It represents a circle with  center at $s=s_t$ and a radius $|s-s_t|$. We assume  that 
${|s-s_t|}/(4M_\pi^2)$ is small, such that the zeros of $D(s)$ are avoided. 
The variable $s$ travels along this circle
from the upper rim of the positive real $s$-axis to its lower rim.

\begin{figure}
\centering
\includegraphics[width=0.5\linewidth]{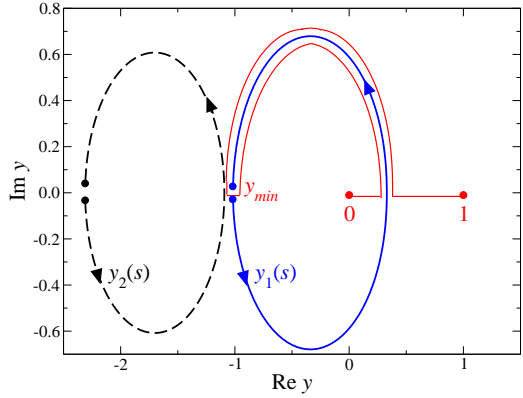}
\caption{The trajectories of $y=y_1(s)$ and $y=y_2(s)$ in the complex $y$-plane, when 
the variable $s$ travels around the generic 
circle shown in Fig.~\ref{fig:circle_s}. 
For $\mbox{Im}\, s<0$ the integration contour has to be deformed as shown,
in order to circumvent the singularity at  $y=y_1(s)$.}
\label{fig:eggstravel}
\end{figure}
\begin{figure}
\centering
\includegraphics[width=0.5\linewidth]{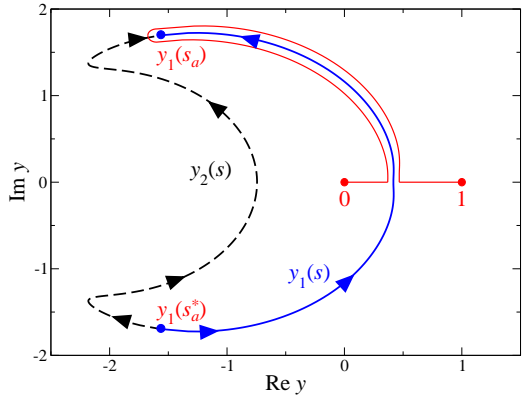}
\caption{The deformation of the integration contour in the variable $y$, 
as $s$ approaches the anomalous threshold $s_a$.}
\label{fig:pinch}
\end{figure}

The roots $y_1(s)$ and $y_2(s)$ then move in the complex plane along the 
trajectories shown in Fig.~\ref{fig:eggstravel}. When $\mbox{arg}\,(s-s_t)$
becomes equal to $\pi$, the root $y_1(s)$ hits the original integration
contour and enforces its deformation. When $s$ approaches the real axis from below,
the deformed path looks as follows (see Fig.~\ref{fig:eggstravel}):
it starts at $y=0$, goes in the negative direction, encircles the singularity
at $y=y_{min}(s)=y_1(s)$ and goes back to $y=1$. To illustrate what happens when a zero of $D$ is encountered, 
we consider $M_a=M_b=M_{\pi^0},~M_c=M_d=M_3=M_\pi$ (case~\ref{item:anomthresh} above), and  envisage 
analytic continuation of $F(s)$  from $s_a^*$ to $s_a$ along the segment
of a circle with  radius $|s_t-s_a|$. The trajectories of $y_{1,2}(s)$ for
the variable $s$ moving along this segment are shown in Fig.~\ref{fig:pinch}.
Since $y_1(s_a^*)=y_2(s_a^*)$ (and equally for $s_a^*\to s_a$), the trajectories
start at the same point and converge to the same point -- the potential generation 
of a pinch singularity at $s=s_a$ is clearly visible. The cases \ref{item:s1} and \ref{item:s1pm} look similar. 
In order to investigate whether these anomalous thresholds really do generate singularities in $F(s)$, 
we consider the function $H(s)$ defined by
\beq\label{eq:funcH(s)}
H(s)\doteq \int_0^1\frac{dy}{\sqrt{y}}\left[\frac{d}{dy}\bar g(y,s)\right]\ln{\bar g(y,s)}\scs\quad
\bar g(y,s)=(y-y_1)(y-y_2)\scs
\eeq
cf.\ the integral representation of $F(s)$ in Eq.~\eqref{eq:Fint}. 
$H(s)$ differs from $F(s)$ only by terms which are irrelevant 
as far as   holomorphic properties are concerned, but is much simpler to discuss. 
We denote by $H_+ (H_-)$ the function $H(s)$, evaluated at the upper (lower) rim of the real $s$-axis, by continuation 
according to Fig.~\ref{fig:circle_s}. 
We perform a partial integration and obtain
\beq\label{eq:Hpmanalytic}
H_-(s)=H_+(s)+\frac{4\pi}{3}\sqrt{-y_1}\left[y_1+3y_2\right]\,,\quad s>s_t\fs
\eeq
Inserting the expressions Eq.~\eqref{eq:verybig} for $y_{1,2}$, it is seen that the second term on the right hand side 
behaves as follows in the vicinity of the anomalous thresholds,
\beq\label{eq:branchpoints}
 \sqrt{-y_1}\left[y_1+3y_2\right]=c_0+\sum_{n=2}^\infty c_n(P_L)^{n/2}\,,\quad c_k\in\mathbb{C}\fs
\eeq
The polynomial $P_L$ is displayed in Eq.~\eqref{eq:PL+P}.
In other words, the function $H_-$ generates a singularity of the square root type at $s=s_{1\pm}$ ($s=s_a$) in the 
case~\ref{item:s1pm} (case~\ref{item:anomthresh}) listed above, whereas it is regular in case~\ref{item:s1}. 
The power of the leading singular behavior $\propto (P_L)^{3/2}$ in Eq.~\eqref{eq:branchpoints} 
agrees with the general analysis of these singularities provided in Ref.~\cite{nakanishi}.

\begin{table}
\renewcommand{\arraystretch}{1.2}
\centering
\addtocounter{footnote}{1}
\caption{The positions of the leading Landau singularities 
generated by the graphs displayed in Fig.~\ref{fig:allLLS} (in MeV).\protect\footnotemark[\value{footnote}]
The quantities $s_a,s_{1\pm}$ denote zeros of the polynomial $P_L$ 
displayed in Eq.~\eqref{eq:PL+P}.
For the mass configuration in Fig.~\ref{fig:allLLS}A (Fig.~\ref{fig:allLLS}D), 
$s_a$ ($s_{1 \pm}$) is given in Eq.~\eqref{eq:sa0} (Eq.~\eqref{eq:s1pm0}).
\label{tab:LLS}}
\vspace{2mm}
\begin{tabular}{cccc}\hline
&$\sqrt{s_{1-}}$&$\sqrt{s_{1+}}$&$\sqrt{s_a}$\\\hline
$A$&&&$339.6-i 25.4$\\
$B$&&&$339.6-i 25.4$\\
$C$&&&$341.6-i 26.7$\\
$D$&308.4&355.8\\
$E$&308.9&359.0\\ 
$F$&308.9&359.0\\
\hline
\end{tabular}
\renewcommand{\arraystretch}{1.0} 
\end{table}
\footnotetext[\value{footnote}]{We use the
following masses: $M_\pi=139.6$~MeV, $M_{\pi^0}=135.0$~MeV, $M_K=M_{K^+}=493.7$~MeV, $M_{K^0}=497.6$~MeV.} 

\begin{figure}
\centering
\includegraphics[width=12cm]{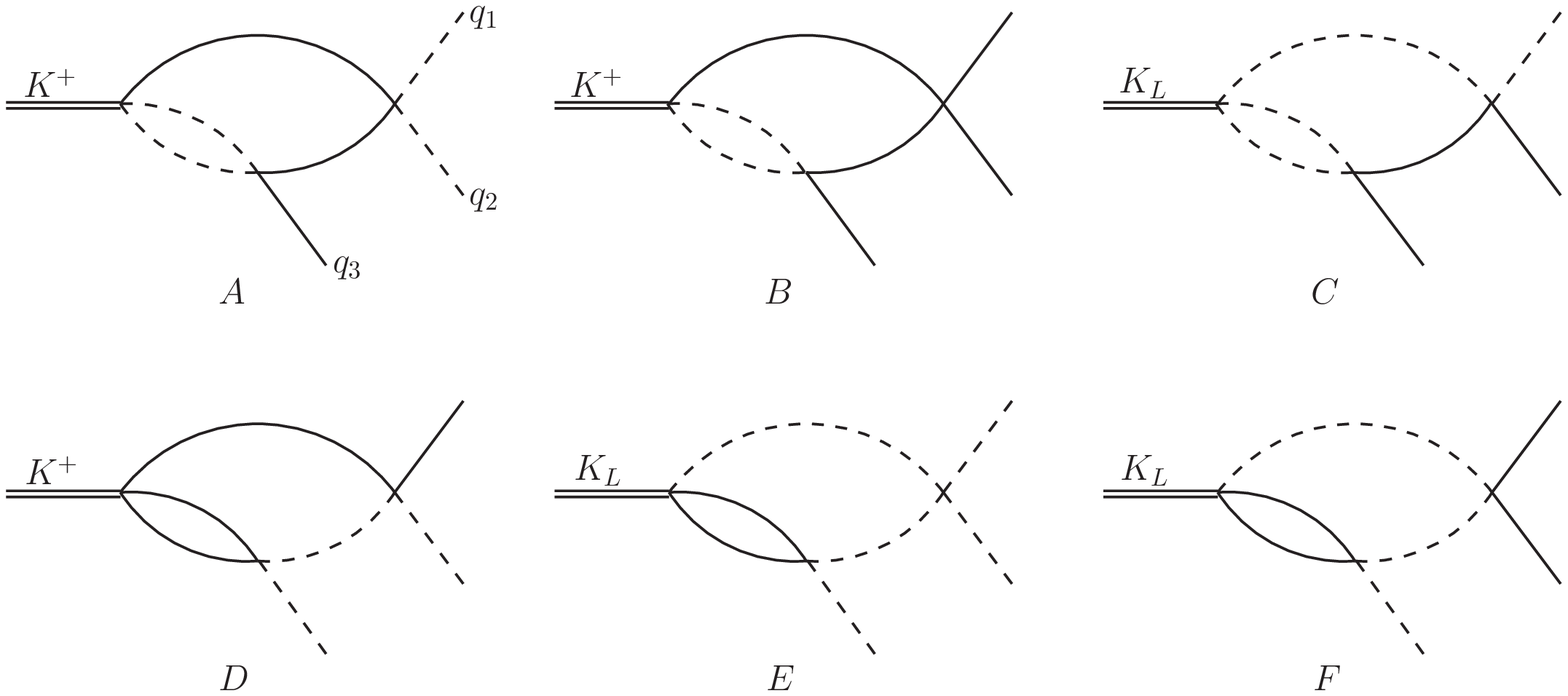}
\caption{The graphs that generate leading Landau singularities in $K\to 3\pi$ decays at two-loop 
order. In Figs.~A--C (D--F) the singularities are located  in the complex plane, 
at $s=s_a$ (on the real axis, 
at $s=s_{1\pm }$), where $s$ denotes the invariant mass of the outgoing two rescattered pions, 
e.g.\ $s=(q_1+q_2)^2$ in graph~A. See Table~\ref{tab:LLS} for numerical values.}
\label{fig:allLLS}
\end{figure}

It turns out that the topologies displayed in Fig.~\ref{fig:analytic} are the generic ones 
for the occurrence of leading Landau singularities, in the following sense. 
At two loops, these singularities occur iff the 
four-pion vertex at the inner loop amounts to a charge-exchange one, 
as is the case in Figs.~\ref{fig:analytic}B, C.  In Fig.~\ref{fig:allLLS} we display all two-loop graphs that
generate leading Landau singularities, 
for both $K^+\to 3\pi$ and $K_L\to 3\pi$ decays. 
The solid (dashed) lines denote charged (neutral) pions, and the double lines incoming kaons. 
The graphs in Fig.~\ref{fig:allLLS}A--C (D--E) generate singularities in the complex plane, at $s=s_a$ 
(on the real axis, at $s=s_{1 \pm}$), see Table~\ref{tab:LLS} for numerical values.  
We conclude that, at two-loop order,  leading Landau singularities do occur in {\it all} decay channels.

The results  Eqs.~\eqref{eq:Hpmanalytic}, \eqref{eq:branchpoints} have important consequences for the decomposition 
of the amplitudes into regular and singular parts, as proposed in Ref.~\cite{cabibboisidori}, 
see \ref{app:decomposition}.

\subsubsection{Singularity at the pseudothreshold}
Finally, we discuss singularities at the pseudothreshold $s=s_p\doteq (M_K-M_3)^2$. 
As can be seen from the explicit expression in Eq.~\eqref{eq:verybig}, 
$\lim_{s\to s_p}y_1(s)=-\infty$, whereas $y_2$ stays finite.
 From Eq.~\eqref{eq:Hpmanalytic}, we conclude that $H_-$ therefore tends to infinity 
as $s\to s_p-i\epsilon$. The same is therefore true for $F(s)$.

\bigskip
In summary, we conclude that, 
in the cases \ref{item:s1} and \ref{item:s1pm}, the function $F(s)$ 
is analytic in the low-energy region of the complex $s$-plane,
cut along the positive axis. The cut starts at the normal threshold $s=s_t=(M_c+M_d)^2$
 and extends beyond the pseudothreshold.
 In the case \ref{item:anomthresh}, the cut must be deformed in order to encompass the anomalous threshold at $s=s_a$.
We refer the reader to \ref{app:relequal} and \ref{app:reldiff} 
for more detailed discussions concerning the relativistic diagrams.
The physical values of $F(s)$ are obtained by approaching
 the real axis from above.

\subsection{The discontinuity of $F$}
\label{app:discontinuity}

The origin of the discontinuity $\Delta F(s)=F(s+i\eps)-F(s-i\eps)$ of the function $F(s)$ 
across the cut was identified above, in case of the function $H(s)$: the integration contours for $s$ located
on the upper and on the lower rims of the cut are not the same. In principle, in the same manner, $\Delta F(s)$
could be evaluated from the integral representation Eq.~\eqref{eq:Fint}.
For actual calculations this procedure is not very convenient, because
$ {\cal F}(y,s)$ is only known in form of the series Eq.~\eqref{eq:calFn}.
For this reason, we now evaluate $\Delta F(s)$   by using a different 
method.

Since $F(s)$ and $\bar {\cal M}(s)$ only differ by a constant, see Eq.~\eqref{eq:F-barM}, 
we calculate
the discontinuity of $\bar {\cal M}(s)$ in the vicinity of the elastic threshold
$s=s_t$ and then analytically continue it in $s$.
The discontinuity of $\bar {\cal M}(s)$ in the vicinity of threshold 
$q_0^2\ll \Delta^2$
can be obtained directly from Eq.~\eqref{eq:nowexpand}, since only the first
denominator is singular:
\begin{align}\label{eq:disc_NR}
\Delta\bar {\cal M}(s) &= \frac{\pi i}{\sqrt{s}}\,
\int\frac{d^dl}{(2\pi)^d}\,\frac{d^dk}{(2\pi)^d}\,
\delta({\bf k}^2-q_0^2)
\frac{N(x)}{({\bf l}^2+\frac{(1+\delta)^2}{4}\,{\bf Q}^2+x-\Delta^2)} \nnnl
&=-\frac{iq_0}{32\pi^2\sqrt{s}}\,\int_{-1}^{1} dy\,
N(\hat x)\biggl(\frac{(1+\delta)^2}{4}\,{\bf Q}^2+\hat x-\Delta^2\biggr)^{1/2}+\Order(d-3)\, , \\
\hat x&=q_0^2+y^2\frac{q_0^2{\bf Q}^2}{s}+y\frac{q_0|{\bf Q}|Q^0(1+\delta)}{\sqrt{s}}
\, , \quad
\frac{(1+\delta)^2}{4}\,{\bf Q}^2+\hat x=
\biggl(\frac{Q^0(1+\delta)}{2}+y\frac{q_0|{\bf Q}|}{\sqrt{s}}\biggr)^2
-M_c^2\, . \nonumber
\end{align}
Calculating the integral over $y$, we obtain
\beq\label{eq:DiscM}
\Delta{\cal M}(s)=-\frac{1}{128\pi^2\sqrt{s}}\,
\Big(\Phi(q_0)-\Phi(-q_0)\Big)\, ,
\eeq
where
\begin{align}\label{eq:def-Phi}
\Phi(y)&=
a\ln\Bigl(\sqrt{b-c_-a-y}+\sqrt{b-c_+a-y}\Bigr)
-\sqrt{(b-c_-a-y)(b-c_+a-y)}
\nnnl
&+2a\sqrt{c_+c_-}\ln\frac{\sqrt{c_+(b-c_-a-y)}+\sqrt{c_-(b-c_+a-y)}}{\sqrt{b-y}}\, ,
\end{align}
and
\beq\label{eq:def-ab}
a=\frac{2(M_a^2+M_b^2)\sqrt{s}}{2M_K|{\bf Q}|}\, ,\quad
b=\frac{(M_K^2+M_c^2-M_KQ_0(1+\delta))\sqrt{s}}{2M_K|{\bf Q}|}\, , \quad
c_\pm=\frac{(M_a\pm M_b)^2}{2(M_a^2+M_b^2)}\, .
\eeq
Closer inspection of the trajectory of $y_1(s)$, see Fig.~\ref{fig:eggstravel}, shows that,
in order to perform the analytic continuation consistently, 
the variable $s$ in the discontinuity should have an infinitesimal 
{\em negative} imaginary part, $s\to s-i\eps$.

\section{Relativistic two-loop integrals -- equal pion masses}
\label{app:relequal}

The evaluation of the two-loop graphs  Fig.~\ref{fig:2loop_text}A  in the 
non-relativistic theory is presented in Sect.~\ref{sec:2loops}, 
and the holomorphic properties of the pertinent loop functions
are discussed in \ref{app:analytic}.
The  procedure is quite complex and non-standard. 
We therefore wish to compare  the
result with the calculation in relativistic quantum field theory,
 where the propagators in
Fig.~\ref{fig:2loop_text}A are taken to be the relativistic ones. 
In this and the following Appendix, we discuss
this more standard calculation, and, in particular, the holomorphic properties of the pertinent loop functions. 
We stick here to the three topologies shown in Fig.~\ref{fig:analytic}.
In \ref{app:relequal} (\ref{app:reldiff}), we consider
 the case of equal (unequal) pion masses running in the loops, see  Fig.~\ref{fig:analytic}A (Figs.~\ref{fig:analytic}B, C).
We use scalar vertices throughout, because derivative couplings
 do not change the holomorphic properties of the integrals.
 A  comparison between the two approaches  is provided 
in \ref{app:compare_nrrel}, where we show that the non-relativistic calculation approximates the relativistic one 
in a well-defined manner -- as it must be for the present framework to make sense.

\subsection{Notation }
\label{app:notation}
 We use dimensional regularization and put
\beq
\omega=\frac{D}{2}-2\scs
\eeq
where $D$ denotes the dimension of space-time.
Loop integrations are symbolized by a bracket,
\beq
\la\ldots\ra_l = \int \frac{d^Dl}{i(2\pi)^D}(\ldots)\scs\quad
\la\la\ldots\ra\ra_{lk} =\int\frac{d^Dl}{i(2\pi)^D}\int\frac{d^Dk} {i(2\pi)^D}(\ldots)\fs
\eeq
 We abbreviate 
Feynman parameter integrals by
\beq
\{\ldots\}_{1}=\int_0^1 dx_1\{\ldots\}\scs\quad
\{\ldots\}_{12}=2\int_0^1 dx_1\int_0^1 x_2 dx_2\{\ldots\}\fs
 \eeq
Furthermore,
 we use the measure
\beq
[d\sigma]=\frac{C(\omega)\Gamma(3/2)}
{\Gamma(3/2 + \omega)}\frac{\sigma^\omega}{4^\omega}\left(1-\frac{4M_\pi^2}{\sigma}\right)^{\omega+1/2}\;d\sigma ~,
\eeq
with
\beq
C(\omega)=\frac{1}{(4\pi)^{2+\omega}}\fs
\eeq
Further, in order to simplify the notation, we will suppress the dependence of the various loop functions  on the meson masses.

\subsection{The loop function $\bar{K}(s)$}
\label{app:loopfunctionF}

As mentioned, we first consider 
the case where the masses of all pions running in the loops, as well as of the
outgoing pion with momentum $q_3$, are 
taken to be equal [cf.\  Figs.~\ref{fig:2loop_text}A and \ref{fig:analytic}A],
\beq
M_a=M_b=M_c=M_d=M_\pi\scs \quad q_3^2=M_\pi^2\fs
\eeq
The fact that we consider two neutral pions in one of the vertices does not change the holomorphic properties of the graph, 
as these pions enter the final expression only 
through the square of their  total four momentum, $s=(q_1+q_2)^2$. 
  The pertinent loop integral is
\beq\label{eq:feynman}
G(s)=\bla\bla\frac{1}{D_1D_2D_3D_4}\bra\bra_{lk}\scs
\eeq
with
\beq
D_1=M_\pi^2-k^2\,,\quad D_2=M_\pi^2-(Q-k)^2\,,\quad
D_3=M_\pi^2-l^2\,,\quad D_4=M_\pi^2-(P_K-l-k)^2\,,
\eeq
and
\beq
Q^\mu = (P_K-q_3)^\mu\,,\quad P_K^2=M_K^2\,,\quad Q^2=s\,.
\eeq
The integral $G(s)$ is proportional to the quantity $V_{121}$ investigated in 
great detail for real external momenta in
Ref.~\cite{passeraetal}. In particular, these authors have set up codes to evaluate
 the ultraviolet  finite part of $G(s)$ by numerical integration. Here, 
the purpose is quite different: we wish 
to investigate the holomorphic properties of $G(s)$ as well. Our method to evaluate
these integrals is therefore necessarily very different from the one developed 
in Ref.~\cite{passeraetal}.

 We proceed  in two steps. First, we identify the part of  $G(s)$ 
that stays finite as $D\rightarrow 4$ by subtracting the
divergent subintegral and by removing the remaining overall divergence. In the
second step, we express  the finite part through a once subtracted dispersion
relation in the variable $s$.

To start with, we consider the inner loop
\beq\label{eq:J(t)}
J( t)=\bla\frac{1}{D_3D_4}\bra_l=C(\omega)\Gamma(-\omega)
\big\{z^\omega\big\}_1\,,\quad
z=M_\pi^2-x_1(1-x_1) \, t\,, \quad t=(P_K-k)^2\,.
\eeq
It is useful to write $J(t)$ in  a dispersive manner~\cite{pipi2loops},
\beq\label{eq:J(t)disp}
J( t)= 
\int_{4M_\pi^2}^\infty\frac{[d\sigma]}{\sigma- t}\,,
\eeq
and to single out its divergence at $D=4$,
\beq
J( t)=J(0)+\bar J( t)\,,\nnnl
\eeq
such that $G(s)$ in Eq.~\eqref{eq:feynman} becomes
\beq
\label{eqa:defF(s)}
G(s)=J(s)J(0)+K(s),\quad
K(s)=
\int_{4M_\pi^2}^\infty\frac{ \left[d\sigma\right] }{\sigma}
\bla\frac{  t}{D_1D_2(\sigma- t)}\bra_k\fs
\eeq
This integral has still an overall divergence at $D=4$. 
It can be made finite by subtracting its value at $s=0$,
\beq\label{eqa:asymptF}
K(s)=K(0)+{\bar K}(s)\fs
\eeq
As a result, one has
\beq
G(s)=J(s)J(0)+K(0)+\bar K(s)\fs
\eeq
We now discuss the finite part $\bar K(s)$.

\subsection{Peculiarities of   $\bar{K}(s)$}
\newcommand{\discK}{\Delta\bar{\cal K}}
To illustrate the difficulties one is faced with when investigating the holomorphic properties of $\bar K(s)$,
 we  first consider  the one-loop integral 
 $J(t)$ in Eq.~\eqref{eq:J(t)}, in $D$ dimensions. 
The $i\epsilon$ prescription 
for the square of the masses 
in the propagators is equivalent to providing the variable $t$ with a positive
imaginary part: the integral Eq.~\eqref{eq:J(t)} defines a function ${\cal J}(t)$
which is holomorphic in the 
upper half plane  ${\rm Im}\,t>0$,
and real on the real axis for $t <4M_\pi^2$. It may therefore be holomorphically continued to the lower half plane through the 
Schwarz reflection principle. The denominator $z$ vanishes at two points $x_{1,2}\in(0,1)$ for $t > 4 M_\pi^2$.
Therefore, ${\cal J}(t)$ is holomorphic in the complex plane, cut along the real axis for ${\rm Re}\,t\geq 4 M_\pi^2$,
and $J(t)$ is the boundary 
value $J(t)=\lim_{{\rm Im}\,t\to 0^+} {\cal J}(t)$. 
The dispersive representation in Eq.~\eqref{eq:J(t)disp} 
is based on these observations.

\begin{figure}
\centering
\includegraphics[width=4.5cm]{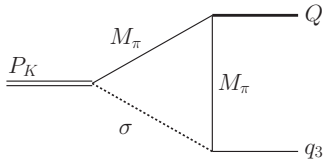}
\caption{The triangle graph which corresponds to the amplitude $T(s)$ in 
Eq.~\eqref{eq:defTs}. We use  $P^\mu_K=(Q+q_3)^\mu$, $Q^2=s$, $q_3^2=M_\pi^2$, $P_K^2=M_K^2$.}
\label{fig:triangle1}
\end{figure}

This chain of arguments cannot be carried over to  $\bar K(s)$ without further ado.
  To show why this is so, we consider the integrand
\beq\label{eq:defTs}
T(s)=\bla\frac{1}{D_1D_2(\sigma-t)}\bra_k\scs \sigma\geq 4M_\pi^2\scs t=(P_K-k)^2
\eeq
in Eq.~\eqref{eqa:defF(s)} [we drop $t$ in the numerator, because this does not 
  affect the analytic properties of $T(s)$]. 
The quantity $T(s)$ corresponds to the triangle diagram 
displayed in Fig.~\ref{fig:triangle1}. 

The integral is ultraviolet convergent and can thus
be evaluated at $D=4$. After integration over the momentum $k$, 
we are left with the Feynman parameter integral
\beq\label{eq:Tfeynman}
T(s)=\big\{z^{-1}\big\}_{12}\,,\quad
z=x_2^2M_\pi^2+\sigma(1-x_2) 
-x_2x_1\left\{s(1-x_1)x_2+(1-x_2)
(M_K^2-M_\pi^2)\right\}-i\epsilon\,.
\eeq
 We have explicitly displayed the $i\epsilon$ prescription. 
In this case, we cannot replace $i\epsilon$ by providing $s$ with a positive
imaginary part: the denominator $z$ vanishes at two points on
the $x_1=1$ boundary, for any value of $s$, provided that $M_K>\sqrt{\sigma} +M_\pi$, 
a condition which is met for the  physical value of the kaon mass. 
In other words, at $\epsilon=0$, the
integral Eq.~\eqref{eq:Tfeynman} is not well defined for real, physical values of 
$M_\pi^2, M_K^2$, at any $s$, and cannot straightforwardly be considered 
to be the boundary 
value of  a holomorphic function ${\cal T}(s)$. This 
renders the discussion  more 
complicated than in the case of $J(s)$~\cite{barton1}. A similar remark applies to  $\bar K(s)$.

A way out of the problem is based on the following observation. 
First, examining Eq.~\eqref{eq:Tfeynman}, one notes that $T(s)$ has 
a unique extension ${\cal T}(s)$ into the product of the
upper half complex planes of $s$ and $M_K^2$.\footnote{It was already noted
 in Ref.~\cite{kaellen} that $T(s)$ is holomorphic when all three variables 
 $M_K^2$, $s$, and $q_3^2$ lie in the same half planes.}
One then considers the kaon mass to be an adjustable 
parameter~\cite{anisovich,barton1,fronsdal}. 
 For sufficiently small value of
$M_K^2$,  the above-mentioned problem does not exist -- the function ${\cal T}(s)$ satisfies an
unsubtracted dispersion relation in the variable $s$, with threshold
$s_t=4M_\pi^2$. For physical values of the kaon mass, ${\cal T}(s)$ is then 
defined by analytic continuation of the dispersive representation 
through the upper complex $M_K^2$-plane, 
and 
$T(s)=\lim_{\,{\rm Im}\,s\to 0^+,\,{\rm Im}\,M_K\to 0^+}{\cal T}(s)$~\cite{anisovich,barton1,fronsdal}.

This procedure may now be directly carried over to $\bar K(s)$ -- the final
integration over  $\sigma$ in Eq.~\eqref{eqa:defF(s)} does 
not affect the arguments just
outlined, aside from the necessity to use a subtracted dispersion relation.
 Concerning the continuation in  $M_K^2$, we consider the Cauchy representation
\beq\label{eqa:dispF}
\bar{\cal K}(s) = \frac{s}{2\pi i}
\int_{4M_\pi^2}^{\infty\hspace{-.6cm}{{\cal C}}}\frac{dx\, \Delta\bar{\cal K}(x)}{x(x-s)}\scs
\eeq
where ${\cal C}$ denotes the path of integration in the complex $s$-plane, and  
$\discK$
stands for the discontinuity of $\bar {\cal K}$ across ${\cal C}$.
 For small kaon masses, ${\cal C}$ may be taken to run along the
real axis from $4M_\pi^2$  to infinity, and $\bar {\cal K}$ is holomorphic in the complex $s$-plane, cut along ${\cal C}$. The
discontinuity  is itself a holomorphic function  $\discK (s)$ in a certain region of the complex $s$-plane, with singularities $s_i(M_K,M_\pi)$   whose positions depend on the pion and kaon masses. These singularities generate  singularities in the function $\bar{\cal K}$ on higher Riemann sheets.   
During the process of increasing the
value of the kaon mass to its physical mass, one has to make sure 
that the singularities $s_i$ do not
cross the path ${\cal C}$, by eventually deforming it 
 properly~\cite{anisovich,barton1,fronsdal} -- in other words, the singularities in $\bar{\cal K}$ may eventually move to the first Riemann sheet.  Once the continuation has
been accomplished, it is straightforward to read off the analytic
properties of ${\bar{\cal K}}(s)$, together with e.g.\ its threshold behavior, which is of 
particular interest in our case. Finally, one has
$\bar K(s)=\lim_{\,{\rm Im}\,s\to0^+,\,{\rm Im}\,M_K\to 0^+}\bar{\cal K}(s)$.

\subsection{The discontinuity  $\Delta\bar{\cal K}$}
It remains to evaluate the discontinuity in the dispersive
representation Eq.~\eqref{eqa:dispF}. For sufficiently small values of the kaon
mass, unitarity of the $S$-matrix fixes it unambiguously.
It is given by the angular integral\footnote{Here and below, we  denote the independent variable in the discontinuity with $s$.}
\begin{align}\label{eq:angular}
\discK(s)&= \frac{iv(s)}{256\pi^3}
\int_{-1}^{1}dz \,\left\{\beta(z)\ln\frac{\beta(z)-1}{\beta(z)+1}+2\right\}\,,\quad s\in [4M_\pi^2,\infty]\,;\quad
v(s)= \left(1-\frac{4M_{\pi}^2}{s}\right)^{1/2}\scs \nnnl
\beta(z) &= \left(1-\frac{4M_{\pi}^2}{A+Bz}\right)^{1/2}\scs \quad
A = \frac{1}{2}\left[M_K^2+3M_\pi^2-s\right]\,\scs \quad
B=\frac{v(s)}{2}\lambda^{1/2}\left(s,M_K^2,M_\pi^2\right)\fs
\end{align}
After the substitution $(\beta(z)-1)/(\beta(z)+1)=u$, the integral  can straightforwardly be 
performed. For the following discussions, it is useful to introduce 
the functions
 \begin{align}\label{eq:PhiFGdef}
\Phi(x,y;z_+,z_-;m)&= m^2(r_+^2- r_-^2)-(x+y)z_+ r_+ + (x-y)z_- r_-\scs &
r_\pm&= \ln(1+z_\pm)-\ln(z_\pm-1)\scs\nnnl
F(x,y;z_+,z_-;m)&= \left.\Phi(x,y;z_+,z_-;m)\right|_{r_\pm\rightarrow R_\pm} \scs &
R_\pm&= \ln(1+z_\pm)-\ln(1-z_\pm)\scs \nnnl
G(x,y;z_+,z_-;m)&= 2m^2(R_- +R_+) - (x+y)z_+ -(x-y)z_-\fs
\end{align}
The discontinuity for small values of the kaon mass is given by
\beq\label{eq:discI}
\Delta\bar{\cal K}(s)= \frac{iv(s)}{256\pi^3}
  \left[6+B^{-1}D(s)\right]\,,\quad
D(s)= \Phi(A,B;W_+,W_-;M_\pi)\scs\quad
W_\pm= \left(1-\frac{4M_\pi^2}{A\pm B}\right)^{1/2}\fs
\eeq
According to the discussion above, this representation must be continued
analytically  to the physical value of $M_K^2$. Potential singularities in the
discontinuity occur for those values of $s$ where the arguments of $W_\pm$
vanish, or  $B$ vanishes. To investigate the first possibility, 
we observe that  the quantity
 $(A-4M_\pi^2)^2-B^2$
agrees up to a factor  with the polynomial $P_{\ref{fig:analytic}A}$ in Eq.~\eqref{eq:polynomialFig6ACL} that determines 
the location of potential leading Landau singularities in the graph Fig.~\ref{fig:analytic}A,
\beq
\label{eq:wminuszero}
(A-4M_\pi^2)^2-B^2=\frac{4M_\pi^2}{s}(s-s_1)^2\scs \quad s_1=\frac{1}{2}(M_K^2-M_\pi^2)\fs
\eeq
The zeros of $B$ occur at $s= 4M_\pi^2$, $(M_K\pm M_\pi)^2$.
 Because $M_K^2$ is equipped with a positive imaginary part,  the zeros at 
$s=s_1$, $(M_K \pm M_\pi)^2$ are located above the path ${\cal C}$ of integration.
At $s=4M_\pi^2$, the discontinuity vanishes.

There are thus four intervals to be distinguished, 
\beq\label{eq:regions}
\begin{tabular}{lcccc}
\word{Interval}&\word{IIIb}&\word{IIIa}&\word{II} &\word{I}\\
&$[s_t,s_1]$&
$[s_1,s_p]$&
$[s_p,s_c]$&
$[s_c,\infty)$
\end{tabular}
\eeq
where
\beq\label{eq:sts1spsc}
s_t = 4M_{\pi}^2\scs \quad
s_p = (M_K-M_{\pi})^2\scs \quad
s_c=(M_K+M_\pi)^2\fs
\eeq
We now list the discontinuity in the 
 four intervals for physical values of the kaon mass,  and use the notation
\beq\label{eq:barB}
\bar B=\frac{v(s)}{2}|\lambda(s,M_K^2,M_\pi^2)|^{1/2}\fs
\eeq

\vskip3mm

\noindent\underline{Interval I} \ [$\discK$ is imaginary] \nopagebreak

\vskip2mm \noindent
Here, $B=\bar B$, and $W_\pm-1>0$.
The discontinuity may therefore  
simply be evaluated from the expression given in Eq.~\eqref{eq:discI}. 
As a check we have verified that the expression for 
$\discK$, evaluated at  $M_K=M_\pi=1$, 
agrees with the discontinuity of the two-loop function $\bar V_1$ 
worked out in Ref.~\cite[Appendix~A.2] {pipi2loops}.

\vskip3mm

\noindent\underline{Interval II} \ [$\discK$ is imaginary]\nopagebreak

\vskip2mm \noindent
The function $D(s)$ is evaluated as in interval I, with $B=-i\bar B$.

\vskip3mm

\noindent\underline{Interval III} \ [$\discK$ is complex]\nopagebreak

\vskip2mm \noindent
In this interval, $B=-\bar B$, and  the discontinuity is obtained from
\beq\label{eq:discequal}
D(s)= i\pi G(A,B;W_+,W_-;M_\pi)+F(A,B;W_+,W_-;M_\pi)\scs\quad
W_\pm= c_\pm\left|1-\frac{4M_\pi^2}{A\pm B}\right|^{1/2}\fs
\eeq
The numbers $c_\pm$ are 
\beq
\begin{tabular}{lcc}
\word{Interval}&\word{IIIa}&\word{IIIb}\\[2mm]
$c_+$&$1$&$-1$\\[2mm]
$c_-$&$1$&$1$
\end{tabular}
\eeq
 The discontinuity develops a singularity at the
pseudothreshold $s=s_p$. This is due to the fact that the angular integral
in Eq.~\eqref{eq:angular} must be deformed to infinity in the complex plane as $s$ 
approaches $s_p$~\cite{anisovich}, and the angular integral finally diverges
 at $s=s_p$. On the other hand, there is no singularity at $s=s_1$ [second order zero  
in Eq.~\eqref{eq:wminuszero}].

\begin{figure}
\centering
\includegraphics[width=0.5\linewidth]{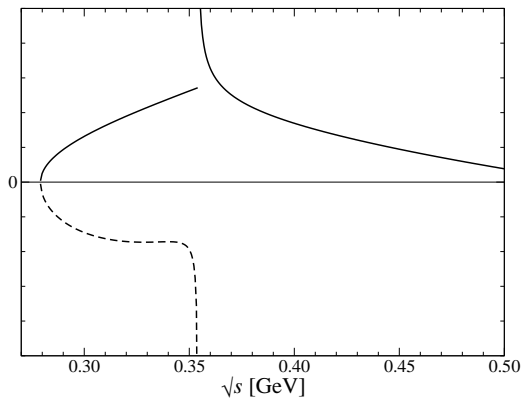}
\caption{The discontinuity $\discK$(s) as a function of 
 $\sqrt{s}$, in the equal mass case, in arbitrary units.
 The solid (dashed) curve stands for the imaginary
 (real) part
  of $\discK$. The singularity at the pseudothreshold
  $\sqrt{s_p}$ is clearly seen.}\label{fig:disc}
\end{figure}

In Fig.~\ref{fig:disc}, we display the discontinuity
$\discK(s)$ as a 
function of $\sqrt{s}$. The solid (dashed) 
curve stands for the imaginary  (real) part of $\discK$. 
The singularity at the pseudothreshold  $s_p$ is clearly visible.
From the explicit expression of the discontinuity, one furthermore concludes 
that it generates a square root singularity at threshold,
\beq
\discK(s) = v(s) d(s)\scs
\eeq
with $d(s)$ holomorphic at $s=4M_\pi^2$. 

\subsection{Holomorphic properties of   $\bar{\cal K}(s)$}
\begin{figure}
\centering
\includegraphics[width=0.45\linewidth]{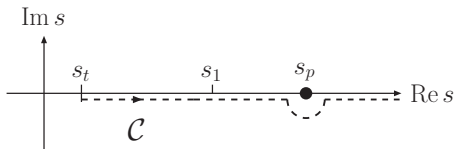}
\caption{The path ${\cal C}$  in the dispersive integral Eq.~\eqref{eqa:dispF}, for physical kaon mass. 
The symbols $s_t$, $s_1$, and $s_p$ are defined in Eq.~\eqref{eq:sts1spsc}. }
\label{fig:patheq}
\end{figure}

\begin{figure}
\centering
\includegraphics[width=0.5\linewidth]{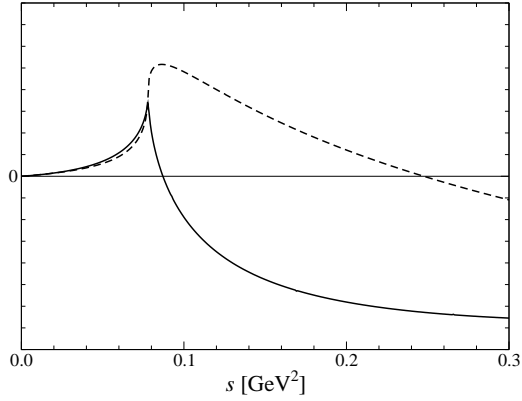}
\caption{The function $\bar{\cal K}(s)$ evaluated numerically from 
 Eq.~\eqref{eqa:dispF}, in arbitrary units, with $s$ at the upper rim of the cut. 
The solid (dashed) line denotes the real (imaginary) part of $\bar{\cal K}(s)$.}
\label{figa:functionF}
\end{figure}

We deform  the path ${\cal C}$ in the representation
Eq.~\eqref{eqa:dispF} such that the singularity 
of the discontinuity
at the pseudothreshold $s_p$ is avoided in the manner 
 indicated in Fig.~\ref{fig:patheq}, as is requested by the 
continuation in $M_K^2$.
 The function $\bar {\cal K}(s)$ is thus holomorphic in
 the complex $s$-plane, cut along the real axis for $s\geq
 4M_\pi^2$. Furthermore, on the upper rim of the cut, it is 
holomorphic as well. The singularity of the
 discontinuity at $s=s_p$ shows up in $\bar {\cal K}(s)$ 
only on the lower rim of the cut, where it diverges 
when $s_p$ is approached from below.
We may perform the dispersion integral Eq.~\eqref{eqa:dispF} numerically.
The result is displayed in Fig.~\ref{figa:functionF}, with the variable $s$ at the upper rim of the cut.

\section{Relativistic two-loop integrals -- unequal pion masses}
\label{app:reldiff}

We now investigate the graphs Fig.~\ref{fig:analytic}B, C, see also Fig.~\ref{fig:2loop_text}A.
The corresponding ultraviolet finite loop functions $\bar{\cal K}_{B,C}$  are again written as a dispersive integral,
\beq\label{eq:dispKBC}
\bar{\cal K}_I(s) = \frac{s}{2\pi i}
\int_{(M_c+M_d)^2}^{\infty\hspace{-.6cm}{{\cal C}}}\frac{dx\, \Delta\bar{\cal K}_I(x)}{x(x-s)}\scs \quad  I=B,C\fs
\eeq
The  discontinuities are given by angular integrals
\beq\label{eq:angularintegral}
\discK_I (s) =
\frac{iv_{cd}(s)}{256\pi^3}\int_{-1}^{1}
\left\{\beta_I(z)
\ln\frac{\beta_I(z)-1}{\beta_I(z)+1}+2\right\}
dz\,
;\quad s \in [(M_c+M_d)^2,\infty]\scs \nnnl 
\eeq
where
\begin{align}
v_{cd}(s)&=\frac{\lambda^{1/2}(s,M_c^2,M_d^2)}{s} \scs \quad
\beta_B(z)= \left(1-\frac{M_\pi^2}{\hat A+\hat Bz}\right)^{1/2} \scs \quad
\beta_C(z)= \left(1-\frac{M_{\pi^0}^2}{\hat A+\hat Bz}\right)^{1/2}\scs\nnnl
\hat A&= M_K^2+M_c^2-\frac{1}{2s}(s+M_c^2-M_d^2)(M_K^2+s-q_3^2)\scs \quad
\hat B = \frac{v_{cd}(s)}{2}\lambda^{1/2}(s,M_K^2,q_3^2) \fs
\end{align}
For later use,  we introduce
\beq
{\bar{B}} = \frac{v_{cd}(s)}{2s} |\lambda(s,M_K^2,q_3^2)|^{1/2}\fs
\eeq

\subsection{Case I}\label{subsec:caseI}
We first consider the graph in Fig.~\ref{fig:analytic}B,
\beq\label{eq:unequalI}
M_a=M_b=M_c=M_\pi\,,\quad M_d=M_{\pi^0}\,,\quad q_3^2=M_{\pi^0}^2\fs
\eeq
Singularities are  generated by the zeros in
\beq
\hat W_{\pm}=\left(1-\frac{4M_\pi^2}{\hat A\pm \hat B}\right)^{1/2}\fs
\eeq
The quantity $(\hat A-4M_\pi^2)^2-\hat B^2$ agrees up to a factor with the polynomial $P_{\ref{fig:analytic}B}$ in Eq.~\eqref{eq:polynomialFig6ACL}, which shows that corresponding 
singularities in $\discK_B$ occur at 
\beq\label{eq:s1Fig6B}
s_{1\pm}= \frac{1}{2}\left[M_K^2+2M_{\pi^0}^2-3M_\pi^2\right]
\pm\frac{1}{2M_\pi}
 \left [(M_{\pi}^2-M_{\pi^0}^2)\lambda(M_K^2,M_\pi^2,4M_\pi^2) \right ]^{1/2}\fs
\eeq
We have to distinguish the following intervals:
\beq
\begin{tabular}{lccccc}
\word{Interval}&\word{IIIc}&\word{IIIb}&\word{IIIa}&\word{II} &\word{I}\\
&$[s_t,s_{1-}]$&
$ [s_{1-},s_{1+}]$&
$[s_{1+},s_p]$&
$[s_{p},s_c]$&
$[s_c,\infty)$
\end{tabular}
\eeq
where
\beq\label{eq:s1pm}
s_t= (M_{\pi^0}+M_{\pi})^2\,,\quad
s_p=(M_K-M_{\pi^0})^2\,,\quad
s_c=(M_K+M_{\pi^0})^2\fs
\eeq
The discontinuity is
\beq\label{eq:discII}
\discK_{B} = \frac{iv_{+0}(s)}{256\pi^3}\left[6+ \hat B^{-1}\hat D_B(s)\right]
\scs 
\eeq
where $\hat D_B(s)$ is given by the following expressions in the intervals I--III:

\vskip3mm

\noindent
\underline{Interval I} \ [$\discK_B$ is imaginary] \nopagebreak

\vskip2mm
\noindent 
Here, one has $\hat B=\bar B$, and
\beq
      \hat D_B(s)= \Phi(\hat A,\hat B;W_+,W_-;M_\pi)\scs\quad
W_\pm = \left(1-\frac{4M_{\pi}^2}{\hat A\pm \hat B}\right)^{1/2}\fs
\eeq

\vskip3mm

\noindent
\underline{Interval II} \ [$\discK_B$ is imaginary] \nopagebreak

\vskip2mm
\noindent 
The discontinuity is evaluated as in interval I, with $\hat B=-i\bar B$.

\vskip3mm

\noindent
\underline{Interval III} \ [$\discK_B$ is complex] \nopagebreak

\vskip2mm
\noindent
In this interval, one uses $\hat B=-\bar B$, and
\beq\label{eq:GFI}
\hat D_B(s)= i\pi G(\hat A,\hat B;W_+,W_-;M_\pi)+F(\hat A,\hat B;W_+,W_-;M_\pi)\scs\quad
W_\pm= c_\pm\left|1-\frac{4M_{\pi}^2}{\hat A\pm \hat B}\right|^{1/2}\scs
\eeq
where
\beq\label{eq:tableI}
\begin{tabular}{lccc}
\word{Interval}&\word{IIIa}&\word{IIIb}&\word{IIIc}\\[2mm]
$c_+$&1&$-i$&$-1$\\[2mm]
$c_-$&1&1&1
\end{tabular}
\eeq

\subsection{Holomorphic properties of $\bar {\cal K}_B$}
\begin{figure}
\centering
\includegraphics[width=0.45\linewidth]{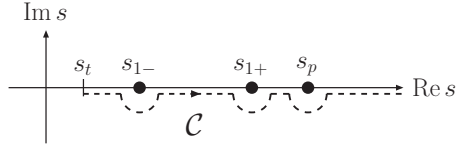}
\caption{The path ${\cal C}$  in the dispersive integral 
Eq.~\eqref{eq:dispKBC}, with mass assignments as displayed in
  Eq.~\eqref{eq:unequalI}, for physical kaon mass. The symbols $s_t$, $s_{1-}$, $s_{1+}$, and $s_p$ denote threshold,
  two anomalous thresholds, and the pseudothreshold, respectively, and are given
  explicitly in Eqs.~\eqref{eq:s1Fig6B}, \eqref{eq:s1pm}.}
\label{fig:pathuneq}
\end{figure}

The discontinuity $\discK_B$ develops a singularity of the
square root type at $s=s_{1\pm}$. The path of integration in the 
dispersive representation Eq.~\eqref{eq:dispKBC} 
 is indicated in Fig.~\ref{fig:pathuneq}. 
 The function $\bar{\cal K}_B$ is holomorphic in the complex $s$-plane, 
 cut along the positive real axis
for $s\geq (M_{\pi^0}+M_\pi)^2$. It is as well holomorphic at the upper rim 
of the cut, and diverges at the pseudothreshold $s=s_p$
when approached from below the cut. Further, at $s=s_{1\pm}$, 
it has singularities of
the square root type on the lower rim of the cut. Its precise behavior there is indicated in Eq.~\eqref{eq:branchpoints}.

\subsection{Case II}\label{subsec:caseII}
\begin{figure}
\centering
\includegraphics[width=0.45\linewidth]{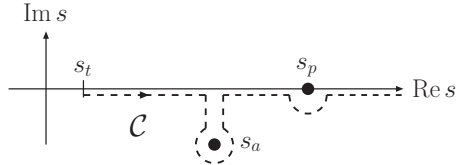}
\caption{The path ${\cal C}$  in the dispersive integral 
Eq.~\eqref{eq:dispKBC}, with mass assignments as displayed in
Eq.~\eqref{eq:unequalII}, for physical kaon mass. The symbols $s_t$, $s_a$, and $s_p$ 
are defined in Eqs.~\eqref{eq:sts1spsc} and \eqref{eq:thresholdslandau}.}
\label{fig:pathanom}
\end{figure}

Finally, we consider the case where the pions in the inner loop are neutral
 and all the others charged,
\beq\label{eq:unequalII}
M_a=M_b=M_{\pi^0}\,,\quad M_c=M_d=M_\pi\,,\quad q_3^2=M_\pi^2\scs
\eeq
see Figs.~\ref{fig:2loop_text}A, \ref{fig:analytic}C.
This case is
the most intriguing one. 
\begin{figure}
\centering
\includegraphics[width=0.45\linewidth]{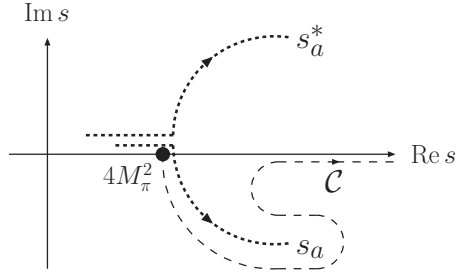}
\caption{The singularities $s_a, s_a^*$ 
in the discontinuity $\discK_C(s)$, for the mass
  assignments Eq.~\eqref{eq:unequalII}, as a function of the kaon mass 
(dotted lines). The symbol $s_a$ is defined in 
  Eq.~\eqref{eq:thresholdslandau}, $s_a^*$ denotes its complex conjugate, and
  the arrows indicate the direction of increasing kaon masses. The singularity $s_a$ 
 intrudes into the original path of integration, such that
  ${\cal C}$ must be deformed [dashed line], see also Fig.~\ref{fig:pathanom}.}
\label{fig:singanom}
\end{figure}
Performing the angular integral in Eq.~\eqref{eq:angularintegral} and investigating the singularities as before, 
we find that the ones in the pertinent square roots $\hat W_\pm$ are given by the polynomial $P_{\ref{fig:analytic}C}$ 
displayed in Eq.~\eqref{eq:polynomialFig6ACL}.
We display the locations of the singularities  $s_a$, $s_a^*$ as a function of the kaon mass in Fig.~\ref{fig:singanom} [dotted lines]. 
We see that, as
the kaon mass is increased, the singularity $s=s_a$  intrudes into the original
path of integration. As a result, the path ${\cal C}$ must be modified [dashed line], 
see also Fig.~\ref{fig:pathanom}. 

\subsection{Holomorphic properties of $\bar {\cal K}_C$}
From the above discussion, we conclude that $\bar{\cal  K}_C(s)$ 
is holomorphic in the complex $s$-plane, cut along the path ${\cal C}$ displayed in Fig.~\ref{fig:pathanom}. It is holomorphic at the upper rim of the cut, and diverges at the pseudothreshold $s_p$
when approached from below. Further, at $s=s_a$, it develops a singularity of
the square root type on the lower rim of the cut. Its precise behavior there is the one indicated in Eq.~\eqref{eq:branchpoints}. The singularity at $s_a^*$ is located on the second Riemann sheet.

\section{Leading Landau singularities}
\label{app:landau}

\begin{figure}
\centering
\includegraphics[width=4.5cm]{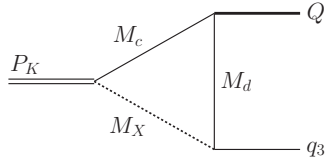}
\caption{The triangle graph with general masses. The positions of its leading Landau singularities agree 
with those of the two-loop graphs in Fig.~\ref{fig:2loop_text}A for $M_X=M_a+M_b$, for general $q_3^2$. 
The notation is   $P^\mu_K=(Q+q_3)^\mu$, $Q^2=s$, $P_K^2=M_K^2$.}
\label{fig:triangle2}
\end{figure}

Some of the graphs that correspond to Fig.~\ref{fig:2loop_text}A 
have the property that they generate singularities that correspond to the solutions of 
the  pertinent leading-order Landau equations. These singularities 
occur because the kaon is unstable. 
The positions of these singularities are given by the two solutions of a
polynomial equation of second order in $s$~\cite{passeraetal},
\beq\label{eq:PL}
P_L(s,q_3^2,M_K^2;M_a,M_b,M_c,M_d)=0\scs
\eeq
with 
\begin{align}\label{eq:polynomialPL}
P_L&=\lambda(s, M_d^2,M_c^2)\lambda\big(q_3^2,(M_a+M_b)^2,M_d^2\big)
-Q^2(s,q_3^2,M_K^2;M_a,M_b,M_c,M_d)\scs\nnnl
Q&=2M_d^2(M_K^2-s-q_3^2)+\Big(q_3^2-(M_a+M_b)^2+M_d^2\Big)(s+M_d^2-M_c^2)\fs
\end{align}
For the graphs Figs.\ref{fig:analytic}A--C,
this polynomial becomes
\begin{align}\label{eq:polynomialFig6ACL}
P_{\ref{fig:analytic}A}&=P_L(s,M_\pi^2,M_K^2;M_\pi,M_\pi,M_\pi,M_\pi) \hspace*{-1.4cm}
            &&=-16M_\pi^4\left[s-s_1\right]^2\scs \nnnl
P_{\ref{fig:analytic}B}&=P_L(s,M_{\pi^0}^2,M_K^2;M_\pi,M_\pi,M_\pi,M_{\pi^0})\hspace*{-1.4cm}
            &&=-16M_\pi^2M_{\pi^0}^2\left[s-s_{1+}\right]\left[s-s_{1-}\right]\scs \nnnl
P_{\ref{fig:analytic}C}&=P_L(s,M_{\pi}^2,M_K^2;M_{\pi^0},M_{\pi^0},M_\pi,M_{\pi}) \hspace*{-1.4cm}
            &&=-16M_\pi^2M_{\pi^0}^2\left[s-s_a\right]\left[s-{s^*_a}\right]\fs
\end{align}
Here, ${s^*_a}$ denotes the complex conjugate of $s_a$.
The thresholds are
\begin{align}\label{eq:thresholdslandau}
s_1&=\frac{1}{2}(M_K^2-M_\pi^2)\scs\nnnl
s_{1\pm}&= \frac{1}{2}\left[M_K^2+2M_{\pi^0}^2-3M_\pi^2\right]
\pm\frac{1}{2M_\pi}
\left[(M_{\pi}^2-M_{\pi^0}^2)\lambda(M_K^2,M_\pi^2,4M_\pi^2)\right]^{1/2}\scs\nnnl
s_a&=\frac{1}{2}\left[M_K^2+3M_\pi^2-4M_{\pi^0}^2\right] 
-\frac{i}{2M_{\pi^0}}\left[(M_\pi^2-M_{\pi^0}^2)\lambda\big(M_K^2,M_\pi^2,4M_{\pi^0}^2\big)\right]^{1/2}\fs
\end{align}
The positions of the leading Landau singularities of the triangle 
graph displayed in Fig.~\ref{fig:triangle2}
are obtained from a polynomial equation $P_T(s,q_3^2,M_K^2;M_X,M_c,M_d)=0$~\cite{ELOP}.  
It is interesting to note that, for  $M_X=M_a+M_b$, the 
polynomials $P_L$ and $P_T$ are proportional to each other, and the positions of the leading Landau singularities 
in the two-loop graphs Fig.~\ref{fig:2loop_text}A and the 
triangle graph Fig.~\ref{fig:triangle2} thus coincide.
This corroborates the statement made in Ref.~\cite[Section~3]{barton1} 
on the origin of the singularities in the graph~\ref{fig:analytic}A. 

The discussions in the previous appendices 
show that the discontinuities of the graphs in Fig.~\ref{fig:analytic}B,~C do have a singularity at $s_{1\pm}$, $s_a$, $s_a^*$, 
while the one of Fig.~\ref{fig:analytic}A is holomorphic at $s_1$. 
As a result of this, the pertinent loop functions generate square root singularities at $s_{1\pm}$, $s_a$ on the lower rim of the cut.

\section{Relativistic vs. non-relativistic approach}
\label{app:compare_nrrel}

As already mentioned in \ref{app:relequal}, the
non-relativistic calculations
should reproduce the relativistic results graph by graph in a systematic
expansion in $\epsilon$ -- otherwise, the whole framework does not make
sense.
In other words, it should be possible to adjust the couplings in the
non-relativistic Lagrangian at order $\epsilon^n$ so that the
relativistic expression of any graph, expanded in powers of $\epsilon$,
up to and including $\Order(\epsilon^n)$
would be reproduced by a uniquely identifiable set of non-relativistic
graphs.

In this Appendix, we check this equivalence for the relativistic
two-loop graph, considered in \ref{app:relequal}. To this end,
we first compare the expression for the non-relativistic discontinuity,
Eq.~\eqref{eq:DiscM},
to the expression obtained in the relativistic theory.
For simplicity,
we restrict ourselves to the case $M_a=M_b=M_c=M_d=M_3=M_\pi$.
The comparison is carried out in the threshold 
region IIIb defined in Eq.~(\ref{eq:regions}) where the 
non-relativistic expression has been originally derived.
The discontinuity in the whole non-relativistic region is obtained by 
using analytic continuation.

The quantities $A$ and $\bar B$, which were defined
in the relativistic theory, see Eqs.~(\ref{eq:angular}) and (\ref{eq:barB}),
are related to the quantities $a$, $b$, $q_0$ through
\beq
a=\frac{4M_\pi^2}{\kappa}\, ,
\quad b=\frac{A}{\kappa}\, ,\quad q_0=\frac{\bar B}{\kappa}\, ,\quad \kappa=\frac{2M_K|{\bf Q}|}{\sqrt{s}}\, .
\eeq
From the above equation we find (cf.\ Eq.~\eqref{eq:discI}))
\beq
\biggl(1-\frac{a}{b\pm q_0}\biggr)^{1/2}=\biggl(1-\frac{4M_\pi^2}{A\pm \bar B}\biggr)^{1/2}=\bar W_\pm 
\eeq
and
\begin{align}\label{eq:phiphi}
\Phi(q_0)-\Phi(-q_0)
&=\frac{\sqrt{s}}{2M_K|{\bf Q}|}\,
\Bigl(2M_\pi^2(\bar R_--\bar R_+)
+(A+\bar B)\bar W_+-(A-\bar B)\bar W_- \Bigr)\, ,
\nnnl
\bar R_\pm&=\ln(1+\bar W_\pm)-\ln(1-\bar W_\pm)\, .
\end{align}
The final expression for the non-relativistic discontinuity is obtained
by substituting Eq.~(\ref{eq:phiphi}) into Eq.~(\ref{eq:DiscM}).
It is seen that  this expression differs from the pertinent relativistic
expression given by Eqs.~\eqref{eq:discI}, \eqref{eq:barB}, 
\eqref{eq:discequal} by 
\beq\label{eq:freedom}
\mbox{Disc}_s(s)\bigr|_{Rel}-\mbox{Disc}_s(s)\bigr|_{NR}=2iv(s)P(s)\, ,
\quad
P(s)=\frac{1}{512\pi^3}\,\biggl(6-\frac{F}{\bar B}\biggr) \, ,
\eeq
where $v(s)=(1-4M_\pi^2/s)^{1/2}$, and $F$ is defined in Eq.~\eqref{eq:PhiFGdef}.
One may ensure that the quantity $P(s)$ is a low-energy polynomial with real coefficients. 
Consequently, the relativistic and non-relativistic amplitudes themselves
differ by $iv(s)P(s)+P'(s)$, where $P'(s)$ is another low-energy 
polynomial.

\begin{figure}
\centering
\includegraphics[width=0.5\linewidth]{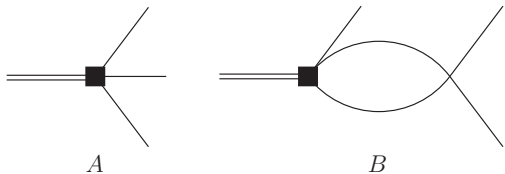}
\caption{Non-relativistic diagrams that remove the right-hand side of Eq.~\eqref{eq:freedom}.
The thick squares stand for the counterterms in the non-relativistic $K\to 3\pi$ Lagrangian at order $a$.}
\label{fig:equivalence}
\end{figure}

In order to establish the origin of the above polynomials, let us first compare
the relativistic and non-relativistic one-loop integrals in the vicinity 
of the threshold. As seen from Eq.~(\ref{eq:Fs}), only the imaginary part of
the one-loop integral
above threshold survives in the non-relativistic approach
(the expression below threshold is obtained through analytic continuation).
This corresponds to the following replacement in the expression of the one-loop
integral $\bar J(s)$ above threshold 
\beq
v(s)\ln\frac{v(s)-1}{v(s)+1}+2 
=v(s)\ln\frac{1-v(s)}{1+v(s)}+2+i\pi v(s) ~\rightarrow~ i\pi v(s)\, .
\eeq
The dropped part is a low-energy polynomial
with real coefficients (in the whole non-relativistic region)
and corresponds to adding the counterterms to the $K\to 3\pi$ 
vertex shown in Fig.~\ref{fig:equivalence}A.
The numerical values of these counterterms are fixed at order $a$.

It can be straightforwardly checked that inserting the vertices 
with the counterterms inside
the pion loop, as shown in Fig.~\ref{fig:equivalence}B,
and evaluating this diagram in the non-relativistic theory
yields exactly the result $iv(s)P(s)$.
Consequently, this term takes 
into account the renormalization in the non-relativistic theory at $\Order(a)$.
Finally, the polynomial
$P'(s)$ corresponds to the renormalization of the $K\to 3\pi$ vertices at $\Order(a^2)$.
At the order we are working, only tree diagrams with these counterterms contribute.
  
To summarize, we have demonstrated that all differences between the relativistic
and non-relativistic amplitudes in the low-energy region 
can be removed by changing the renormalization prescription in the
 $K\to 3\pi$ vertices.
Consequently, relativistic and non-relativistic theories are physically equivalent at this order.
Obviously, this also holds  for the expansion of the pertinent
Feynman graphs in powers of $\epsilon$.

\section{The decomposition ${\cal M}={\cal M}_0 + v_{cd}{\cal M}_1$}
\label{app:decomposition}
One of the crucial ingredients in the 
work of Cabibbo and Isidori~\cite{cabibboisidori} is the 
decomposition of the decay amplitudes ${\cal M}$ into the form
\beq\label{eq:decompCI}
{\cal M}={\cal M}_0+v_{cd}(s){\cal M}_1\fs
\eeq
The amplitudes ${\cal M}_{0,1}$ are assumed 
to be analytic in the physical 
region of the decays, with square root singularities at 
the border of the Dalitz plot,
  associated with different $\pi\pi$ thresholds~\cite{cabibbo}.

In this Appendix, we comment on this decomposition, and start the discussion
with a simpler case, the relativistic one-loop integral
 \beq
J_{cd}(s)=\bla\frac{1}{M_c^2-l^2}\frac{1}{M_d^2-(l-P)^2}\bra_l\,, \quad s=P^2\fs
\eeq
Let
\beq
\bar J_{cd}(s)=J_{cd}(s)-J_{cd}(0)\fs
\eeq
The function $\bar J_{cd}$ is analytic in the complex $s$-plane, cut
along the real axis for $s\geq (M_c+M_d)^2$. Its explicit form  near the 
threshold $s=(M_c+M_d)^2$  is indeed of the form Eq.~\eqref{eq:decompCI}. 
Define ${\cal M}_0$ through 
\beq\label{eq:oneloop}
\bar J_{cd}(s) = {\cal M}_0+\frac{i\,v_{cd}(s)}{16\pi}\fs
\eeq
It then follows that ${\cal M}_0$ has no right hand cut, because the term ${i\,v_{cd}(s)}/{16\pi}$ 
reproduces the discontinuity of $\bar J(s)$ at $s>(M_c+M_d)^2$. On the other hand, 
this term develops singularities at
the pseudothreshold $s=(M_c-M_d)^2$, and at $s=0$.
 Because $\bar J_{cd}$ is analytic there, the
amplitude ${\cal M}_0$ develops the same singularities, with opposite sign,
 such that the
sum is regular at these points. Since $s=0$, $s=(M_c-M_d)^2$ are outside the
kinematic region of interest in the present case, these singularities
  do not matter, and one concludes that $\bar J_{cd}(s)$ indeed has a decomposition 
Eq.~\eqref{eq:decompCI} with amplitudes ${\cal M}_{0,1}$ that  enjoy
the properties proposed in Ref.~\cite{cabibbo}.

For the two-loop graphs discussed above, the situation is more complex.
Let us first discuss the situation  in case of the function $H(s)$ introduced 
in Eq.~\eqref{eq:Hpmanalytic}, with mass assignments that correspond to the 
diagram Fig.~\ref{fig:analytic}B. We start from the discontinuity
\beq \label{eq:discH}
\Delta H(s)\doteq H_+(s)-H_-(s)=-\frac{4\pi}{3}\sqrt{-y_1}(y_1+3y_2)\,,\quad s>(M_\pi+M_{\pi^0})^2\scs
\eeq
worked out in Eq.~\eqref{eq:Hpmanalytic}. 
We note that $y_1=-v_{+0}^2f_1(v_{+0}^2), y_2=-f_2(v_{+0}^2)$ near threshold, with $f_i(z)$ 
 holomorphic at $z=0$, and $f_i(0)>0$.
Now define the amplitude $H_0(s)$ through
\beq 
H(s)=\frac{1}{2}\Delta H(s)+H_0(s)\,,\quad s>(M_{\pi^0}+M_\pi)^2\fs
\eeq
Evaluating the discontinuity on both sides, it is seen that $H_0$ is holomorphic 
near threshold. Therefore,
\beq
H(s)=H_0(s)+v_{+0}H_1(s)\,,\quad H_1=\frac{1}{2v_{+0}}\Delta H(s)\scs
\eeq
with $H_{0,1}$ holomorphic at threshold. However, as is shown in  
\ref{app:subsectionpinch}, $\Delta H$ is singular at $s=s_{1\pm}$. Because $H(s)$ is regular there, $H_0$ must be singular as well. 
A completely analogous argument may be used to discuss the inadequacy of the decomposition~\eqref{eq:decompCI} 
in case of the relativistic integral discussed in \ref{subsec:caseI}. 
Numerical values for the positions of these singularities are provided in Table~\ref{tab:LLS}.

\begin{figure}
\centering
\includegraphics[width=0.45\linewidth]{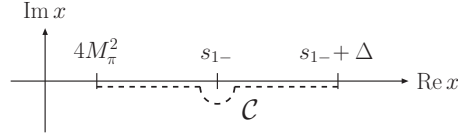}
\caption{The path ${\cal C}$ along which the integral is taken in Eq.~\eqref{eq:decompdisp}.}\label{fig:pathC}
\end{figure}

We find it instructive to display the singular behavior of ${\cal M}_{0,1}$
in a fully explicit manner, in a simplified example inspired by the relativistic integrals discussed in \ref{app:reldiff}. 
Consider the dispersive integral
\beq\label{eq:decompdisp}
{I}(s)=\frac{1}{\Delta}\int^{{\hspace*{-4.2mm}}{ \cal C}}\hspace{2mm}\frac{dx \sqrt{x-4M_\pi^2}\sqrt{s_{1-}-x}}{x-s}\scs 
\quad s\not\in{\cal C}\fs
\eeq
Here, $\sqrt{x-4M_\pi^2}$ stands for the standard phase space factor, $\sqrt{s_{1-}-x}$ is a substitute 
for the square-root singularity in the discontinuity at $s=s_{1-}$, and $\Delta$ denotes a  (mass-scale)$^2$ 
specified below.  The path ${\cal C}$ is indicated in Fig.~\ref{fig:pathC}, compare with Fig.~\ref{fig:pathuneq}.

The function  $I(s)$ is holomorphic in the 
complex $s$-plane, cut along the real axis for $s\in [4M_\pi^2,s_{1-}+\Delta]$. 
Further, $I(s)$ can be holomorphically continued through the upper rim of the cut, 
because the singularity at $s=s_{1-}$ is not effective there -- the path ${\cal C}$ can be suitably deformed. 
On the other hand, approaching the cut from below, it is seen that ${\cal C}$ cannot be deformed any more at $s=s_{1-}$, 
as a result of which one encounters a (pinch) singularity there. 

To perform the integration, we set $\Delta=s_{1-}-4M_\pi^2$, $x=4M_\pi^2+\Delta y$,
\beq\label{eq:functionK(z)}
I(s)= G(z)\doteq \lim_{\eps\to 0^+}\int_0^2\hspace{2mm}\frac{dy\,\sqrt{y}\sqrt{1-y+i\eps}}
{y-z}\,,\quad z=\frac{s-4M_\pi^2}{\Delta}\fs
\eeq
We find
\beq
G(z)= \int_0^1\frac{dy\,\sqrt{y}\sqrt{1-y}}{y-z}+i\int_1^2\frac{dy\sqrt{y}\sqrt{y-1}}{y-z} 
=G_1(z)+iG_2(z)\scs
\eeq
with
\begin{align}
G_1&=i \pi\sqrt{z}\sqrt{1-z}+ p_1(z)\scs &
G_2&=\sqrt{z}\sqrt{1-z}\arctan{\frac{4\sqrt{z}\sqrt{1-z}}{(3z-2)\sqrt{2}}}+p_2(z)\scs\nnnl
p_1&=-\pi\bigg(z-\frac{1}{2}\bigg)\scs &  p_2&=\sqrt{2}+\ln\Big(3+2\sqrt{2}\Big)\bigg(z-\frac{1}{2}\bigg)\scs
\end{align}
where $z$ is located on the upper rim of the cut, near $z=0$.  One concludes that
\beq\label{eq:sumK}
I(s)=i\sqrt{s-4M_\pi^2}I_1(s)+I_0(s)\scs \quad
I_1=\frac{\pi}{\Delta}\sqrt{s_{1-}-s}\scs \quad I_0=p_1(z)+i G_2(z)\scs
\eeq
with $I_{0,1}$ holomorphic at threshold. This is indeed a decomposition of the 
type Eq.~\eqref{eq:decompCI}.
 Note, however,
 that $I_1$  develops a square root singularity at $s=s_{1-}$.
 As was said above, $I(s)$ is holomorphic on the upper rim of the cut. 
Therefore, $I_0$ develops a singularity at $s=s_{1-}$ as well, canceling the one generated by $I_1$. 
Indeed, near $z=1$, one has
\beq
G_2=\sqrt{z}\sqrt{1-z}\,\bigg[\arctan{\frac{4\sqrt{z}\sqrt{1-z}}{(3z-2)\sqrt{2}}}-\pi\bigg]+p_2(z)\scs
\eeq
as a result of which the singularities  in $I_{0,1}$ cancel in the sum \eqref{eq:sumK}. 
On the other hand, on the lower rim of the cut, near $s=s_{1-}$, the singularities add, 
and $I(s)$ shows a square root behavior.

\end{appendix}

\end{document}